\documentclass[fleqn,10pt]{wlscirep}
\usepackage[utf8]{inputenc}
\usepackage[T1]{fontenc}
\usepackage{graphics,graphicx,pgf}
\usepackage{amsmath,amsthm,amssymb}
\usepackage{url}
\usepackage{rotating}
\usepackage{tfrupee}
\usepackage{float}
\usepackage{sidecap}

\definecolor{mygreen}{HTML}{008000}
\definecolor{myorange}{HTML}{FFA500}

\title{Mitigating Financial Risk from Climate-Induced Agricultural Price Volatility}

\author[1]{Sourish Das}
\author[2,3]{Sudeep Shukla}
\author[4]{Abbinav Sankar Kailasam}
\author[1]{Anish Rai}
\author[3]{Sejal Garg}
\author[5,*]{Anirban Chakraborti}

\affil[1]{Chennai Mathematical Institute, Chennai-603103, Tamil Nadu, India}
\affil[2]{AI 4 Water LTD, Orpington, BR6 9QX United Kingdom}
\affil[3]{Institute for Interdisciplinary Research, New Delhi-110074, India}
\affil[4]{University College London, Gower Street, London WC1E 6BT, United Kingdom}
\affil[5]{School of Computational \& Integrative Sciences, Jawaharlal Nehru University, New Delhi-110067, India}
\affil[*]{anirban@jnu.ac.in}

\begin{abstract}
Agricultural price volatility, driven by market dynamics and meteorological factors such as temperature and precipitation, poses challenges for sustainable finance, planning, and policy. This study analyzes the impact of climate on crop price volatility for soybean in Madhya Pradesh (India) and Illinois (US), rice in Assam (India), wheat in North Dakota (US), cotton in Gujarat (India), and corn in Iowa (US). Using CMIP6 climate projections from the Copernicus Climate Change Service, we examine historical climate patterns and evaluate two future scenarios: SSP2-4.5 (moderate) and SSP5-8.5 (severe). We estimate conditional price volatility using the Exponential Generalized Autoregressive Conditional Heteroskedasticity (EGARCH) model, and forecast this volatility with a Seasonal Autoregressive Integrated Moving Average with Exogenous Regressors (SARIMAX) model that incorporates meteorological variables. Finally, we apply the Black-Scholes framework to evaluate the cost of put-option–based insurance, which provides protection to farmers against adverse price drops linked to climate change. Our results highlight the role of meteorological data in improving agricultural risk modelling, enabling better design of insurance mechanisms, price stabilization tools, and sustainable policy interventions under climate uncertainty.
\end{abstract}
\begin{document}

\flushbottom
\maketitle

\section{Introduction}

Complex systems are characterized by non-linear interactions, dynamic feedback loops, and sensitivity to external perturbations, features that give rise to collective behavior and emergent macroscopic patterns across physical, biological, and social settings\cite{PARISI1999557,parisi2007physics,kwapien2012physical,chakrabarti2023data,rind1999complexity}. In socio-economic contexts, such as financial markets and interacting multi-agent systems, such complexity manifests through correlated fluctuations, volatility clustering, and contagion-like spreading, phenomena long recognized in quantitative finance and socio-physics \cite{mandelbrot1997variation,bartram2005another,sociophysics2014,Brockmann_2013}. Environmental and climate systems provide particularly salient examples, where extreme events and shocks can cascade across tightly coupled agricultural, economic, and institutional networks, amplifying systemic risk and long-term vulnerability \cite{Pacelli2023SystemicRisk,chakraborti2020emerging,chakraborti2020phase}. Recent empirical studies show that climate variability induces correlated risks in production, revenues, and financial stability in agricultural systems, underscoring the need for data-driven, complexity-aware frameworks to assess climate risk\cite{BurneyMcIntosh2024}. 

Agriculture is a cornerstone of the global economy, supporting millions of livelihoods and driving both national and international trade \cite{Agricultural,Carleton2017,Burney2014}. The relationship between climate and agriculture represents one such complex system: changes in temperature, precipitation, and humidity influence crop productivity, which in turn affects market prices, farmer incomes, and policy responses~\cite{kumar2011impact,Kumar2025Impact}. Certain crops are particularly sensitive to these variations due to their regional dependence and economic significance. In India, rice in Assam remains central to food security, soybean in Madhya Pradesh is a major oilseed and export crop, and cotton in Gujarat underpins the textile sector. In the United States, soybean in Illinois, wheat in North Dakota, and corn in Iowa represent key staples whose production sustains global markets. Shifts in weather events can disrupt these systems, straining farmer livelihoods, destabilizing supply chains, and amplifying volatility in interconnected markets.

The unpredictability of climate patterns has made price volatility an increasing concern for farmers, policymakers, and investors alike~\cite{mall2006impact,mehta2019climate,singh2020farmers}. Farmers are particularly vulnerable when market prices fall sharply, as such downturns can erode incomes and threaten livelihoods. One way to mitigate this risk is through price insurance mechanisms that provide compensation when market prices drop below a predetermined level. From an economic perspective, crop price insurance can be represented through the framework of a European put option, where farmers secure protection against unfavorable price movements by setting a predetermined strike price. Such option-based instruments are widely used in developed economies to hedge against commodity risk, but their adoption in agriculture, particularly in emerging markets, remains limited. The absence of accessible and well-functioning agricultural derivatives restricts farmers’ ability to manage uncertainty, leaving them more vulnerable to climate-induced shocks. Developing reliable methods to quantify the cost of such protection is therefore essential for designing effective insurance products and strengthening the resilience of agricultural systems.

\begin{figure}
    \centering
    \includegraphics[width=0.75\textwidth]{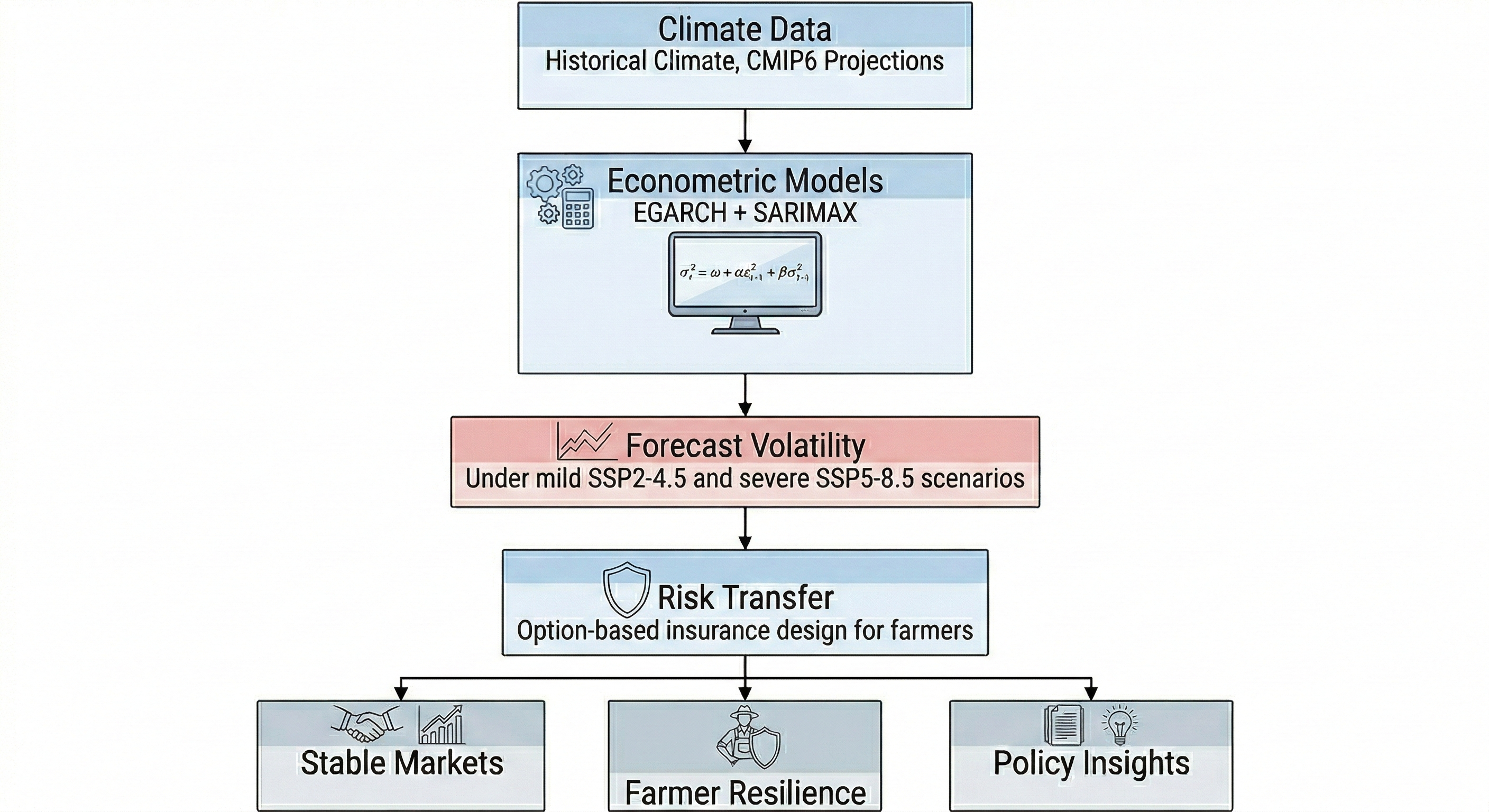}
    \caption{\textbf{Modeling pipeline linking climate variability to agricultural price volatility and risk mitigation.}
    The schematic illustrates the sequential framework used in this study, beginning with historical climate observations and CMIP6 projections, followed by econometric modeling that couples weather variables to price dynamics using EGARCH-SARIMAX formulations. Volatility is then forecast under alternative climate scenarios, enabling the design of option-based risk transfer instruments for agricultural stakeholders. The final stage highlights downstream implications for market stability, farmer resilience, and policy-relevant insights, emphasizing how climate-driven shocks propagate across interconnected environmental, economic, and institutional systems.}
\label{fig:schematic_methodology}
\end{figure}

Due to the increasing variations in climate patterns and their adverse effect on crop price volatility, it is essential to integrate climate data with financial models to assess future risks. Traditional econometric approaches often fail to capture the non-linear and asymmetric effects of extreme weather events on agricultural markets. To address this gap, our study employs the Exponential Generalized Autoregressive Conditional Heteroskedasticity (EGARCH) model~\cite{Bollerslev1986,Nelson1991} to estimate conditional volatility in crop price returns. With estimated conditional volatility as the dependent variable and meteorological data as exogenous predictors, we further apply the Seasonal Autoregressive Integrated Moving Average with Exogenous Regressors (SARIMAX) model~\cite{shumway2017time} to forecast volatility under climate scenarios. The predicted volatility is then used as an input into the Black–Scholes framework~\cite{BlackScholes1973,merton73,Hull2017} to evaluate the premium of put-option–based crop price insurance. By incorporating high-resolution climate datasets from the Copernicus Climate Change Service, we analyze historical patterns and future projections (SSP2-4.5 and SSP5-8.5) to assess and model the potential risks faced by agricultural markets. Figure \ref{fig:schematic_methodology} illustrates the methodology pipeline used to assess agricultural climate risk.

The contribution of this study is twofold. First, it demonstrates how integrating climate variables with advanced econometric models improves the prediction of agricultural price volatility. Second, it shows how these forecasts can be directly linked to the valuation of option-based insurance products, offering a transparent way to quantify the costs of protecting farmers against adverse price movements. By applying this framework to both India and the United States, we highlight common vulnerabilities across distinct agricultural systems while also revealing regional differences. Our findings contribute to the broader goal of designing sustainable risk-management strategies, securing farmer incomes, and strengthening the resilience of agricultural markets in an era of increasing climate uncertainty.

\section*{Data Description}

\subsection*{Crop Price Data}
\begin{figure*}
    \centering
     \includegraphics[width=0.3\linewidth]{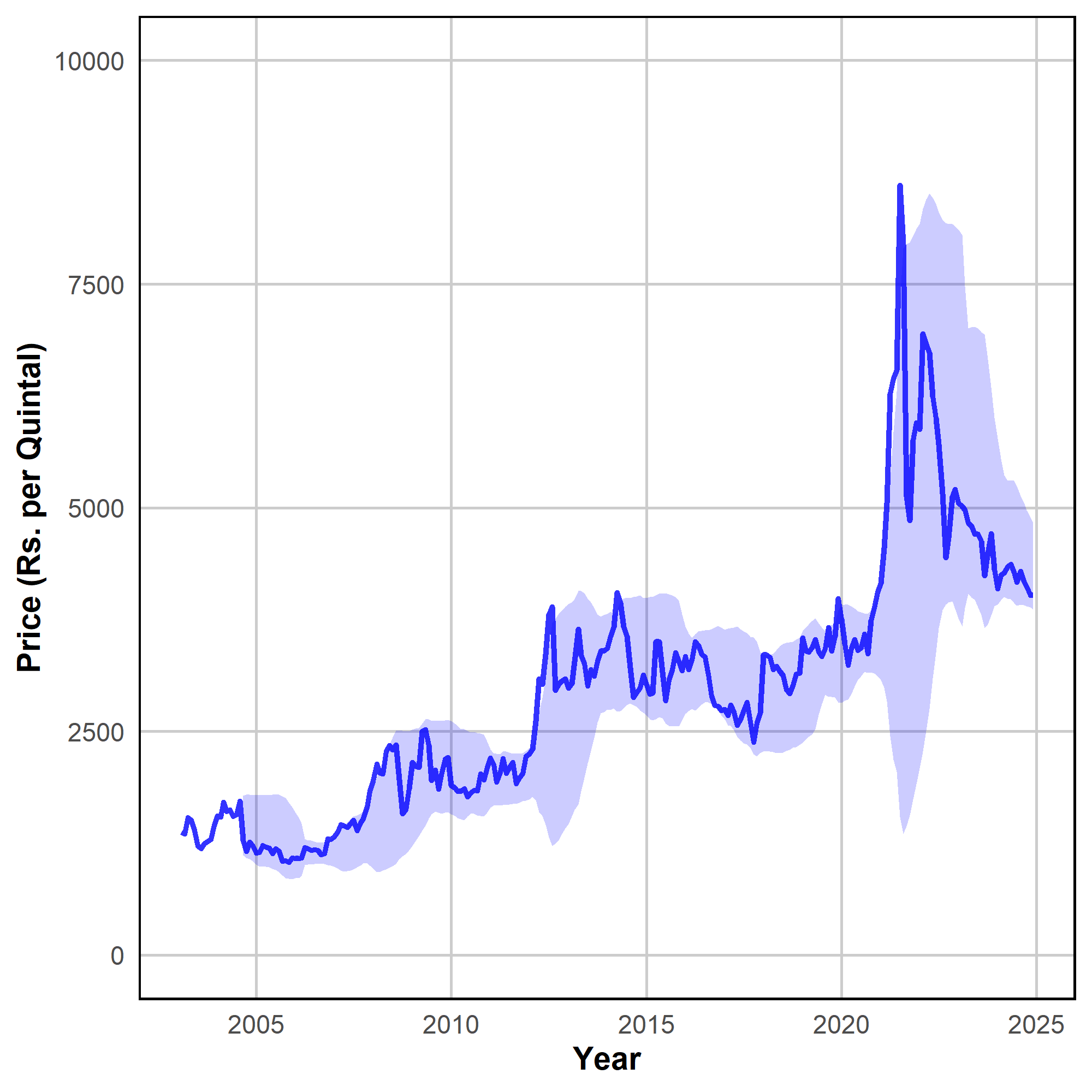}\llap{\parbox[b]{2.0in}{{\large\textsf{A}}\\\rule{0ex}{2.0in}}}
     \includegraphics[width=0.3\linewidth]{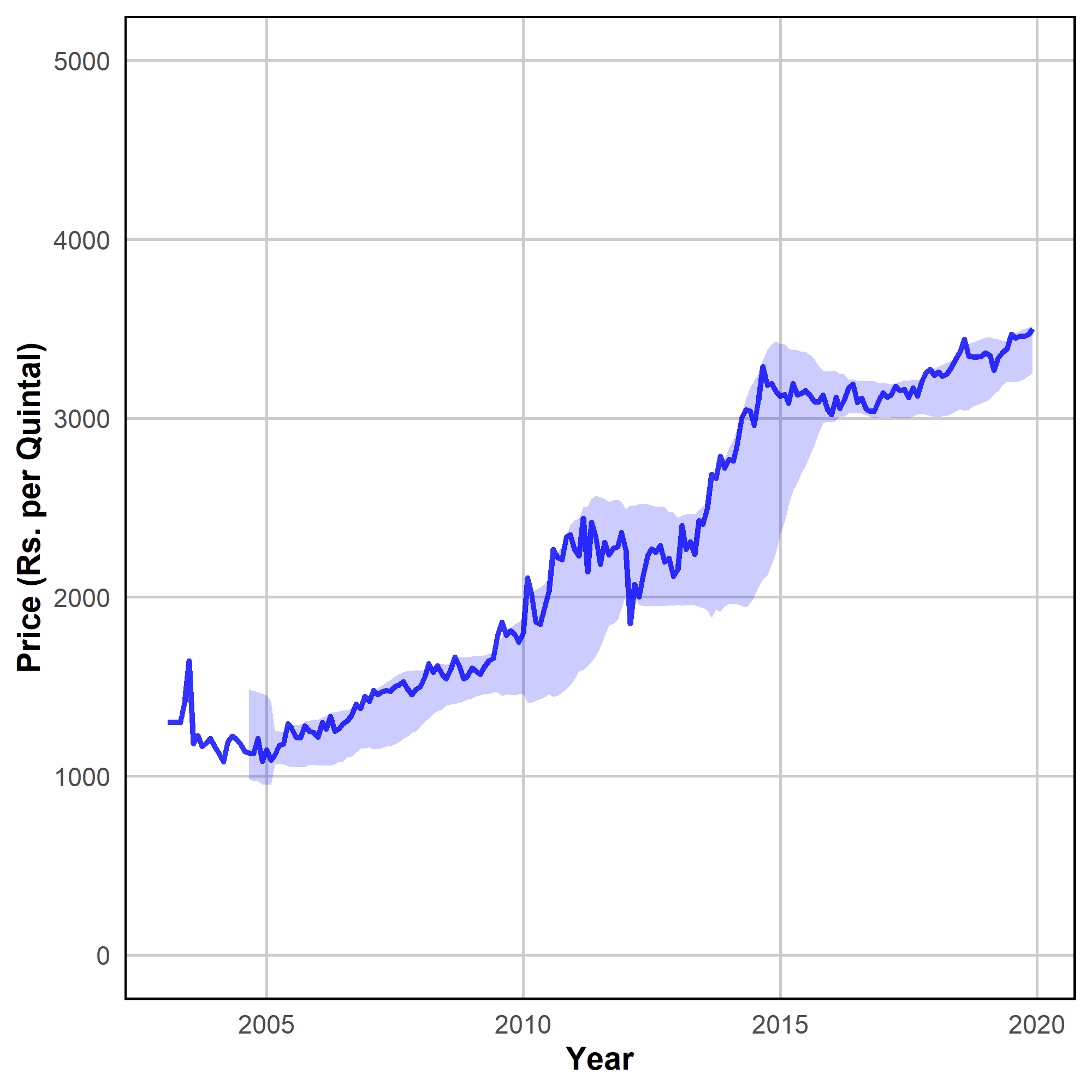}\llap{\parbox[b]{2.0in}{{\large\textsf{B}}\\\rule{0ex}{2.0in}}}
     \includegraphics[width=0.3\linewidth]{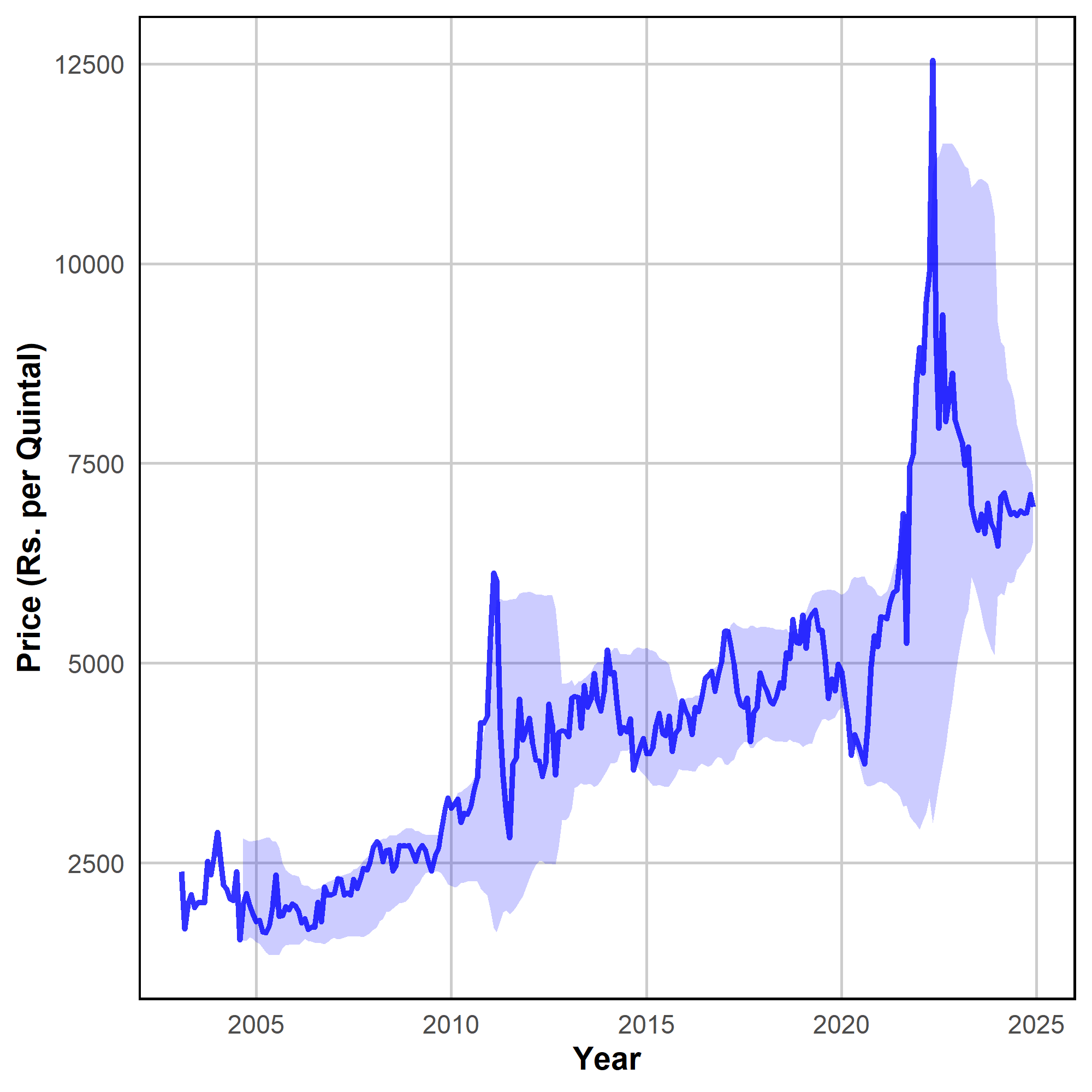}\llap{\parbox[b]{2.0in}{{\large\textsf{C}}\\\rule{0ex}{2.0in}}}
     \caption{\textbf{Plots of price time series, for \textit{(A)} soybean in Madhya Pradesh, \textit{(B)} rice in Assam, and \textit{(C)} cotton in Gujarat.} The blue line indicates the crop prices while the Bollinger Band (blue shaded region) represents the 20-month moving average $\pm 2$  standard deviations.}
    \label{fig:india_price_subplots}
\end{figure*}

The Government of India's Directorate of Marketing and Inspection operates the AGMARKNET portal to establish a nation-wide information network for disseminating agricultural market information, price related information, and infrastructure related information\cite{agmarknet_aboutus}. Real time agricultural market data is accessible to the public through the AGMARKNET portal which provides information on commodity prices and arrivals across India. We retrieved district level monthly crop price data for soybean in  Madhya Pradesh,  rice in Assam, and cotton in Gujarat, for the period February 2003 to December 2024. Using this data, we calculated the monthly mean price for each crop-state combination. Figure~\ref{fig:india_price_subplots} shows the crop price data for each crop-state combination for India. 

\begin{figure*}
    \centering
     \includegraphics[width=0.3\linewidth]{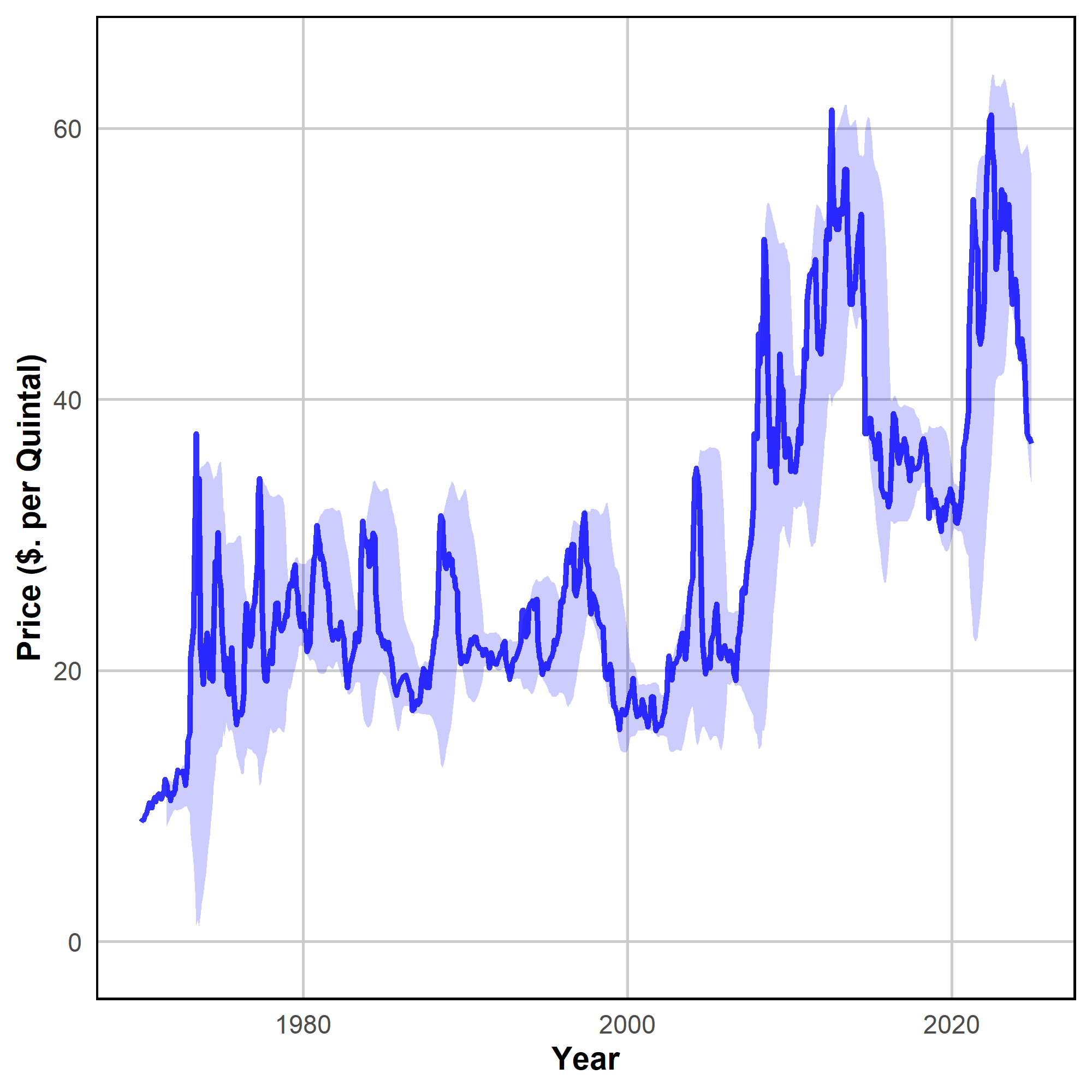}\llap{\parbox[b]{2.05in}{{\large\textsf{A}}\\\rule{0ex}{2.0in}}}
     \includegraphics[width=0.3\linewidth]{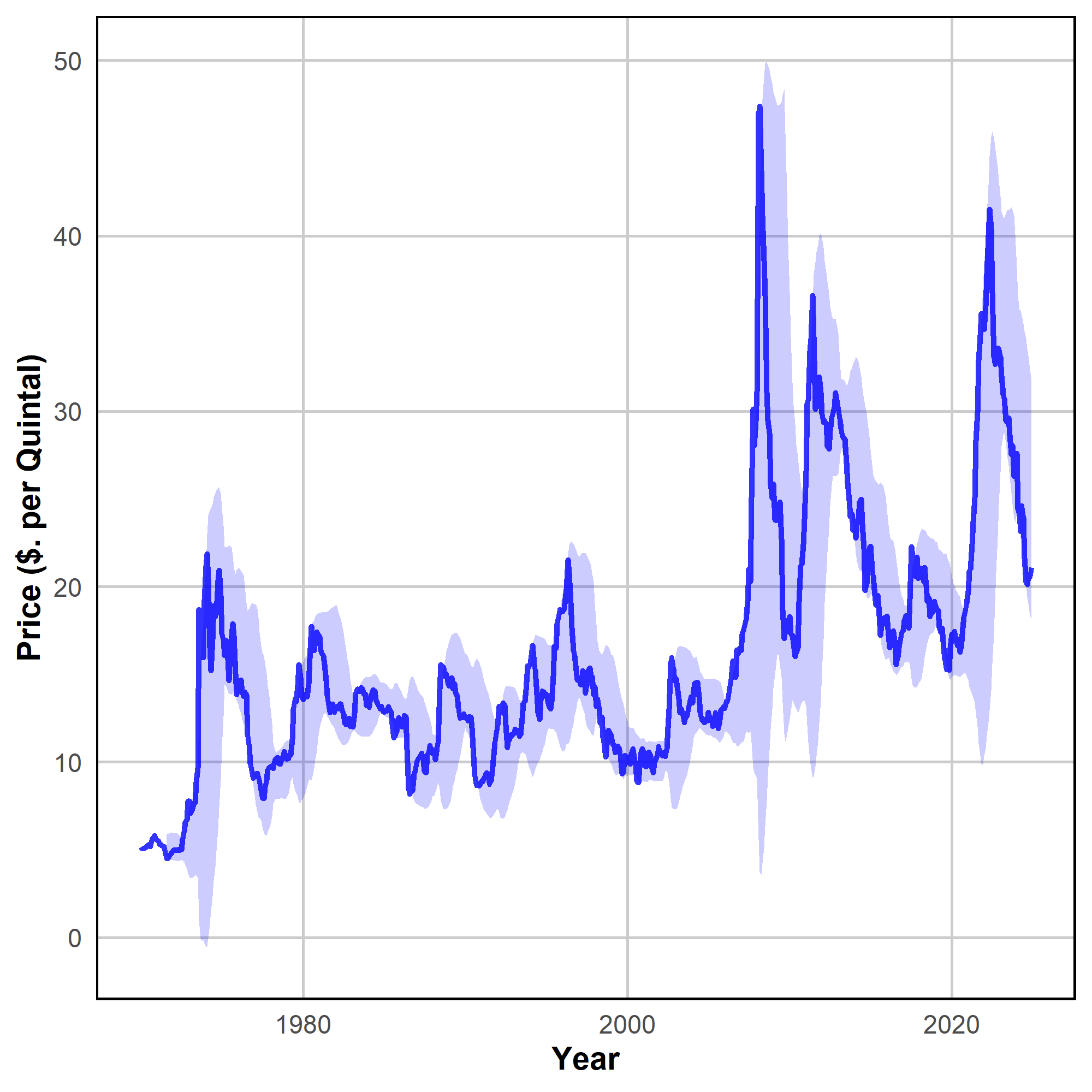}\llap{\parbox[b]{2.05in}{{\large\textsf{B}}\\\rule{0ex}{2.0in}}}
     \includegraphics[width=0.3\linewidth]{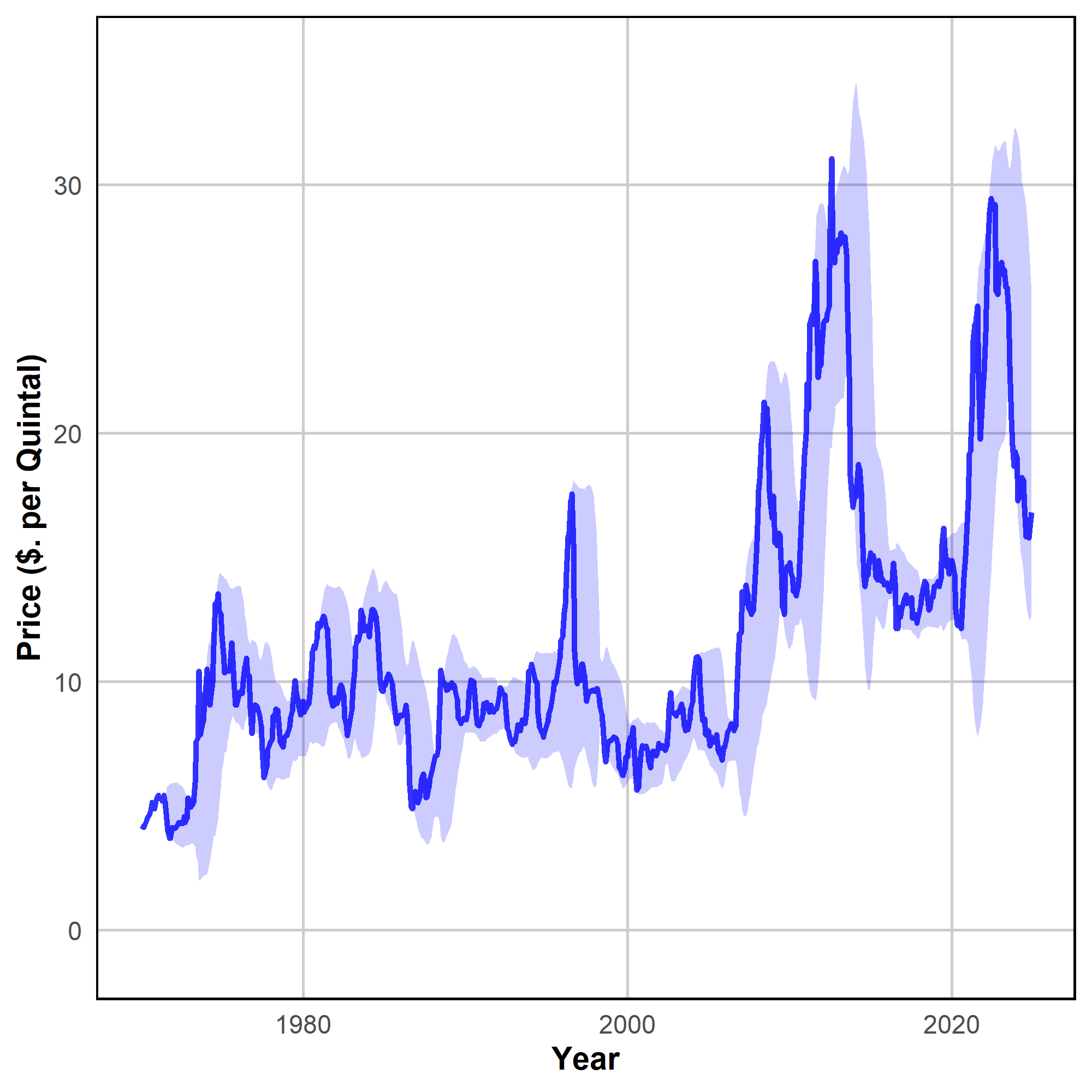}\llap{\parbox[b]{2.05in}{{\large\textsf{C}}\\\rule{0ex}{2.0in}}}
     \caption{\textbf{Plots of price time series, for \textit{(A)} soybean in Illinois, \textit{(B)} wheat in North Dakota, and \textit{(C)} corn in Iowa.} The blue line indicates the crop prices while the Bollinger Band (blue shaded region) represents the 20-month moving average $\pm 2$  standard deviations.}
    \label{fig:us_price_subplots}
\end{figure*}

We sourced monthly crop price data for the United States from the United States Department of Agriculture’s National Agricultural Statistics Service (NASS) via the Quick Stats database which is NASS’s most comprehensive tool for accessing agricultural data\cite{usda_quickstats}. We retrieved monthly crop price data for soybean in Illinois, wheat in North Dakota, and corn in Iowa, for the period January 1970 to December 2024. Figure~\ref{fig:us_price_subplots} shows the crop price data for each crop-state combination for the United States.

\subsection*{Climate Data}

Historical (1970–2014) and projected (2015–2100) monthly climate data for maximum temperature and precipitation were obtained for Madhya Pradesh, Assam, Gujarat, Illinois, North Dakota, and Iowa from gridded outputs of the sixth phase of the Coupled Model Intercomparison Project (CMIP6), which underpins the Intergovernmental Panel on Climate Change Sixth Assessment Report~\cite{copernicus_cmip6}. The data was derived from the Copernicus Climate Data Store (CDS). Climate projections are based on an ensemble mean of six Global Circulation Models (GCMs), listed in Table~\ref{table:GCM}, selected on the basis of data availability, temporal coverage, and relevance to the study regions, rather than ex post performance tuning, in order to avoid selection bias.

\begin{table*}
\centering
%\begin{footnotesize}
    \begin{tabular}{llc}
\toprule
Model & Agency/Institution & Nominal Resolution \\
\midrule
MIROC-ES2L & JAMSTEC, AORI, NIES (Japan) & 2.8\textdegree{} $\times$ 2.8\textdegree{} \\
CNRM-CM6-1 & CNRM, CERFACS (France) & 1.4\textdegree{} $\times$ 1.4\textdegree{} \\
MRI-ESM2-0 & Meteorological Research Institute (Japan)  & 1.125\textdegree{} $\times$ 1.125\textdegree{} \\
MRI-ESM1-2-LR & Max Planck Institute for Meteorology (Germany) & 1.8\textdegree{} $\times$ 1.8\textdegree{} \\
EC-Earth3-CC & EC-Earth Consortium (Europe) & 0.7\textdegree{} $\times$ 0.7\textdegree{} \\
CNRM-ESM2-1 & CNRM, CERFACS (France) & 2.5\textdegree{} $\times$ 2.5\textdegree{} \\
\bottomrule
\end{tabular}
\caption{List of GCMs used.}
%\end{footnotesize}
\label{table:GCM}
\end{table*}

\begin{figure*}   
   \centering
    \includegraphics[width=0.3\textwidth, height=0.3\textwidth]{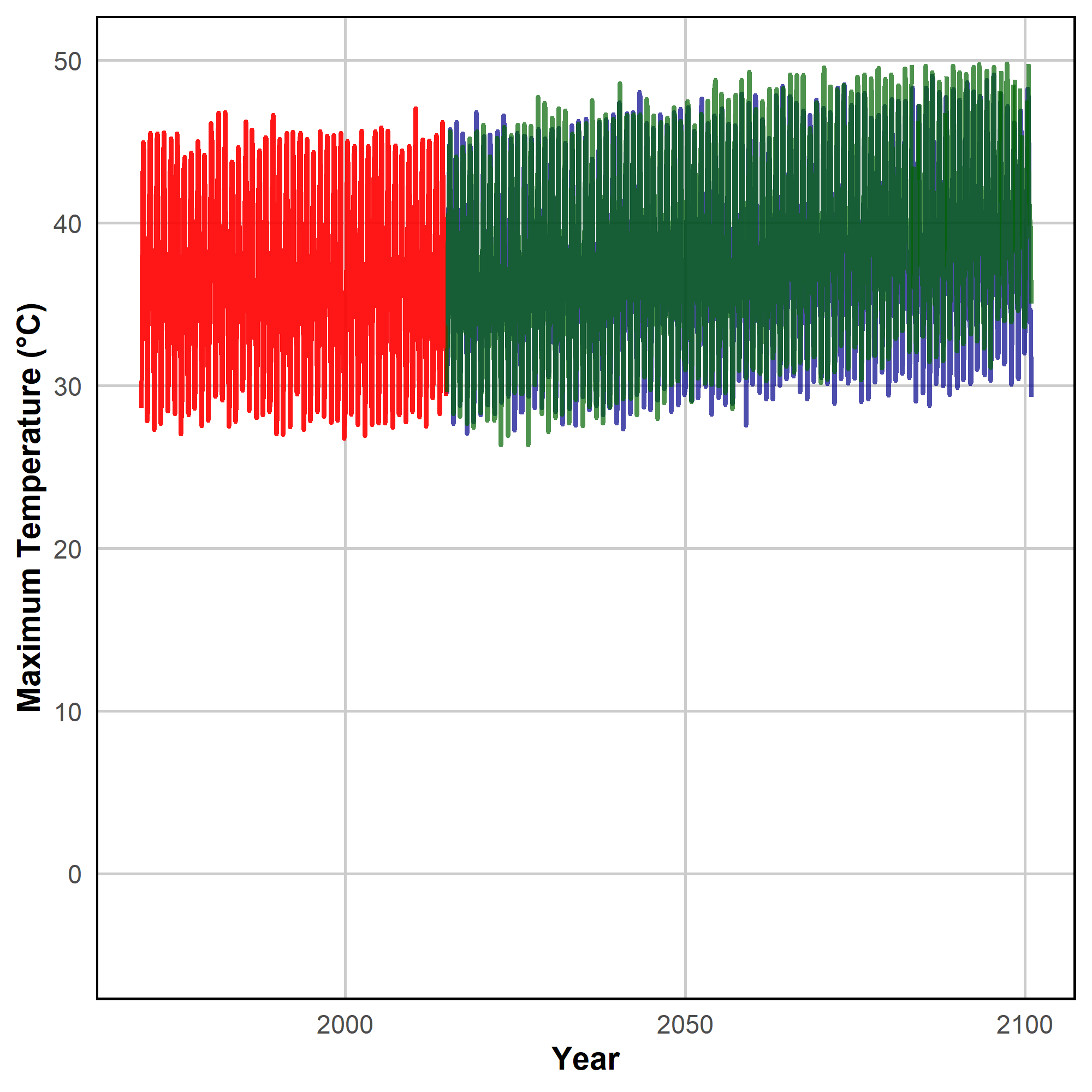}\llap{\parbox[b]{2.1in}{{\large\textsf{A}}\\\rule{0ex}{1.9in}}}
    \includegraphics[width=0.3\textwidth, height=0.3\textwidth]{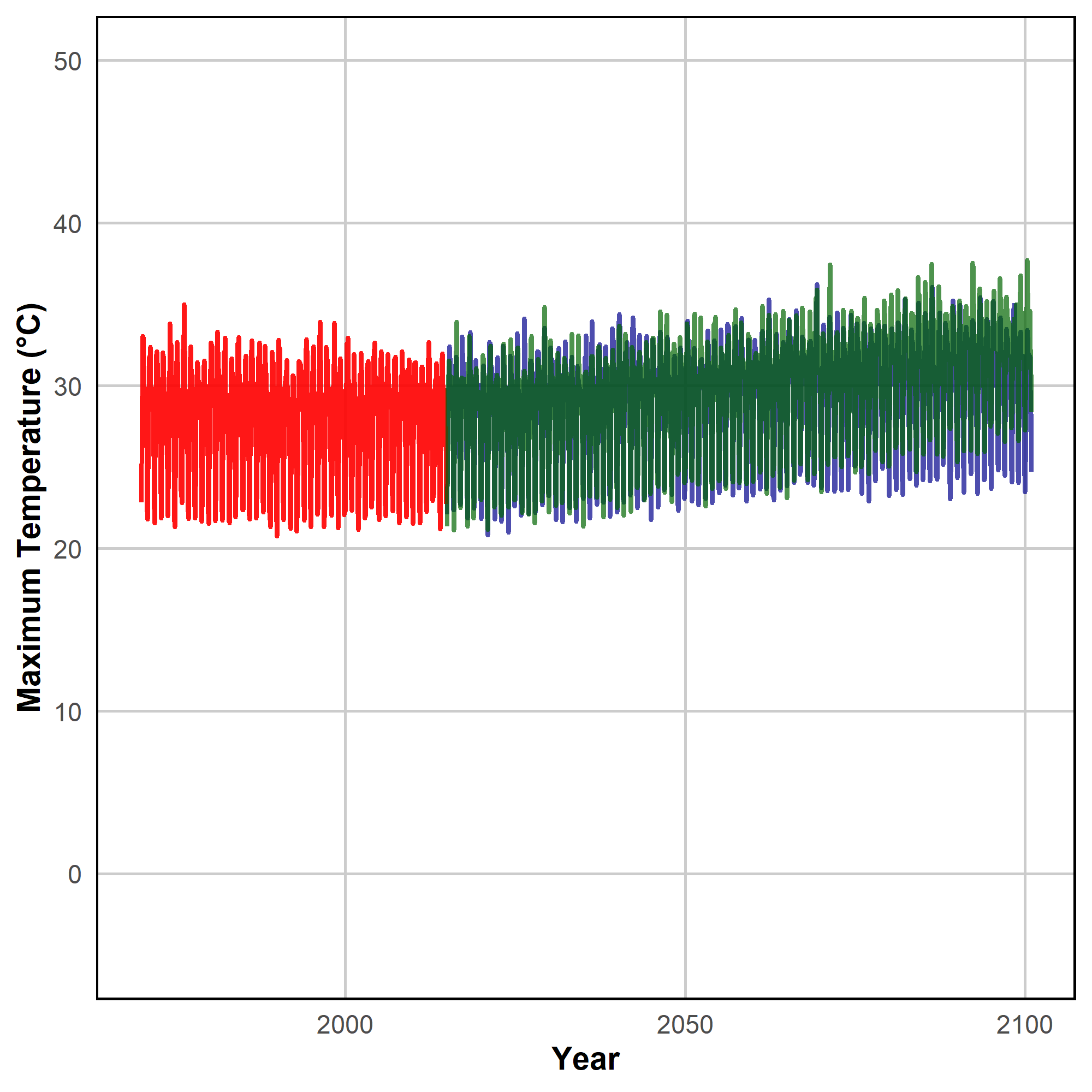}
    \includegraphics[width=0.3\textwidth, height=0.3\textwidth]{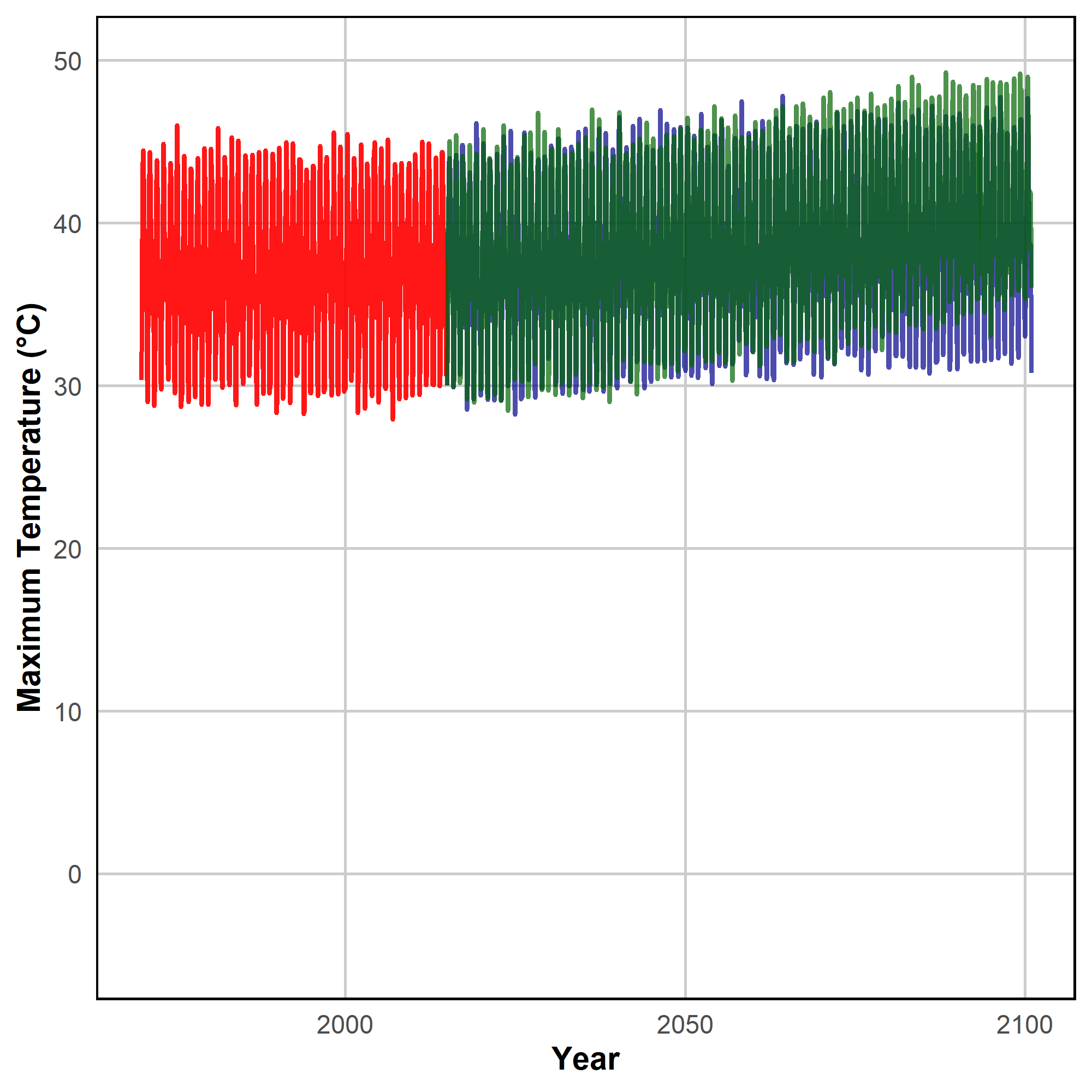}\\
    \includegraphics[width=0.3\textwidth, height=0.3\textwidth]{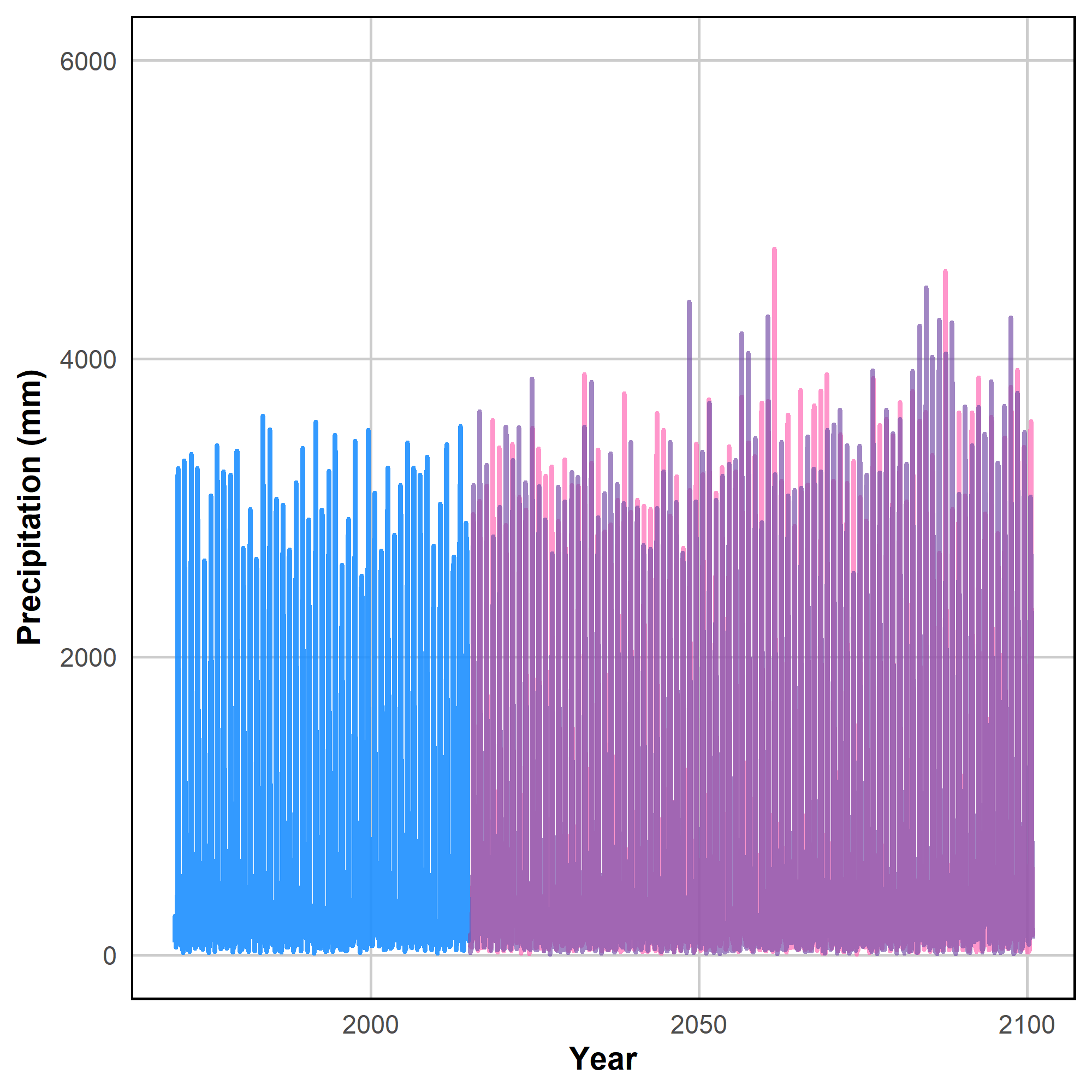}\llap{\parbox[b]{2.1in}{{\large\textsf{B}}\\\rule{0ex}{1.9in}}}
    \includegraphics[width=0.3\textwidth, height=0.3\textwidth]{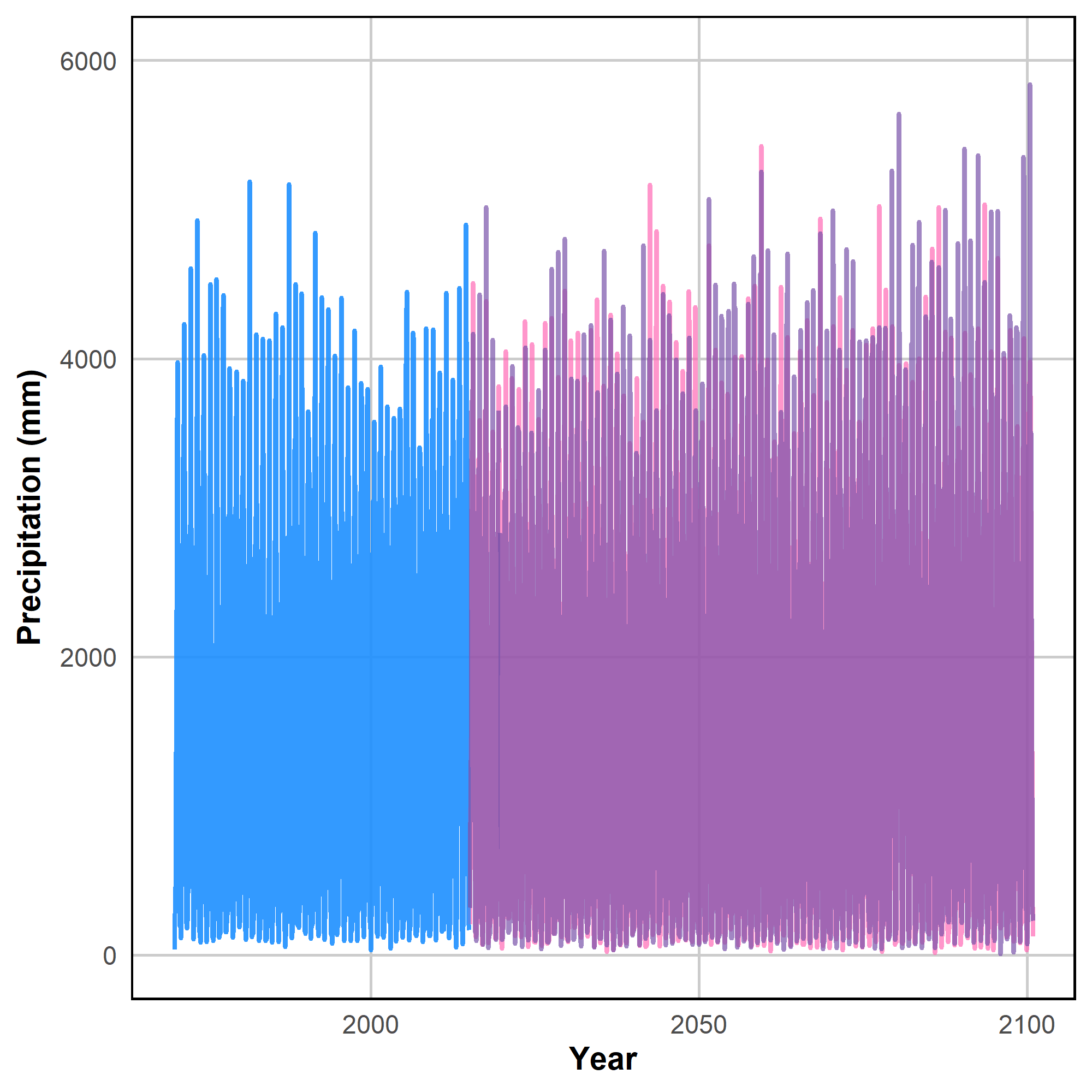}
    \includegraphics[width=0.3\textwidth, height=0.3\textwidth]{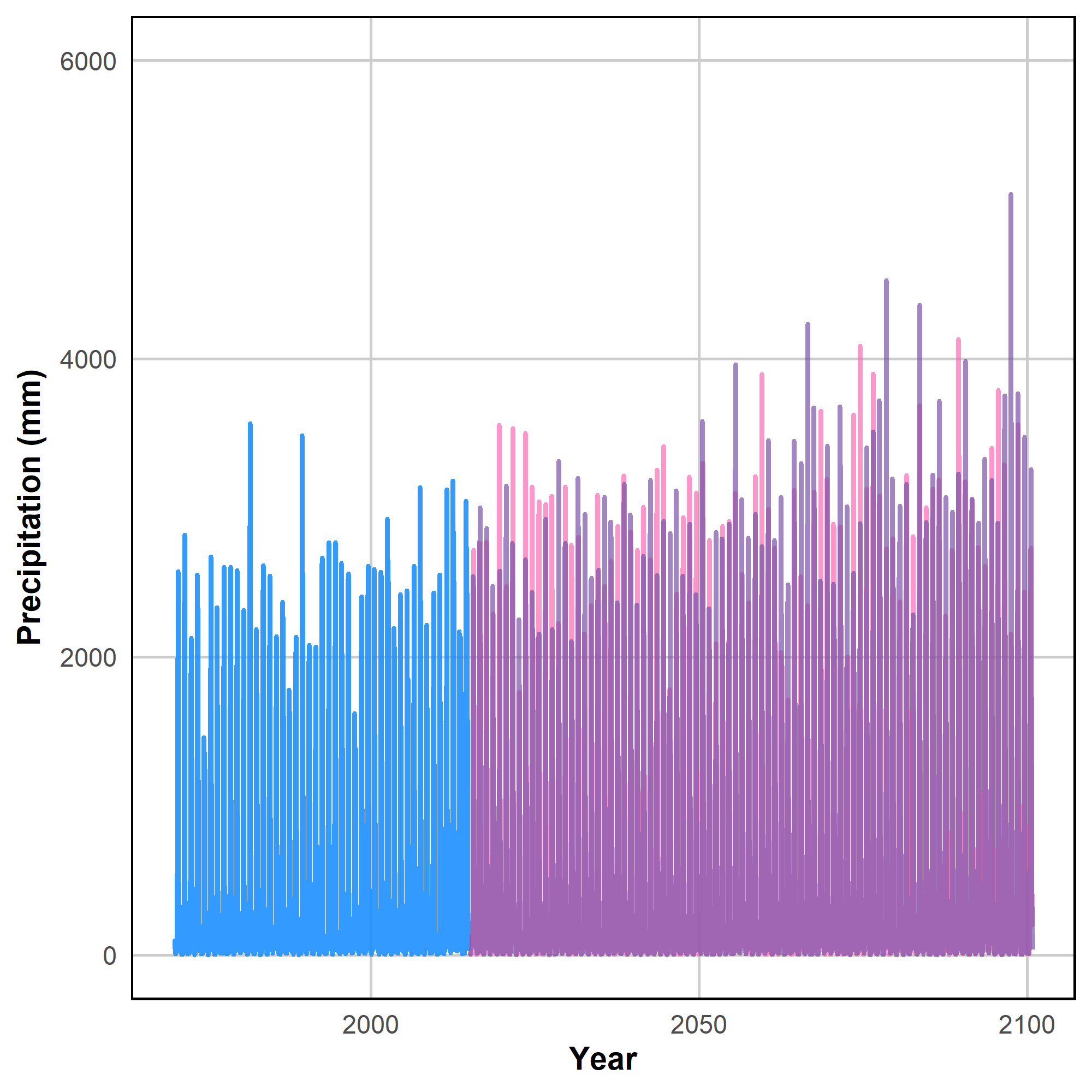}\\
     \caption{\textbf{Plots of historical and projected maximum temperature and precipitation for (left) Madhya Pradesh, (middle) Assam and (right) Gujarat.} \textit{(A)} shows the historical and projected maximum temperature. The red line represents the historical values of maximum temperature, while the dark blue and the dark green lines represent the projected values of maximum temperature under SSP2-4.5 and SSP5-8.5 climate scenarios respectively. \textit{(B)} shows the historical and projected precipitation data under different scenarios. The blue lines represent the historical values of precipitation, while the pink and purple lines represent the projected values of precipitation under SSP2-4.5 and SSP5-8.5 climate scenarios, respectively.}
    \label{fig:india_meteorological_subplots}
\end{figure*}

\begin{figure*}   
   \centering
    \includegraphics[width=0.3\textwidth, height=0.3\textwidth]{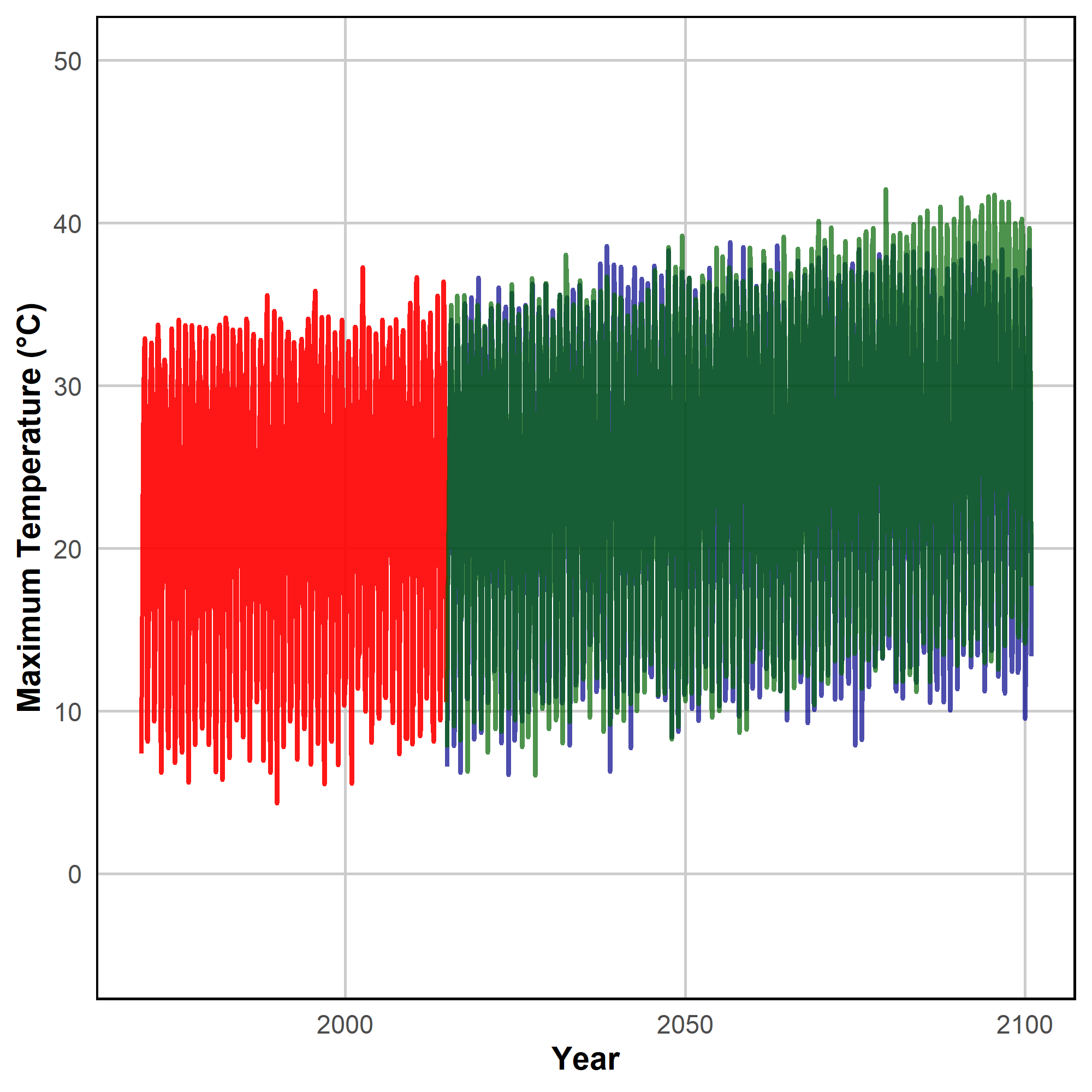}\llap{\parbox[b]{2.1in}{{\large\textsf{A}}\\\rule{0ex}{1.9in}}}
    \includegraphics[width=0.3\textwidth, height=0.3\textwidth]{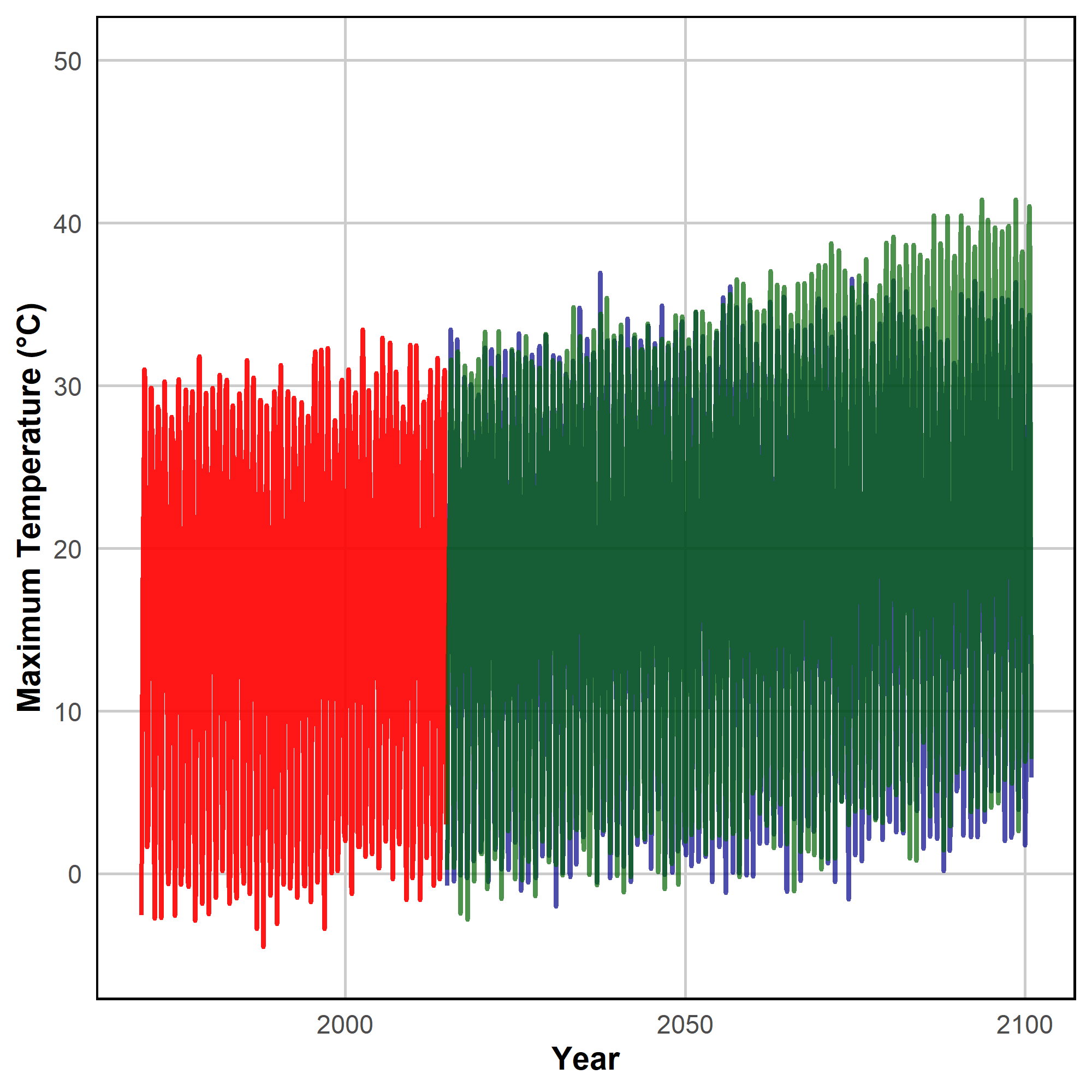}
    \includegraphics[width=0.3\textwidth, height=0.3\textwidth]{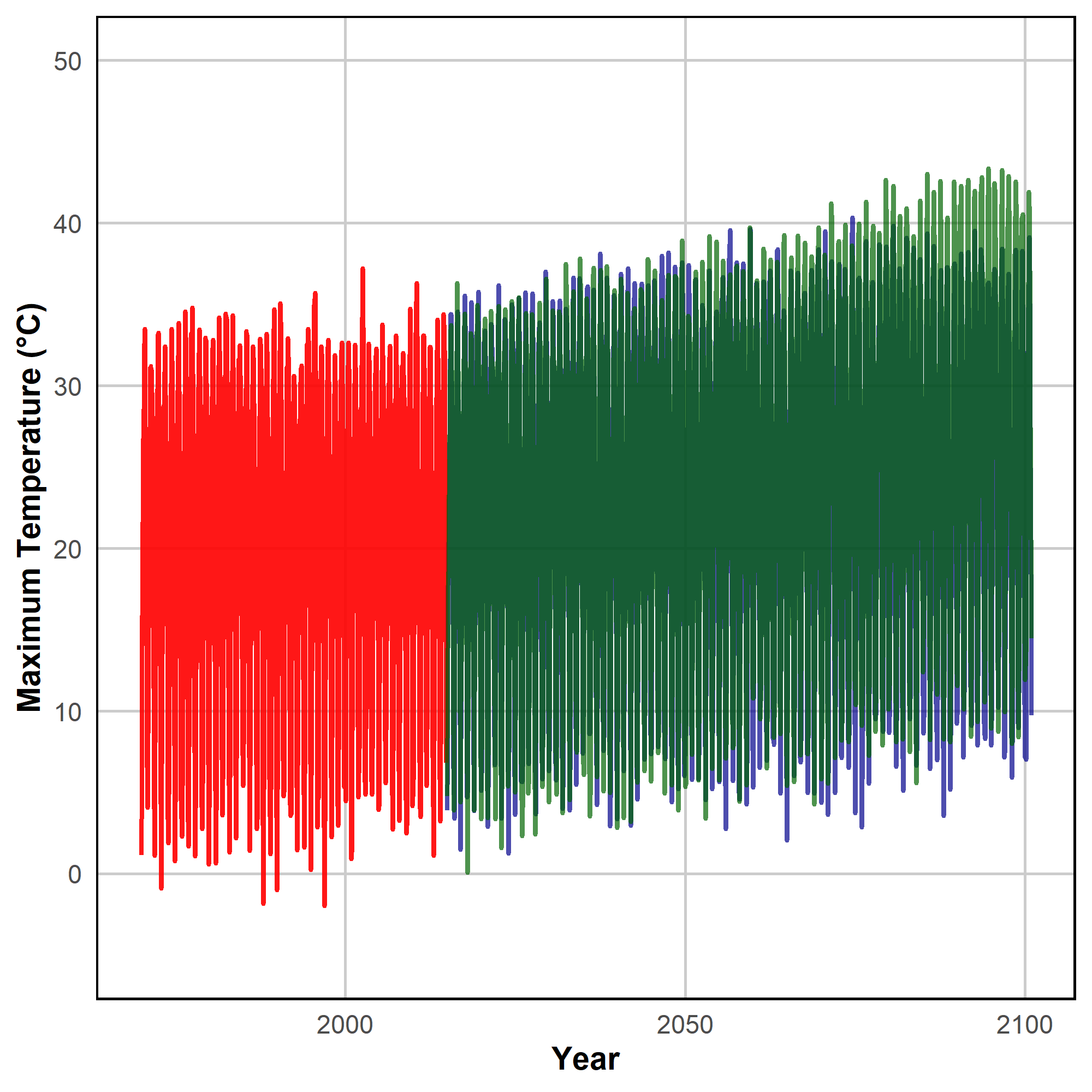}\\
    \includegraphics[width=0.3\textwidth, height=0.3\textwidth]{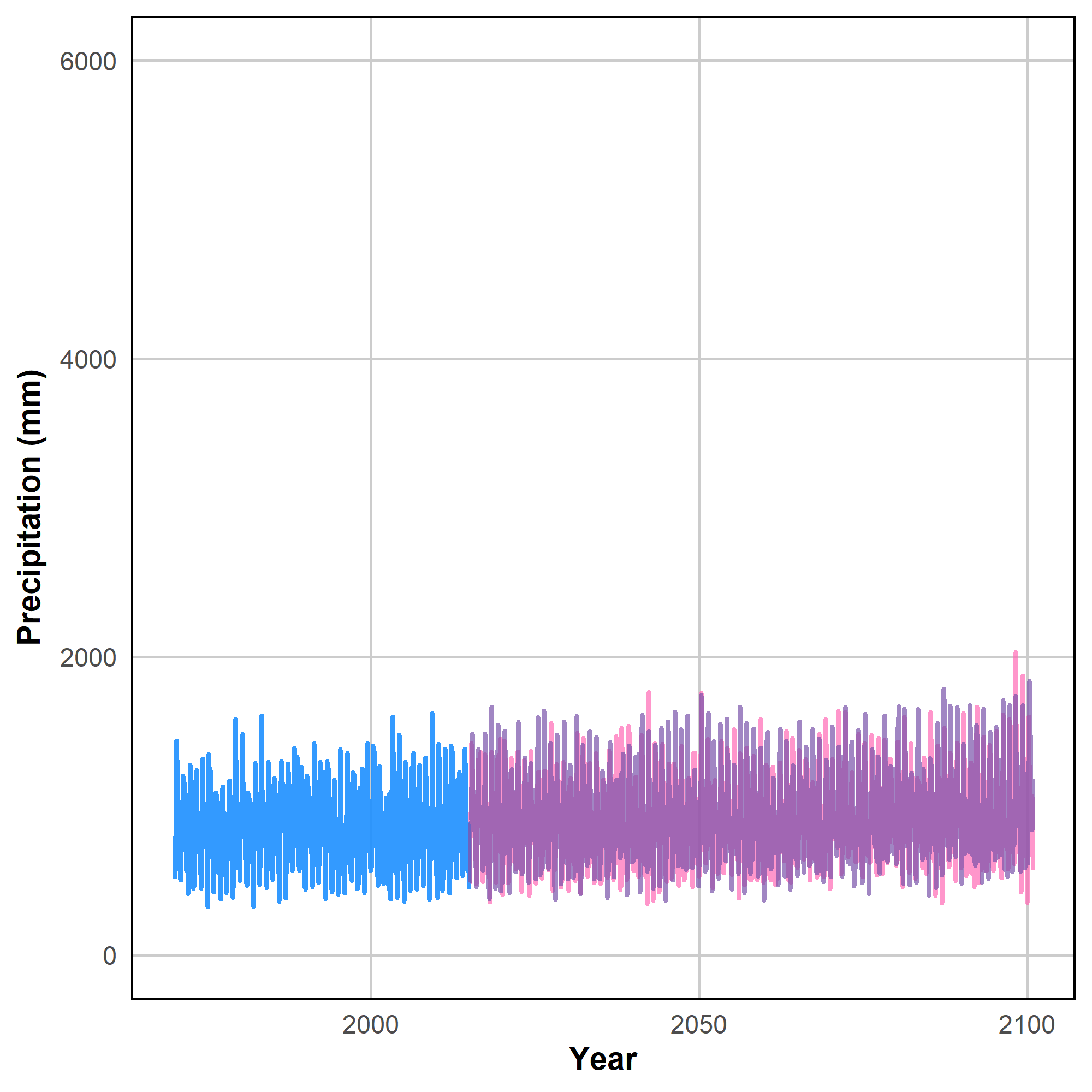}\llap{\parbox[b]{2.1in}{{\large\textsf{B}}\\\rule{0ex}{1.9in}}}
    \includegraphics[width=0.3\textwidth, height=0.3\textwidth]{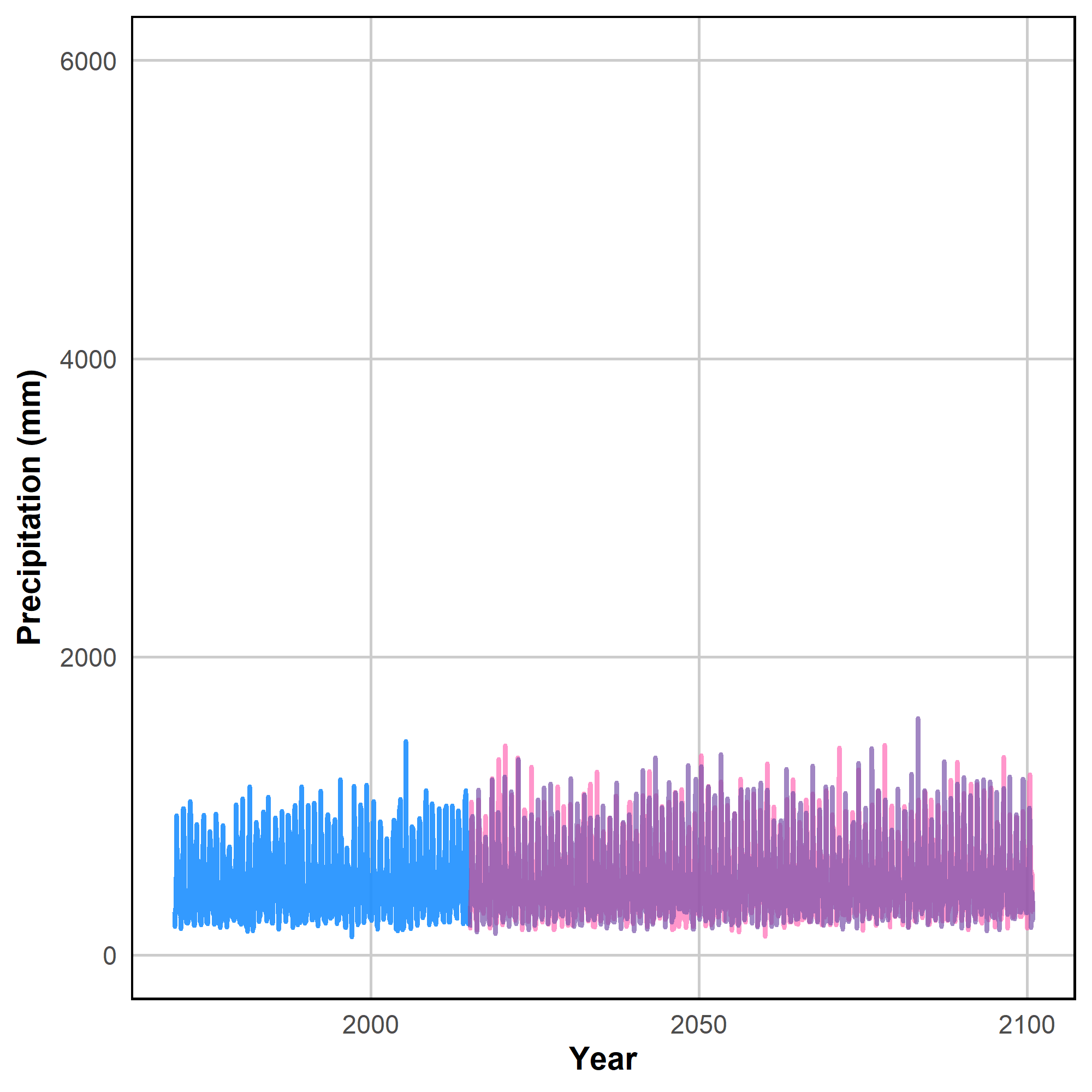}
    \includegraphics[width=0.3\textwidth, height=0.3\textwidth]{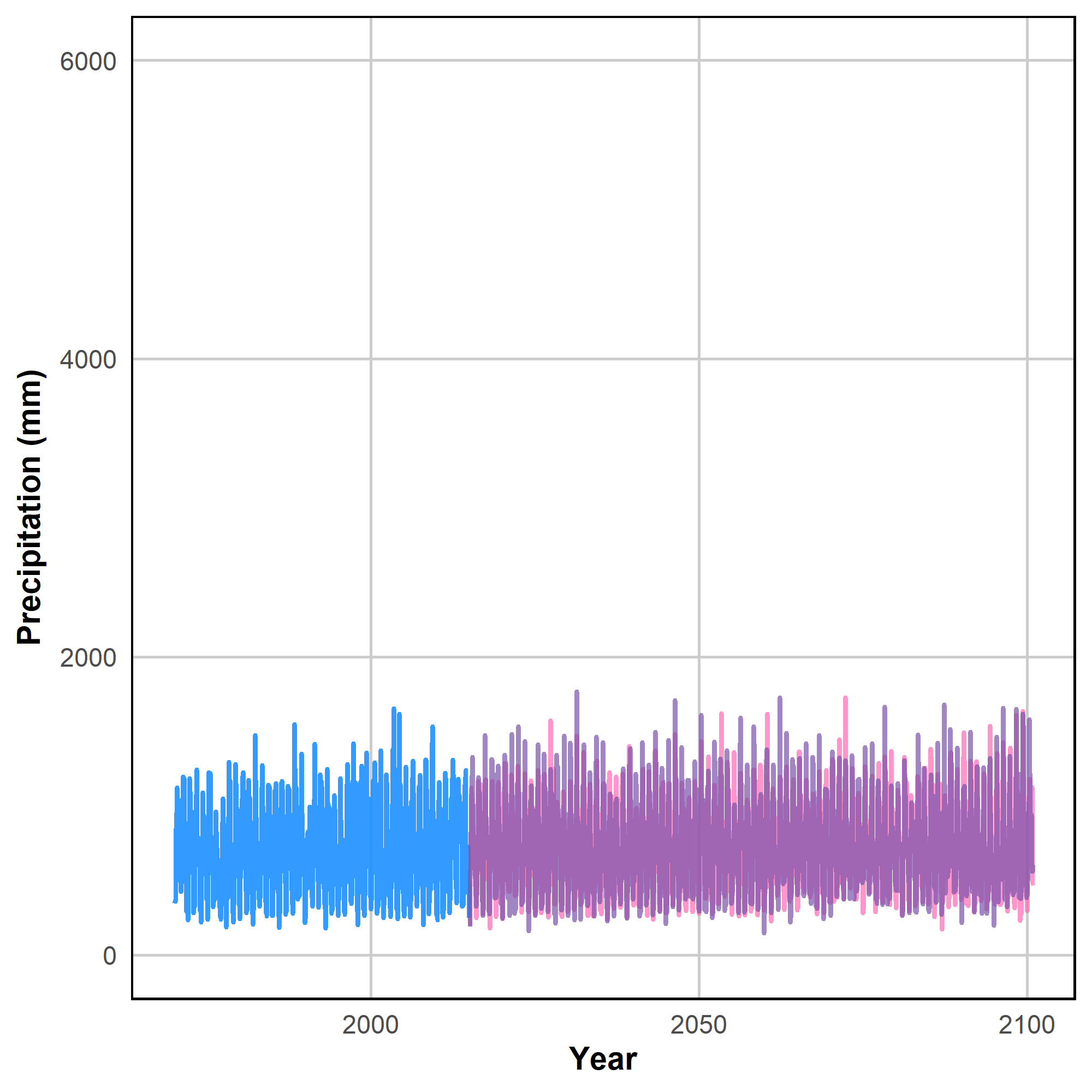}\\
     \caption{\textbf{Plots of  historical and projected maximum temperature and precipitation for (left) Illinois, (middle) North Dakota and (right) Iowa.} \textit{(A)} shows the historical and projected maximum temperature. The red line represents the historical values of maximum temperature, while the dark blue and the dark green lines represent the projected values of maximum temperature under SSP2-4.5 and SSP5-8.5 climate scenarios respectively. \textit{(B)} shows the historical and projected precipitation data under different scenarios. The blue lines represent the historical values of precipitation, while the pink and purple lines represent the projected values of precipitation under SSP2-4.5 and SSP5-8.5 climate scenarios, respectively.}
    \label{fig:us_meteorological_subplots}
\end{figure*}

\noindent\textbf{Shared Socioeconomic Pathways (SSPs)}: Two Shared Socioeconomic Pathways (SSPs) have been used for future climate projections. SSP2-4.5 represents a moderate emissions scenario, which assumes that by 2100, the radiative forcing would have stabilised at 4.5 W/m\textsuperscript{2} due to moderate population increase, technological advancements, and climate regulations, which will result in 2.1 - 3.5\textdegree{}C increase in global temperature over pre-industrial levels. SSP5-8.5 represents a high emission scenario, which assumes that by 2100, the radiative forcing would be 8.5 W/m\textsuperscript{2} due to fossil fuel-driven economic development and lack of climate mitigation, which will result in 3.2 - 5.4\textdegree{}C increase in global temperature over pre-industrial levels~\cite{Tebaldi2021}.
Figure~\ref{fig:india_meteorological_subplots} shows the historical and projected trends of maximum temperature and precipitation in Madhya Pradesh, Assam, and Gujarat under SSP2-4.5 and SSP5-8.5 scenarios, while Figure~\ref{fig:us_meteorological_subplots} shows the corresponding plots for Illinois, North Dakota, and Iowa.

To ensure comparability across models and consistency with observations, historical simulations are evaluated against observational datasets and used to establish a baseline for bias correction. Climate variables are bias-corrected using standard statistical techniques that adjust both mean and variability while preserving temporal structure. Owing to heterogeneity in native model resolutions, all climate fields are regridded to a common spatial grid prior to aggregation, and grid-scale discrepancies are explicitly accounted for during spatial averaging over agricultural regions, thereby preventing spurious effects of resolution differences on downstream volatility estimates.

Uncertainty is quantified through ensemble variability across climate models and retained throughout the modeling pipeline. Rather than collapsing climate inputs into a single deterministic trajectory, the ensemble structure is propagated through subsequent econometric analyses, allowing climate-driven uncertainty to enter the volatility modeling framework via distributions of climate covariates. This approach enables transparent tracing of uncertainty from climate simulations through preprocessing, econometric modeling, and final risk estimates, and allows assessment of how climate-related uncertainty contributes to variability in volatility forecasts and downstream financial risk measures, thereby strengthening the interpretability and robustness of the results.

\section*{Methodology}

Suppose \(\mathcal{P} = \{p_t|t=0,1,2,\cdots,T\}\) is the vector of historical prices of the crop. Then, the log-return is defined as:
\begin{equation}
   r_t =\ln\Big(\frac{p_t}{p_{t-1}}\Big) = \ln (p_t) - \ln (p_{t-1}).  
\end{equation}

\noindent We  model $r_t$ as a mean stationary process with $\mathbb{E}(r_t)=0$ and $\mathbb{V}ar(r_t)=\mathbb{E}(r_t^2) = \sigma_t^2$. Figure~\ref{fig:india_returns_volatility_subplots} and Figure~\ref{fig:us_returns_volatility_subplots} show the log-returns and squared log-returns for each crop-state combination in India and the US, respectively.

\subsection*{EGARCH Model for Conditional Price Volatility}

The \emph{Exponential Generalised Autoregressive Conditional Heteroskedasticity} (EGARCH) model~\cite{Bollerslev1986,Nelson1991} has been used to model conditional price volatility (\(\sigma_t^2\)). The EGARCH model is given by:
\begin{equation}
   \ln \sigma_t^2 = \nu + \sum_{i=1}^{p} \kappa_i \left( \left| \frac{\eta_{t-i}}{\sigma_{t-i}} \right| - \sqrt{\frac{2}{\pi}} \right) + \sum_{j=1}^{o} \delta_j \frac{\eta_{t-j}}{\sigma_{t-j}} + \sum_{k=1}^{q} \phi_k \ln \sigma_{t-k}^2. 
\end{equation}

\begin{itemize}[noitemsep]
    \item \(\nu\) represents the constant term setting the baseline level of the logarithm of the conditional variance,
    \item \(\eta_t\) represents the error term from the mean equation for log-return. Dividing it by $\sigma_t$ standardizes it to make the error term scale-free, 
    \item \(\kappa_i\) (for \(i=1,\dots,p\)) represents how the size of the past conditional volatility impacts the current conditional volatility. It captures the magnitude effect,
    \item \(\delta_j\) (for \(j=1,\dots,o\)) captures the leverage effect. It allows positive and negative shocks to have asymmetric effect on conditional volatility,
    \item \(\phi_k\) (for \(k=1,\dots,q\)) represents how much of the previous conditional volatility impacts the current conditional volatility. It captures the persistence effect.
    \item \(p\), \(o\), and \(q\) denote the number of lagged terms for the magnitude effect, the leverage effect, and the persistence effect, respectively.
\end{itemize}

\noindent Since the model uses the logarithm of variance, it does not require any restrictions on the parameters. The EGARCH model has been used because it captures volatility clustering (periods of high volatility followed by periods of high volatility and, vice versa) and leverage effect (negative and positive shocks have asymmetric effects on future volatility). The best model (p and q) has been chosen using the Akaike Information Criterion (AIC).

\begin{figure}
    \centering
    \includegraphics[width=0.3\textwidth, height=0.3\textwidth]{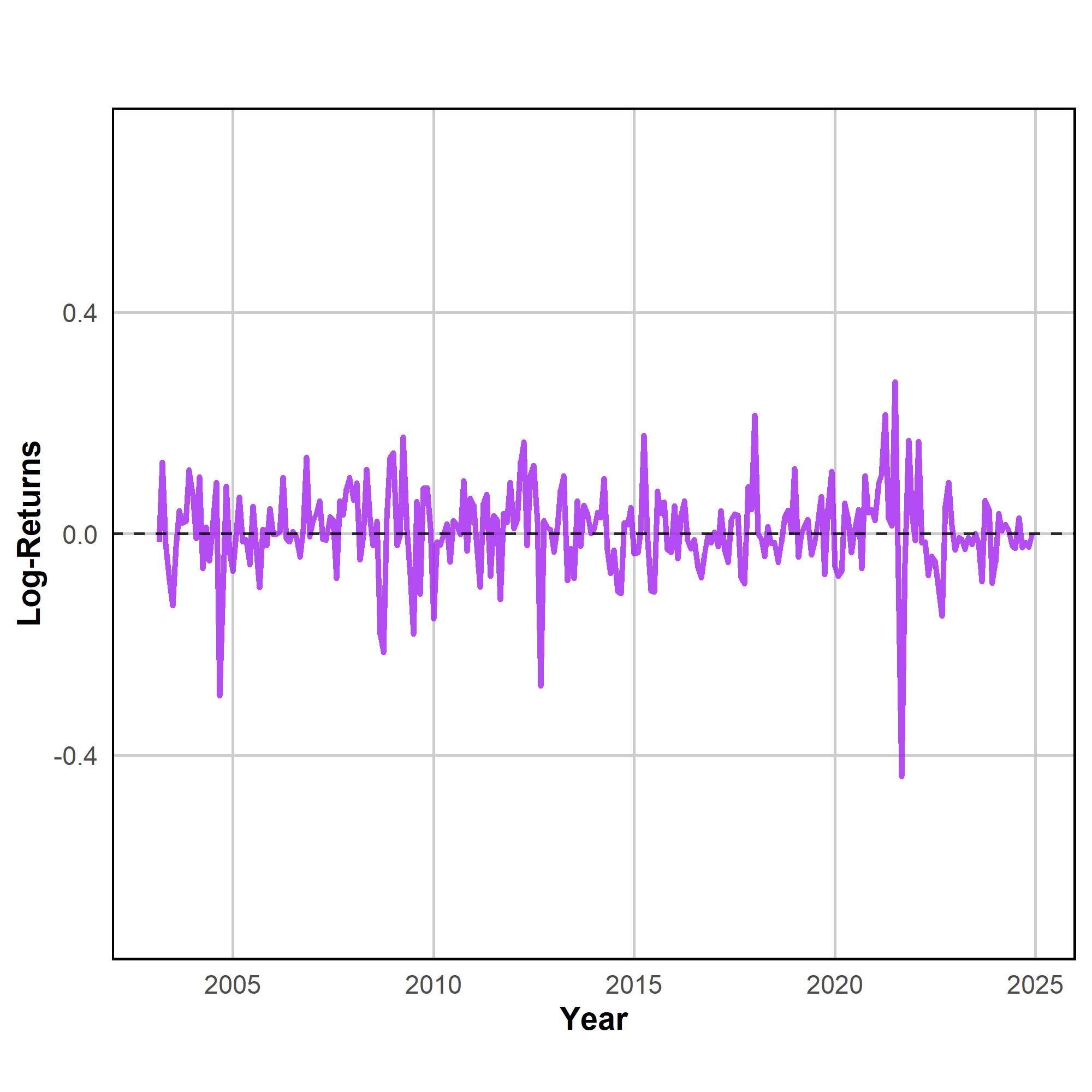}\llap{\parbox[b]{2.1in}{{\large\textsf{A}}\\\rule{0ex}{1.9in}}}
    \includegraphics[width=0.3\textwidth, height=0.3\textwidth]{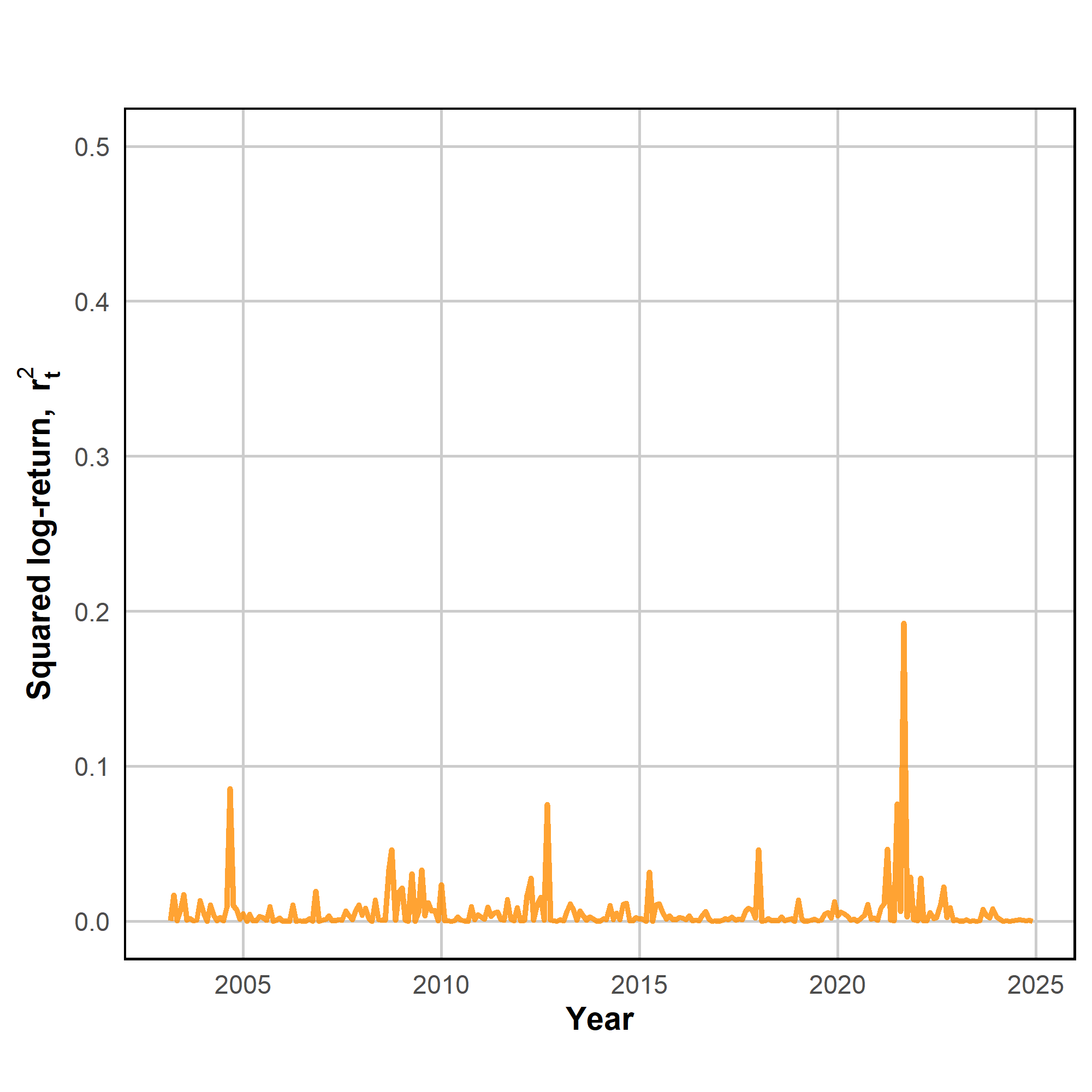}
    \includegraphics[width=0.3\textwidth, height=0.3\textwidth]{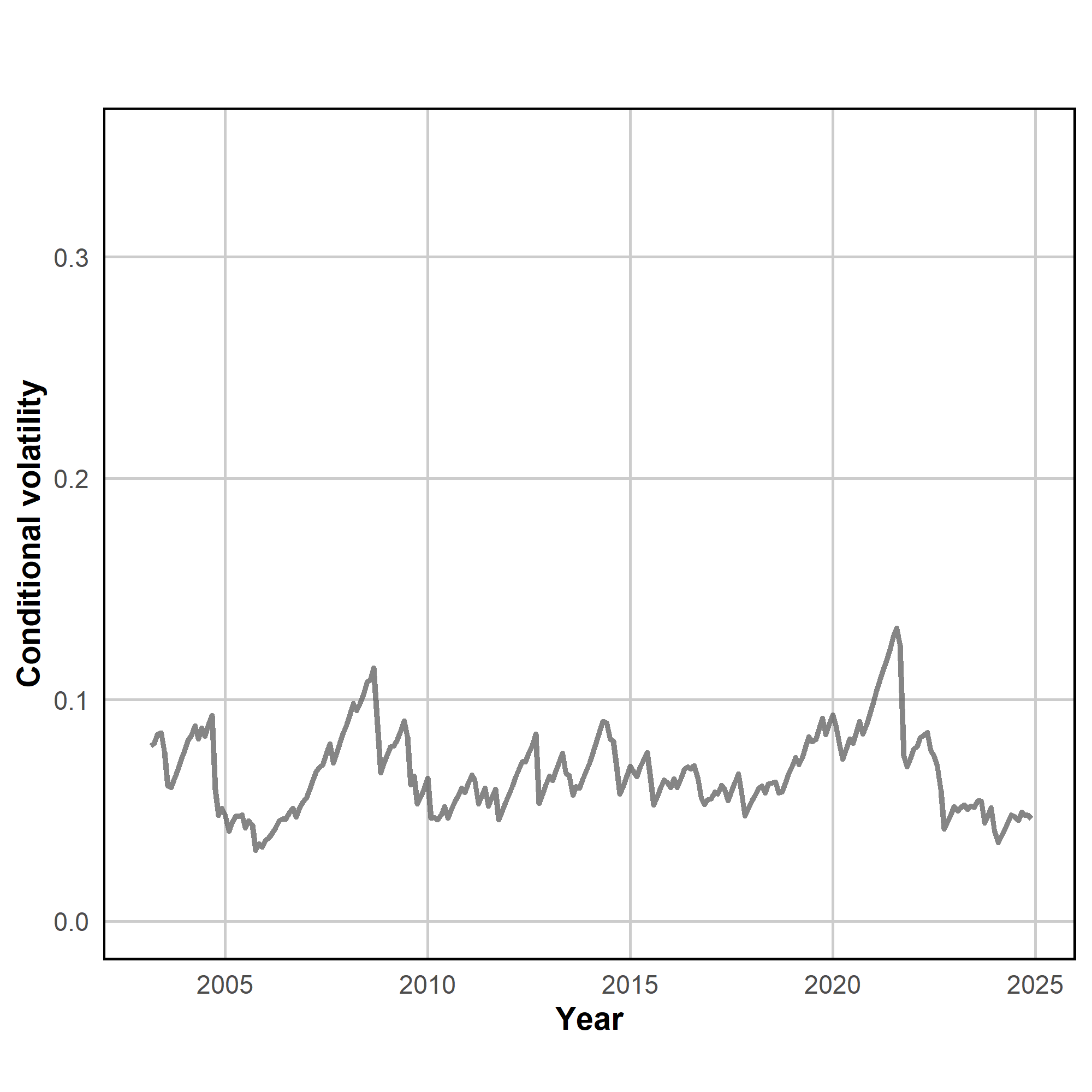}\\
    \includegraphics[width=0.3\textwidth, height=0.3\textwidth]{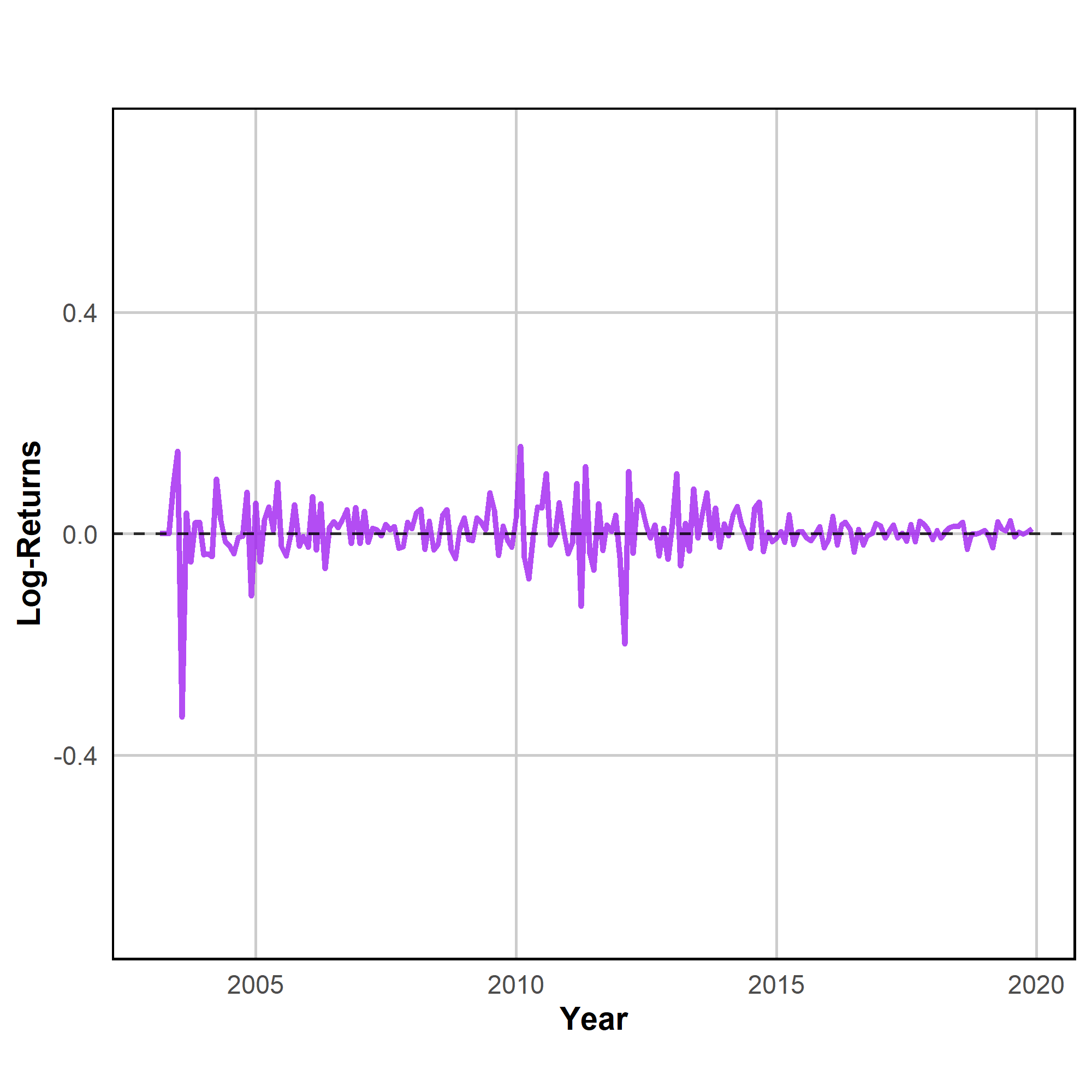}\llap{\parbox[b]{2.1in}{{\large\textsf{B}}\\\rule{0ex}{1.9in}}}
    \includegraphics[width=0.3\textwidth, height=0.3\textwidth]{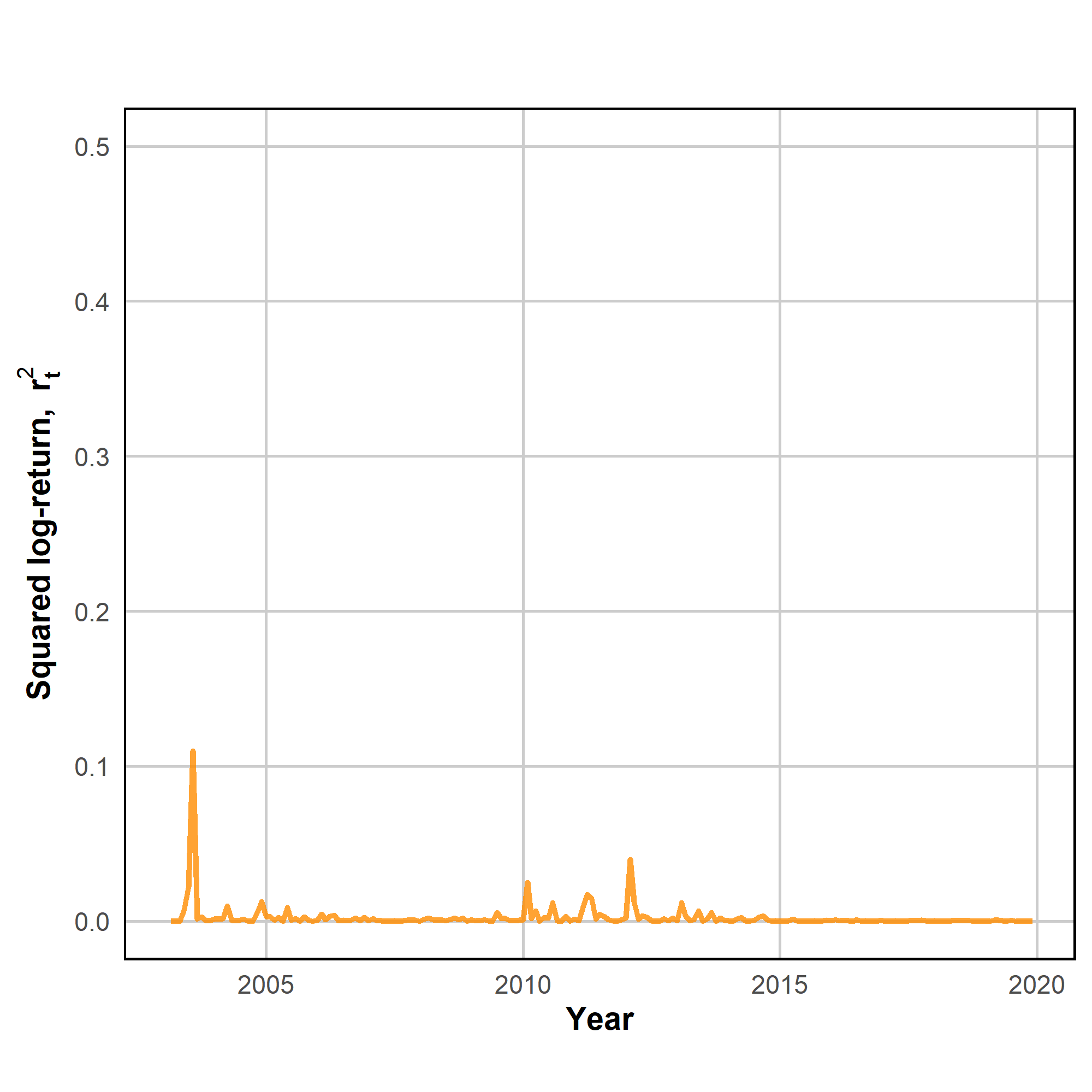}
    \includegraphics[width=0.3\textwidth, height=0.3\textwidth]{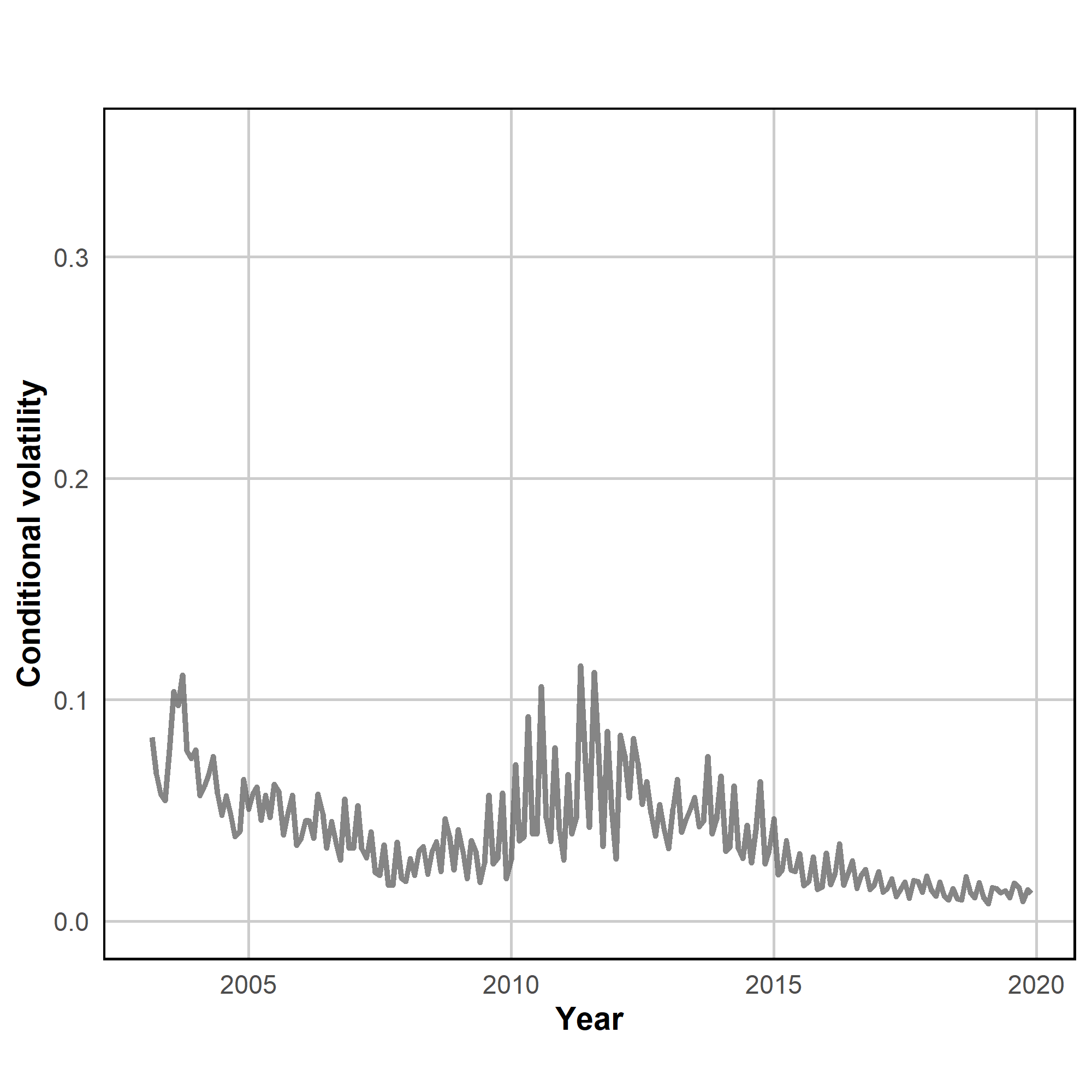}\\
    \includegraphics[width=0.3\textwidth, height=0.3\textwidth]{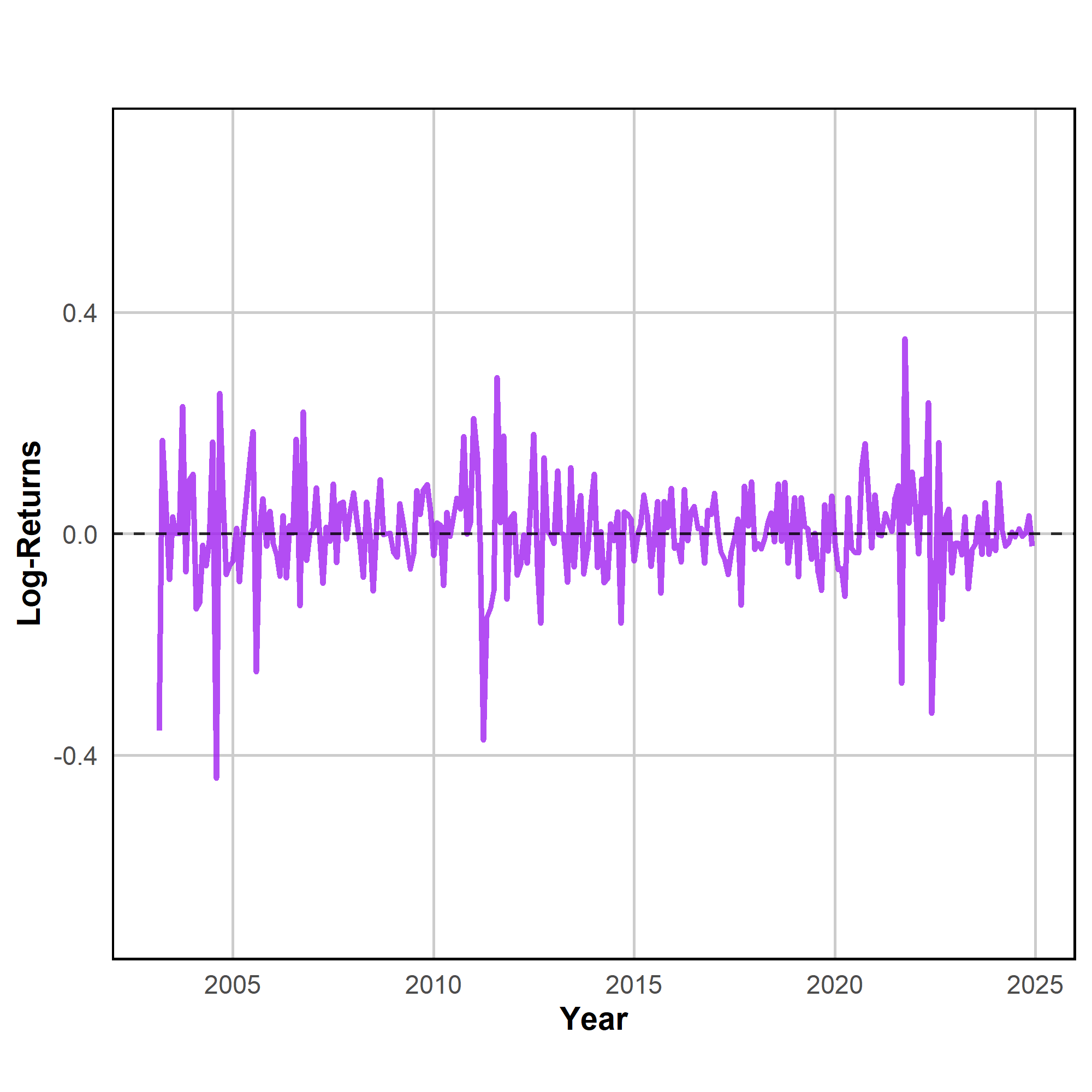}\llap{\parbox[b]{2.1in}{{\large\textsf{C}}\\\rule{0ex}{1.9in}}}
    \includegraphics[width=0.3\textwidth, height=0.3\textwidth]{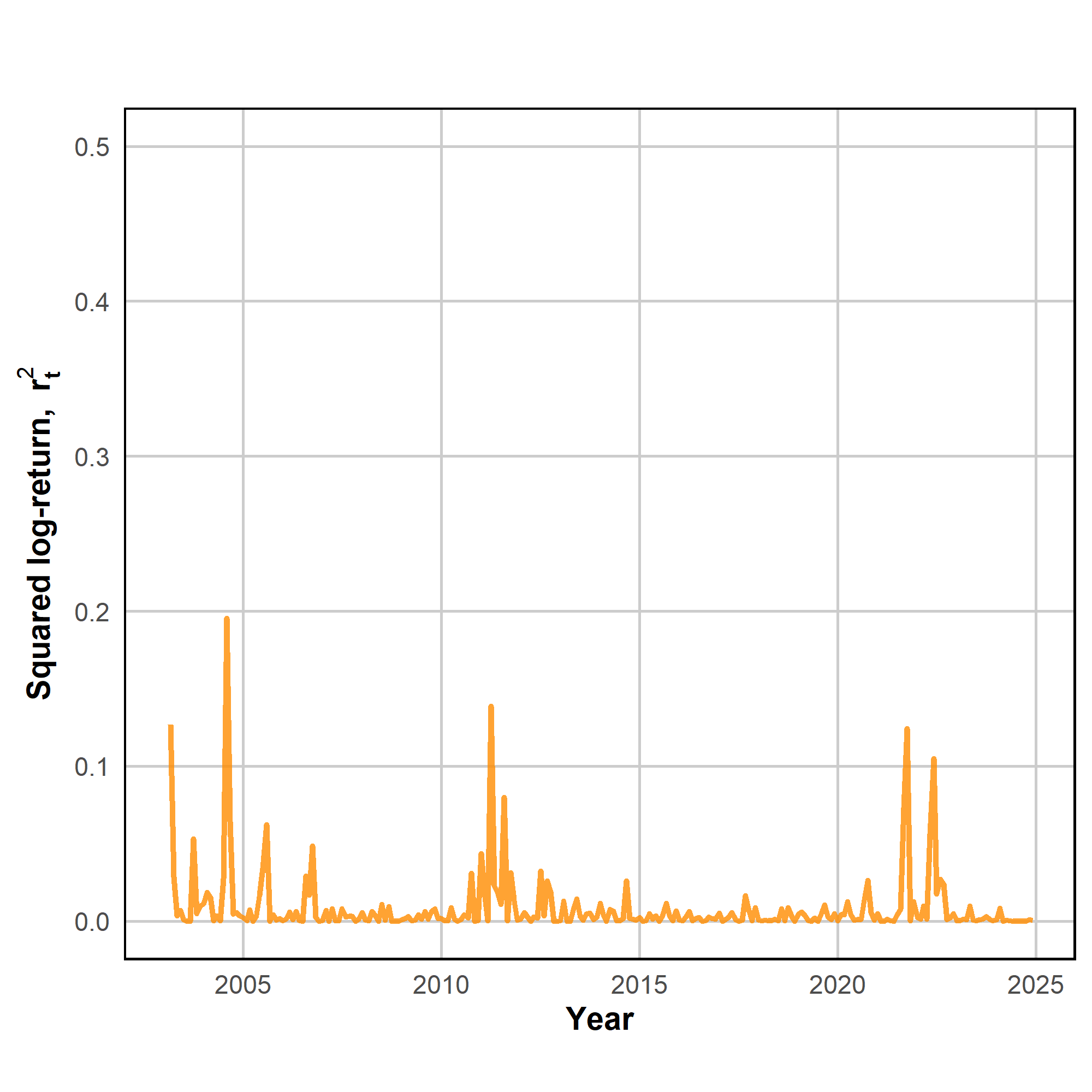}
    \includegraphics[width=0.3\textwidth, height=0.3\textwidth]{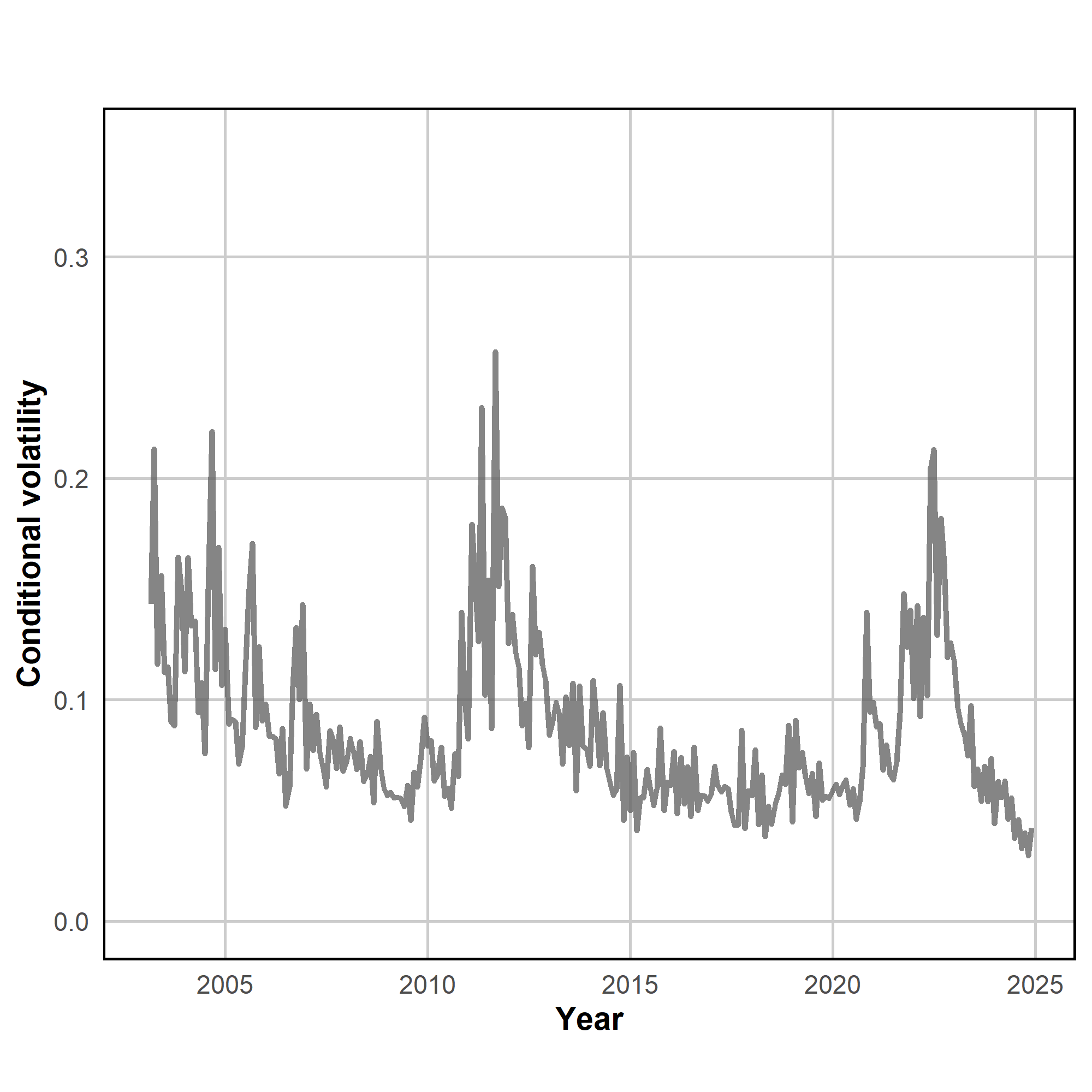}
    \caption{\textbf{Plots of log-returns, squared log-returns and EGARCH conditional volatility, for \textit{(A)} soybean in Madhya Pradesh, \textit{(B)} rice in Assam and \textit{(C)} cotton in Gujarat.} The purple line in the left panel represents the log-returns over time, and the dashed black line represents the mean which is zero. The orange line in the middle panel represents the squared log-returns. The grey line in the right panel represents the conditional volatility estimated using the EGARCH model.}
    \label{fig:india_returns_volatility_subplots}
\end{figure}

\begin{figure}
    \centering
    \includegraphics[width=0.3\textwidth, height=0.3\textwidth]{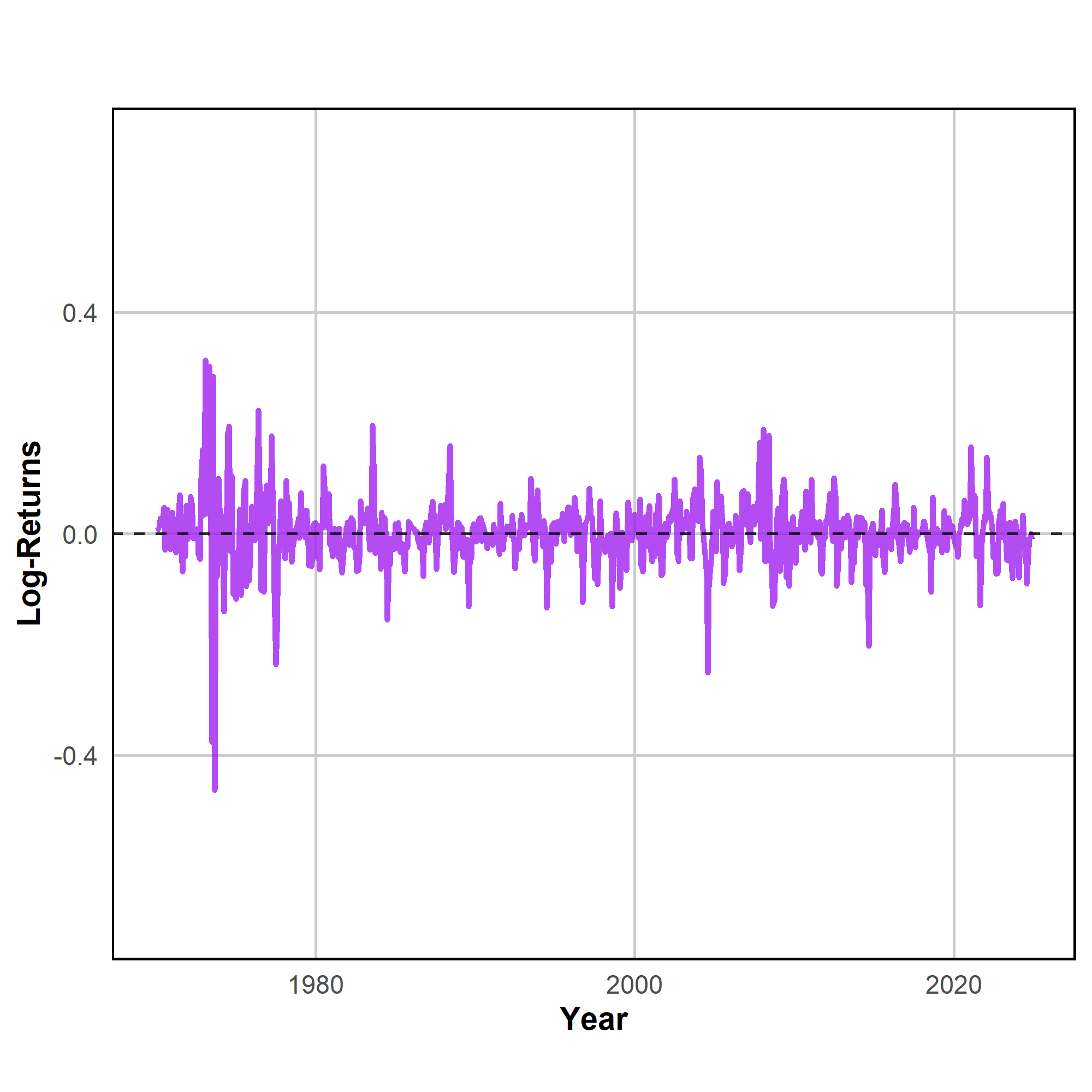}\llap{\parbox[b]{2.1in}{{\large\textsf{A}}\\\rule{0ex}{1.9in}}}
    \includegraphics[width=0.3\textwidth, height=0.3\textwidth]{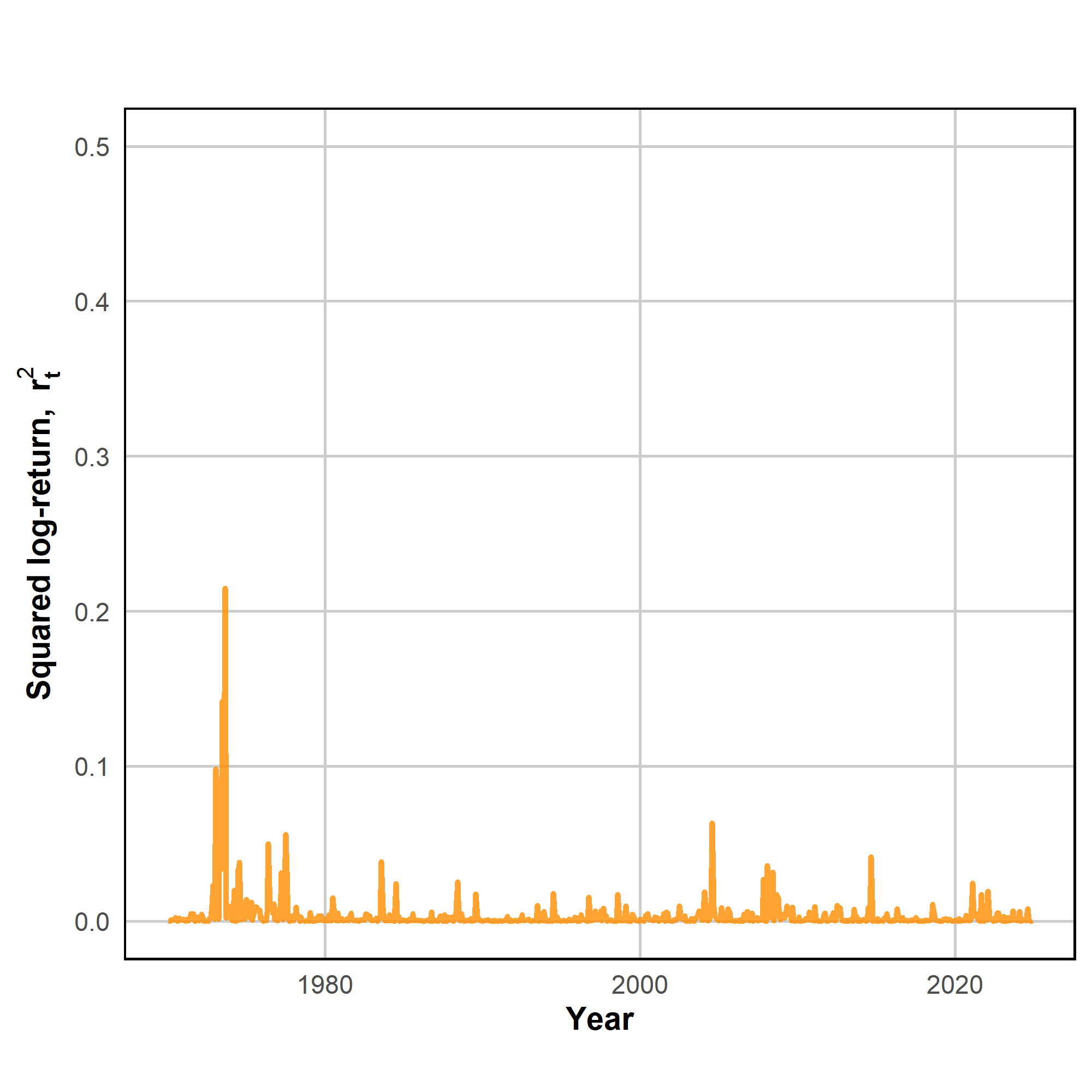}
    \includegraphics[width=0.3\textwidth, height=0.3\textwidth]{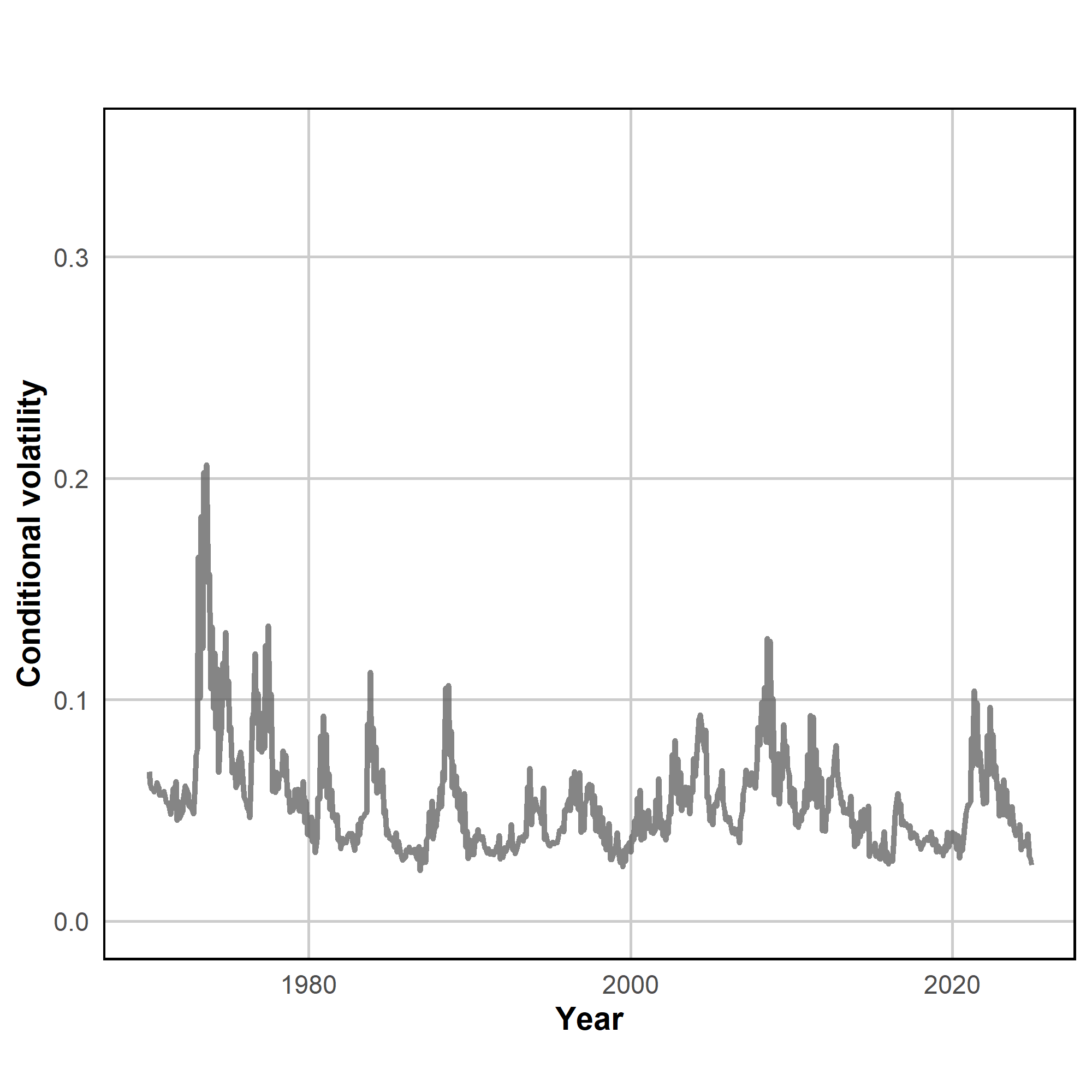}\\
    \includegraphics[width=0.3\textwidth, height=0.3\textwidth]{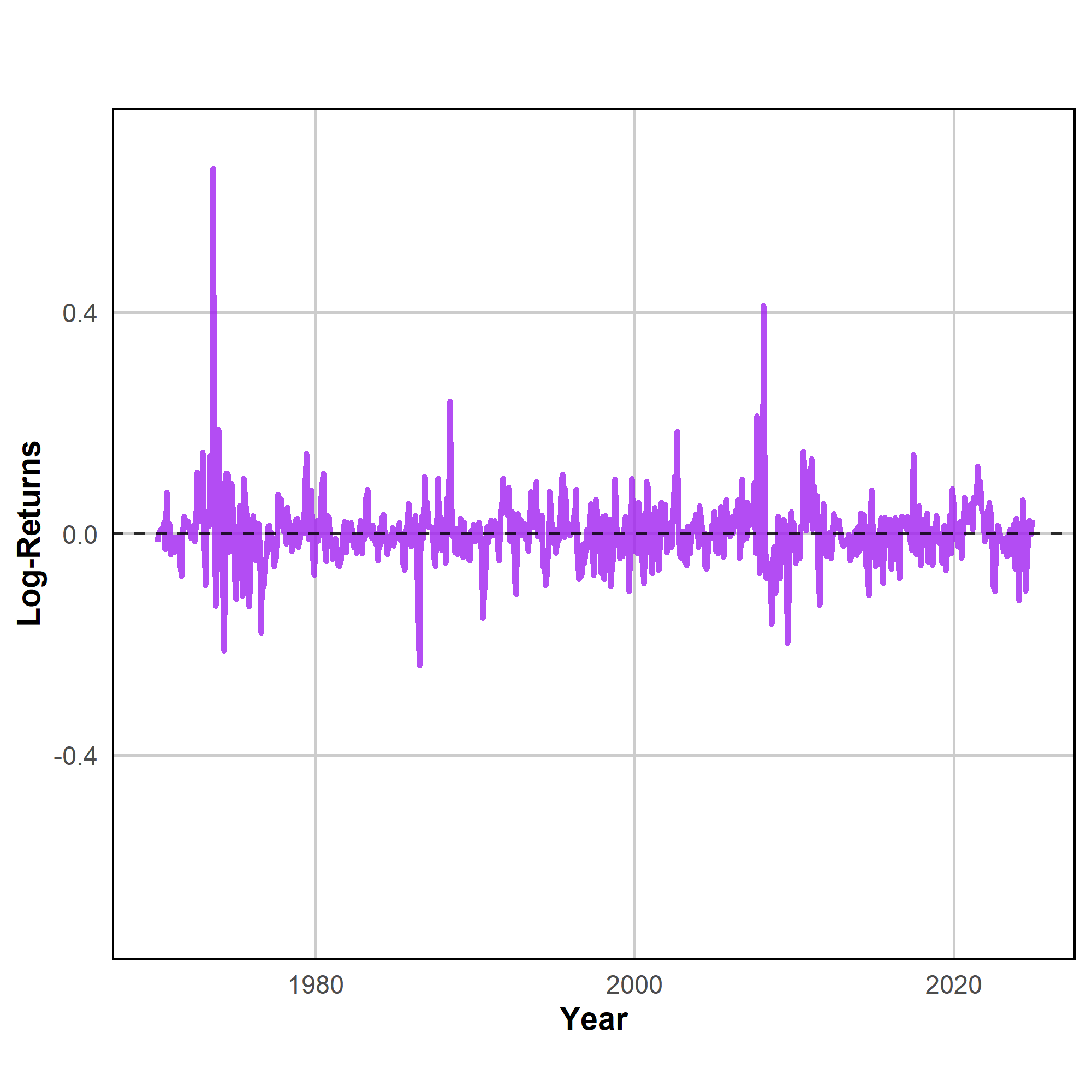}\llap{\parbox[b]{2.1in}{{\large\textsf{B}}\\\rule{0ex}{1.9in}}}
    \includegraphics[width=0.3\textwidth, height=0.3\textwidth]{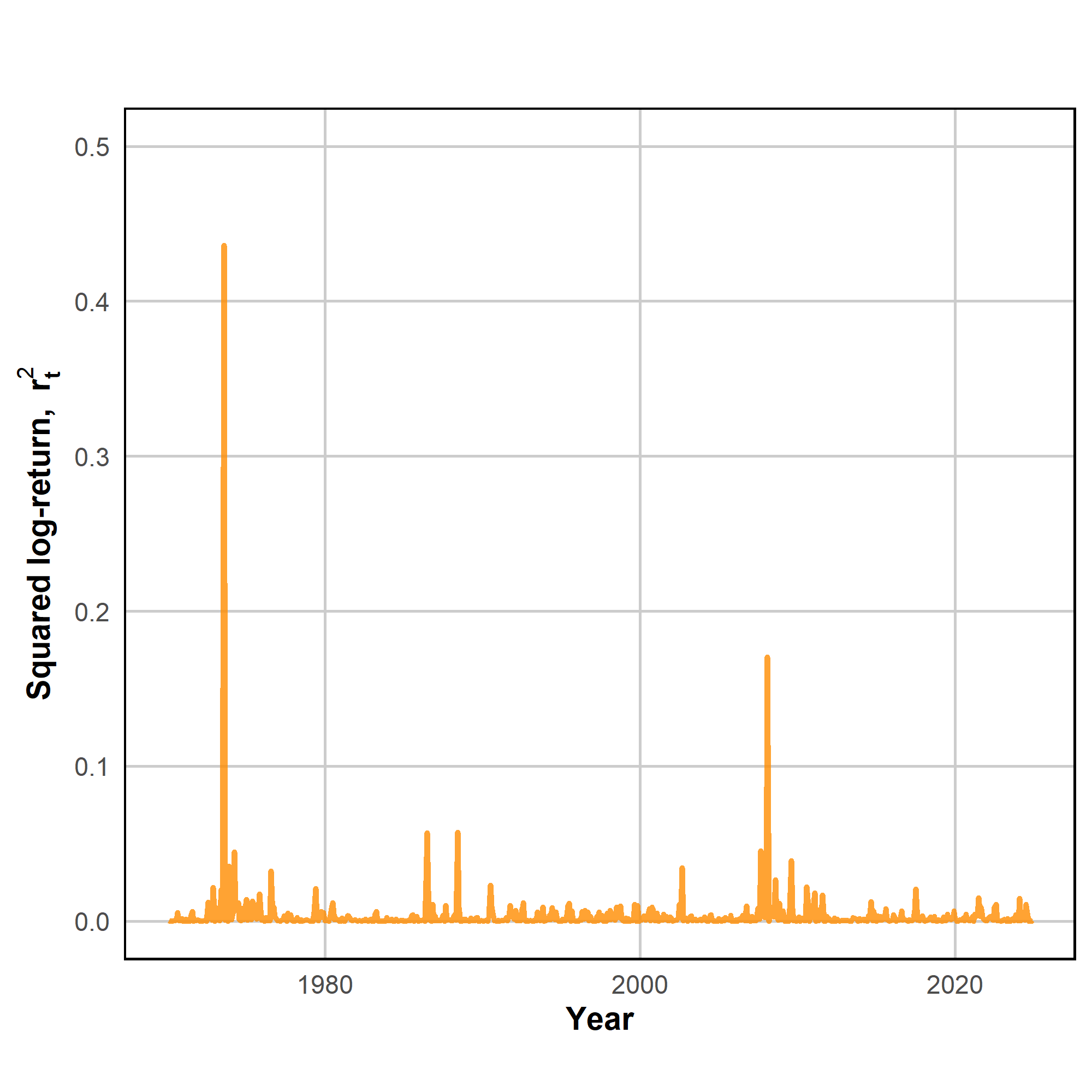}
    \includegraphics[width=0.3\textwidth, height=0.3\textwidth]{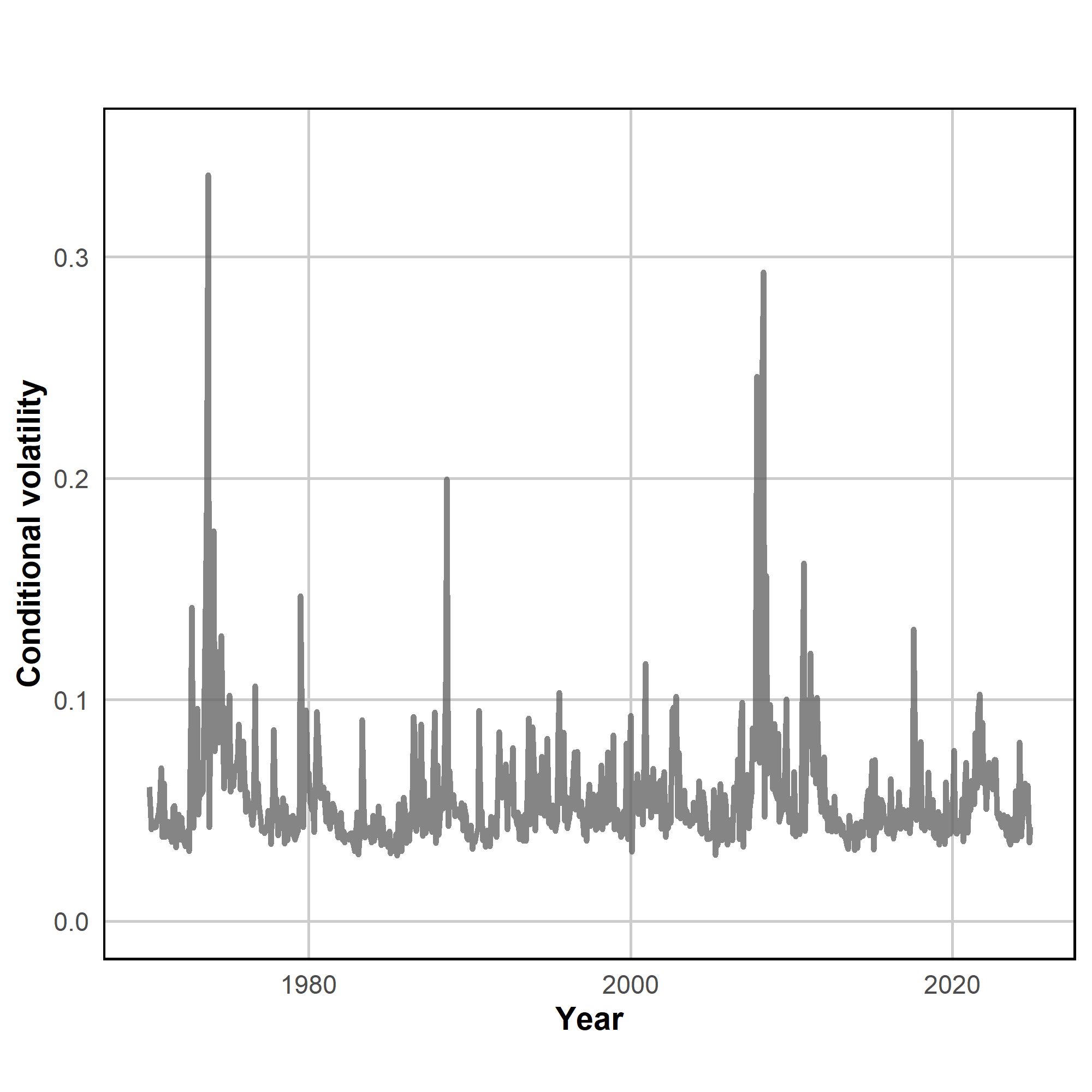}\\
    \includegraphics[width=0.3\textwidth, height=0.3\textwidth]{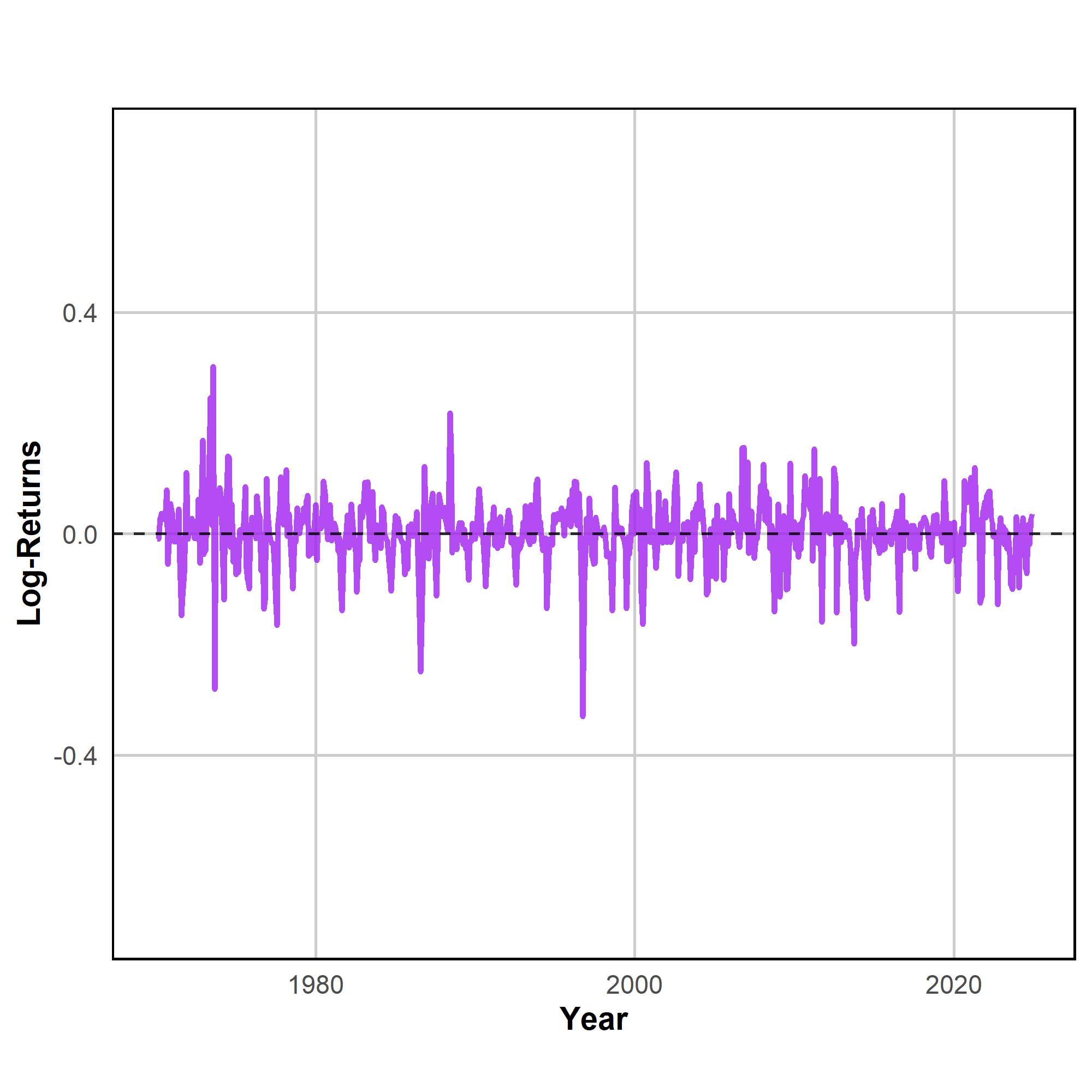}\llap{\parbox[b]{2.1in}{{\large\textsf{C}}\\\rule{0ex}{1.9in}}}
    \includegraphics[width=0.3\textwidth, height=0.3\textwidth]{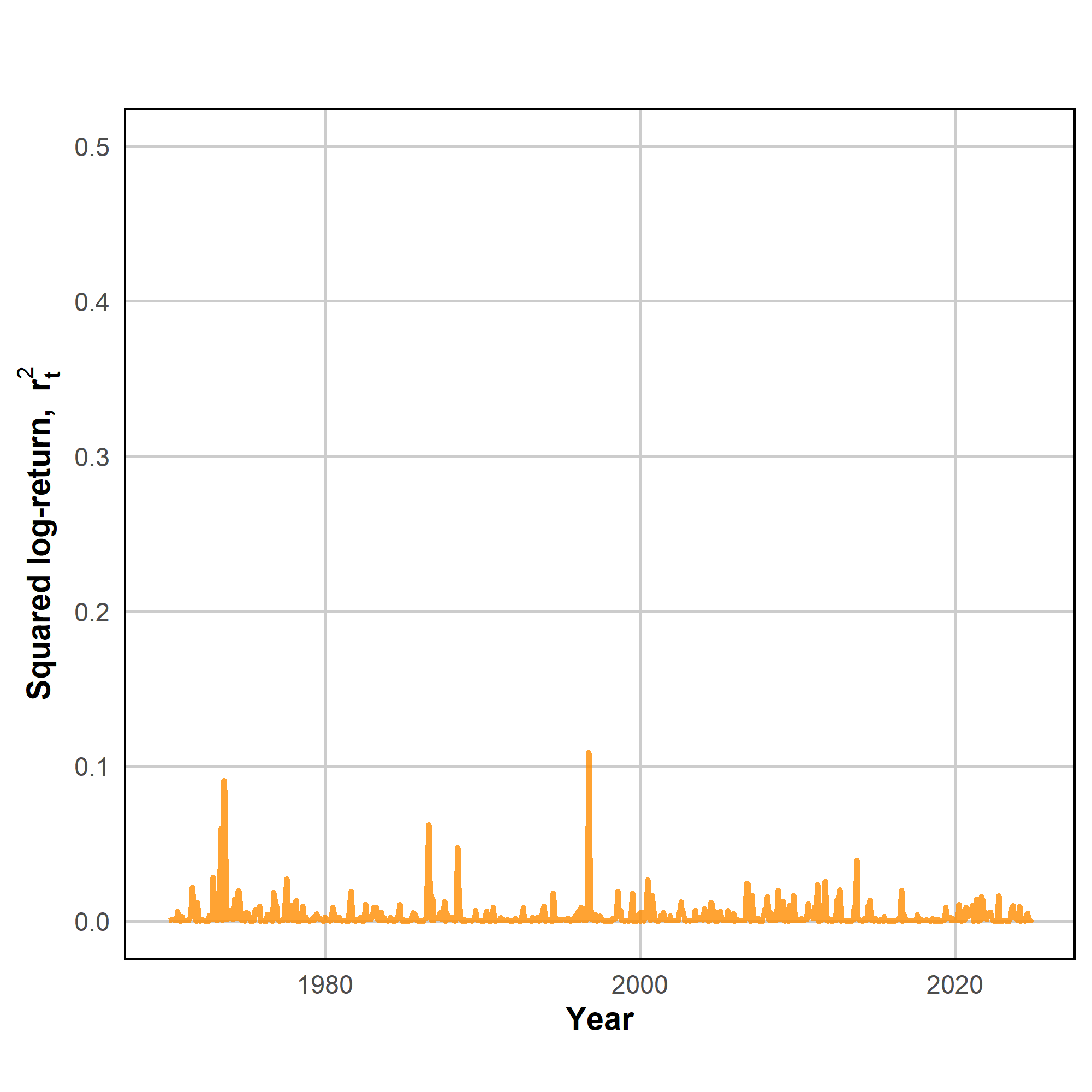}
    \includegraphics[width=0.3\textwidth, height=0.3\textwidth]{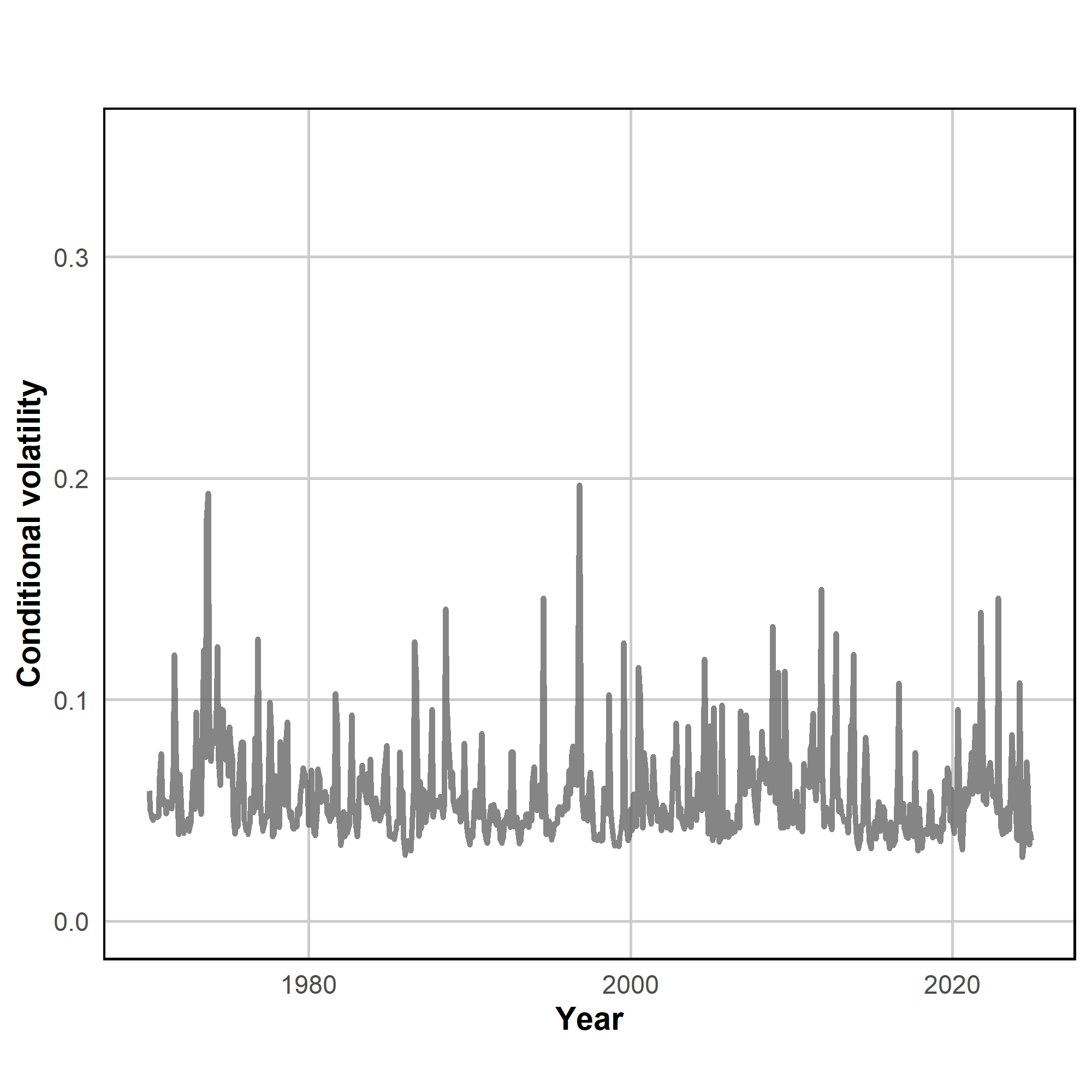}
    \caption{\textbf{Plots of log-returns, squared log-returns and EGARCH conditional volatility, for \textit{(A)} soybean in Illinois, \textit{(B)} wheat in North Dakota and \textit{(C)} corn in Iowa.} The purple line in the left panel represents the log-returns over time, and the dashed black line represents the mean which is zero. The orange line in the middle panel represents the squared log-returns. The grey line in the right panel represents the conditional volatility estimated using the EGARCH model.}
    \label{fig:us_returns_volatility_subplots}
\end{figure}

Figure~\ref{fig:india_returns_volatility_subplots} shows the conditional volatility estimates obtained using the EGARCH model for soybean in Madhya Pradesh, rice in Assam and cotton in Gujarat, while Figure~\ref{fig:us_returns_volatility_subplots} shows the corresponding estimates for soybean in Illinois, wheat in North Dakota, and corn in Iowa.

\subsection*{SARIMAX Model for Conditional Price Volatility Prediction}

The SARIMAX (\textit{Seasonal Autoregressive Integrated Moving Average with Exogenous Regressors}) model extends the classical ARIMA framework by incorporating seasonal components and exogenous variables in time-series analysis.~\cite{shumway2017time} 

\noindent The conditional price volatility (\(\sigma_t\)) estimated using the EGARCH model is used as a dependent variable and maximum temperature and precipitation are used as 

The SARIMAX model is written as:

\begin{equation}
    A_P(L^s)a_p(L)(1-L)^m(1-L^s)^M\sigma_t = B_Q(L^s)b_q(L)\varepsilon_t + \mathbf{z}_t\,\boldsymbol{\gamma},
\end{equation}
where:
\begin{itemize}
    \item \(A_P(L^s)\): Seasonal AR operator of order \(P\): \(A_P(L^s)=1-A_1L^s-A_2L^{2s}-\cdots-A_PL^{Ps}\).
    \item \(a_p(L)\): Non-seasonal AR operator of order \(p\): \(a_p(L)=1-a_1L-a_2L^2-\cdots-a_pL^p\).
    \item \(m\): Order of non-seasonal differencing.
    \item \(M\): Order of seasonal differencing.
    \item \(s\): Seasonal period (e.g., \(s=12\) for monthly data with annual seasonality).
    \item \(B_Q(L^s)\): Seasonal MA operator of order \(Q\): \(B_Q(L^s)=1-B_1L^s-B_2L^{2s}-\cdots-B_QL^{Qs}\).
    \item \(b_q(L)\): Non-seasonal MA operator of order \(q\): \(b_q(L)=1-b_1L-b_2L^2-\cdots-b_qL^q\).
    \item \(\varepsilon_t\): Error term.
    \item \(\mathbf{z}_t\): Vector of exogenous regressors (e.g., \texttt{tasmax} for maximum temperature and \texttt{pr} for precipitation).
    \item \(\boldsymbol{\gamma}\): Coefficient vector for the exogenous regressors.
\end{itemize}

\begin{figure*}
   \centering
    \includegraphics[width=0.3\textwidth, height=0.3\textwidth]{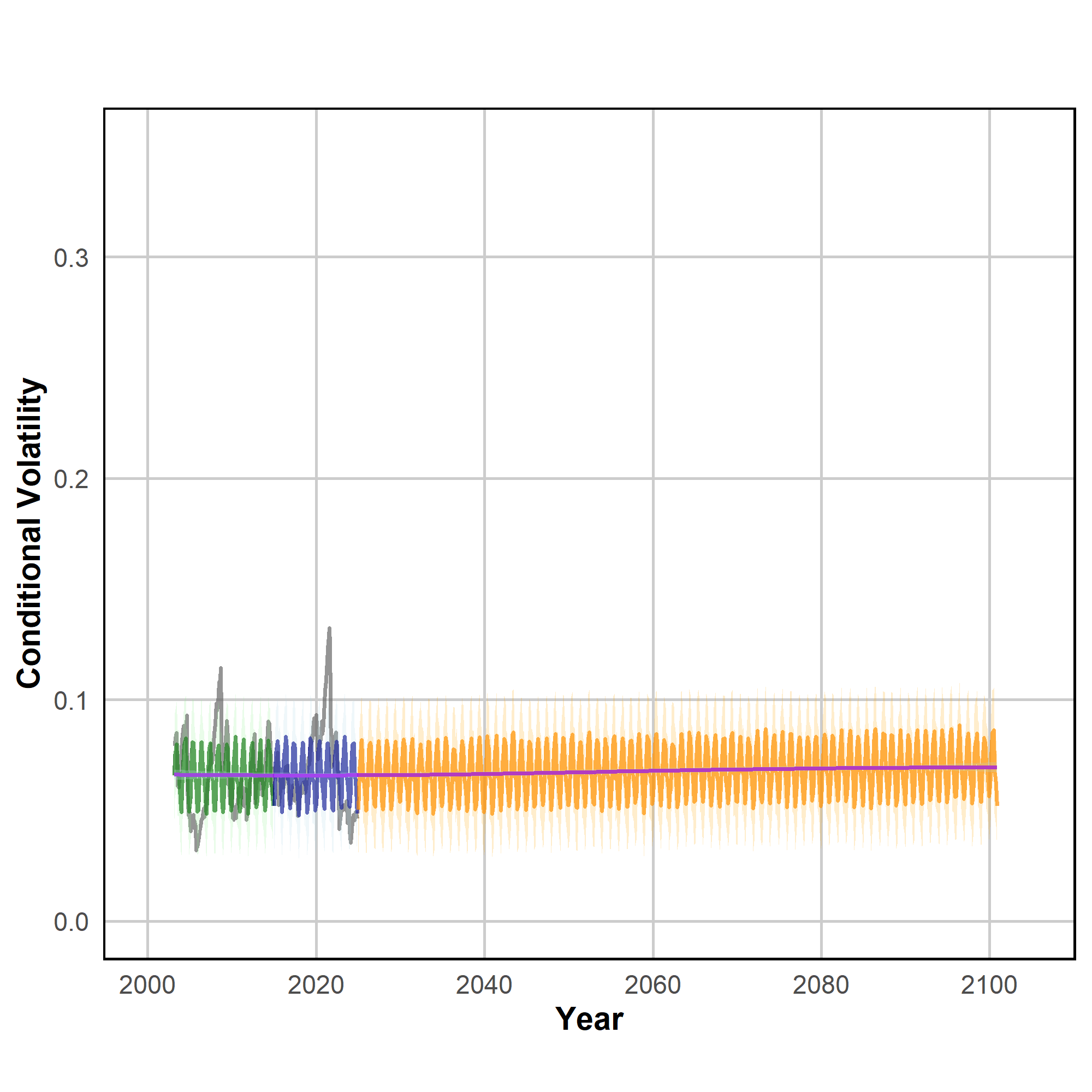}\llap{\parbox[b]{2.1in}{{\large\textsf{A}}\\\rule{0ex}{1.9in}}}
    \includegraphics[width=0.3\textwidth, height=0.3\textwidth]{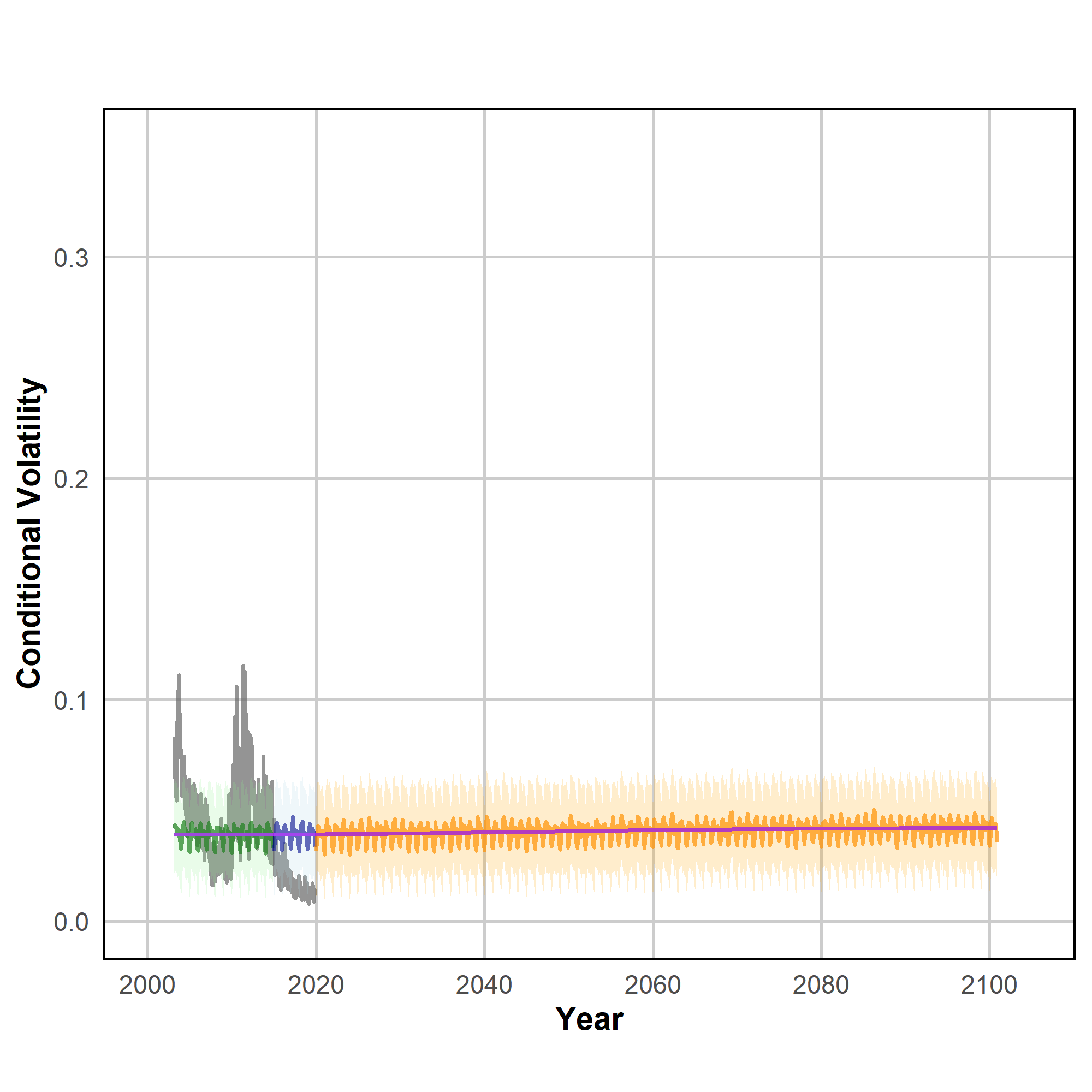}
    \includegraphics[width=0.3\textwidth, height=0.3\textwidth]{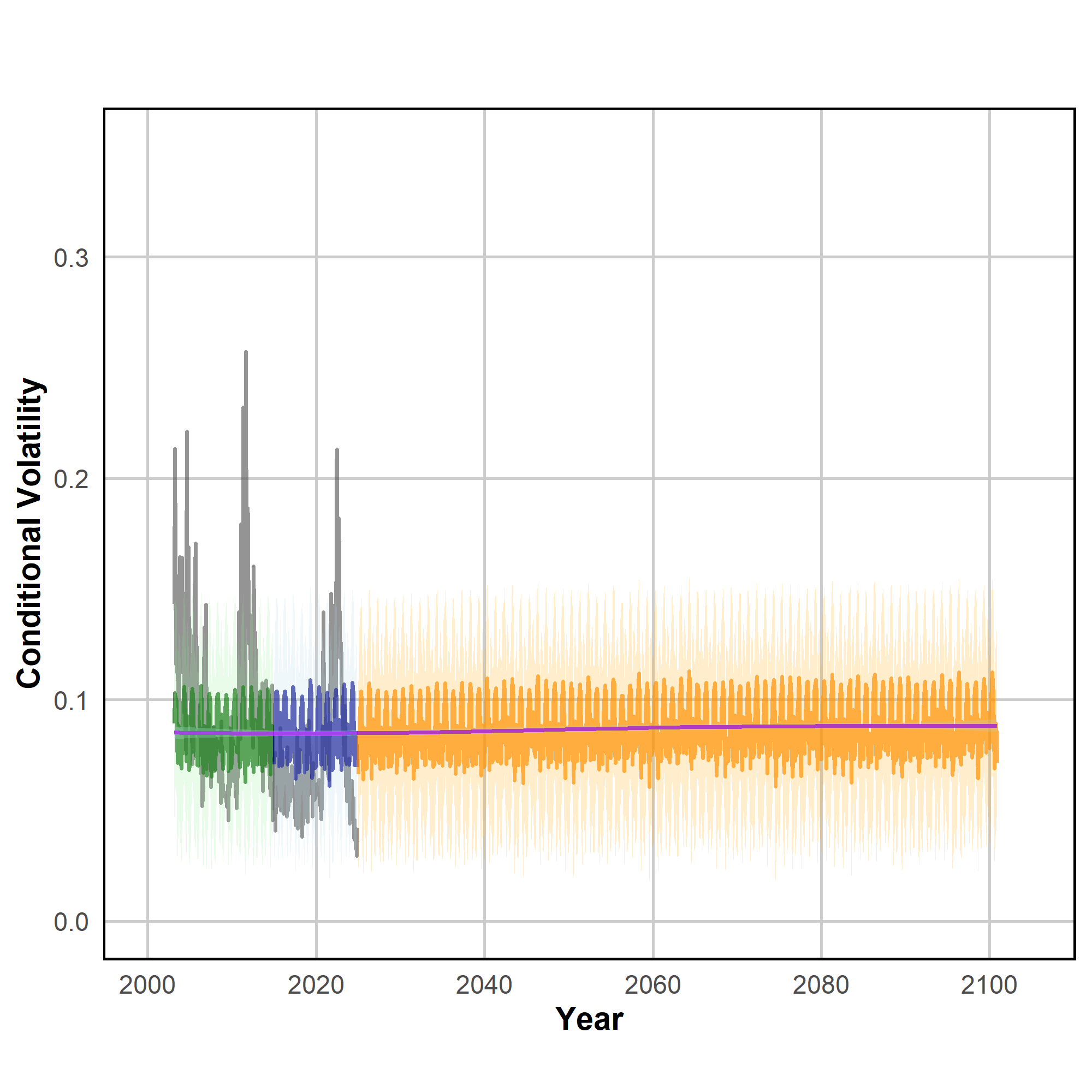}\\
    \includegraphics[width=0.3\textwidth, height=0.3\textwidth]{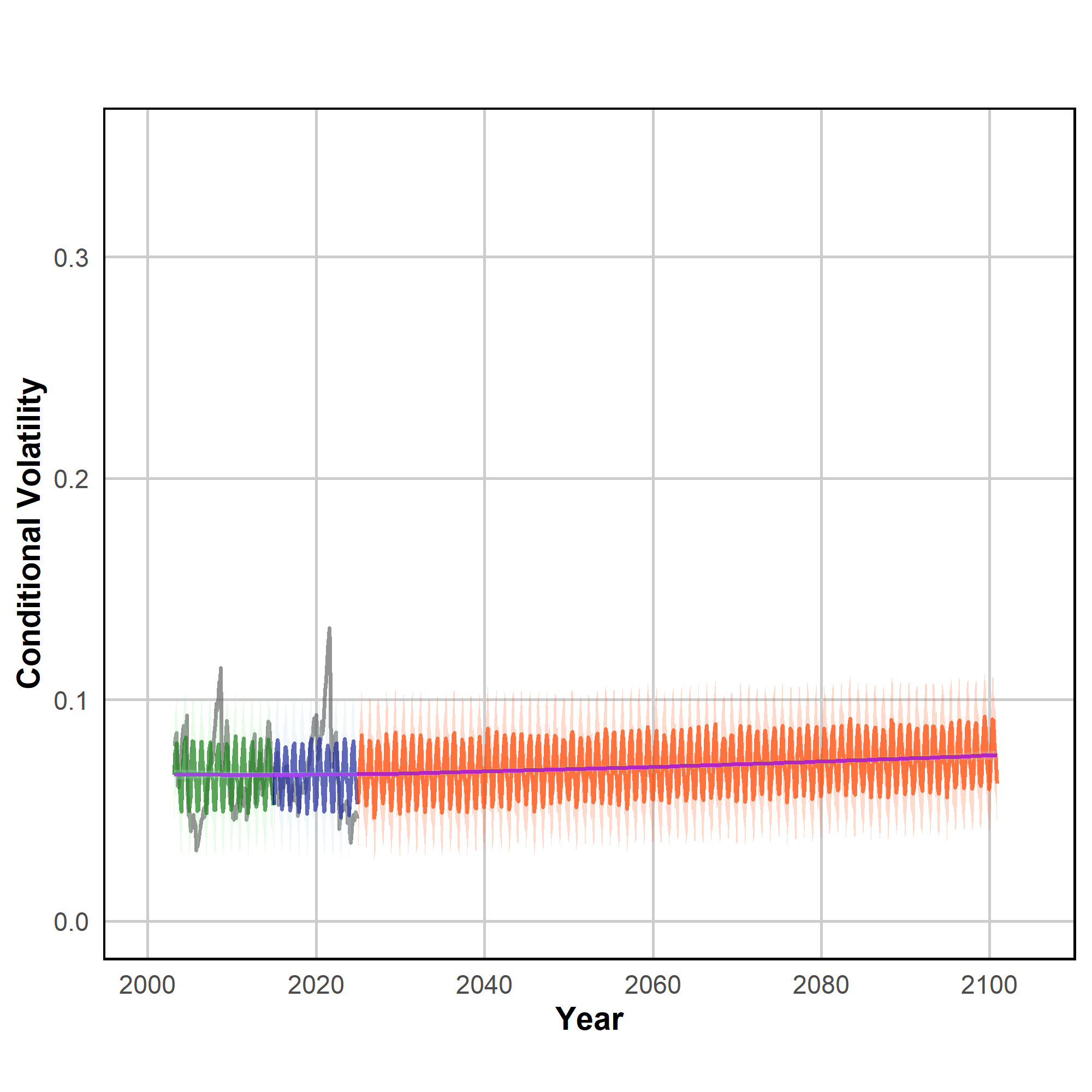}\llap{\parbox[b]{2.1in}{{\large\textsf{B}}\\\rule{0ex}{1.9in}}}
    \includegraphics[width=0.3\textwidth, height=0.3\textwidth]{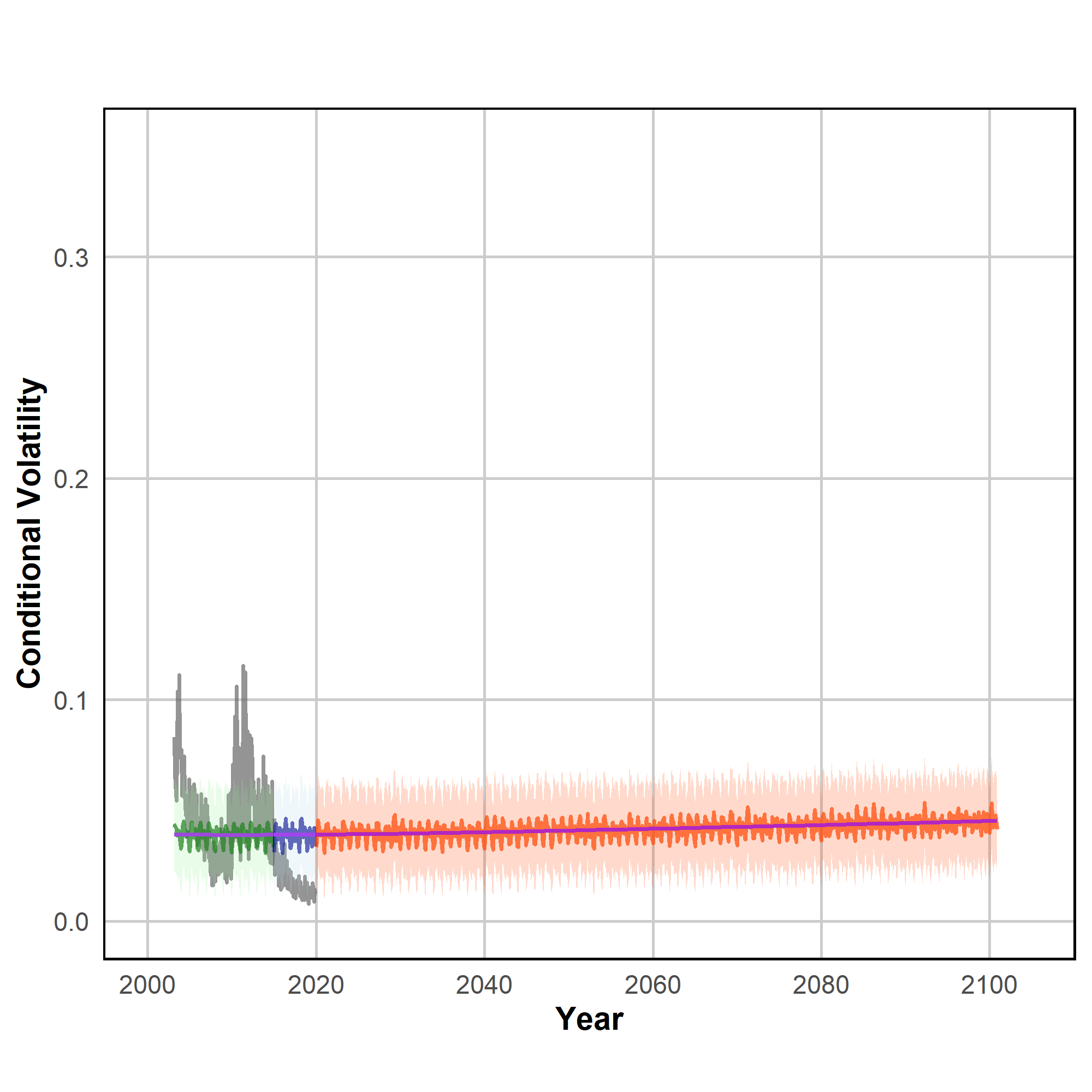}
    \includegraphics[width=0.3\textwidth, height=0.3\textwidth]{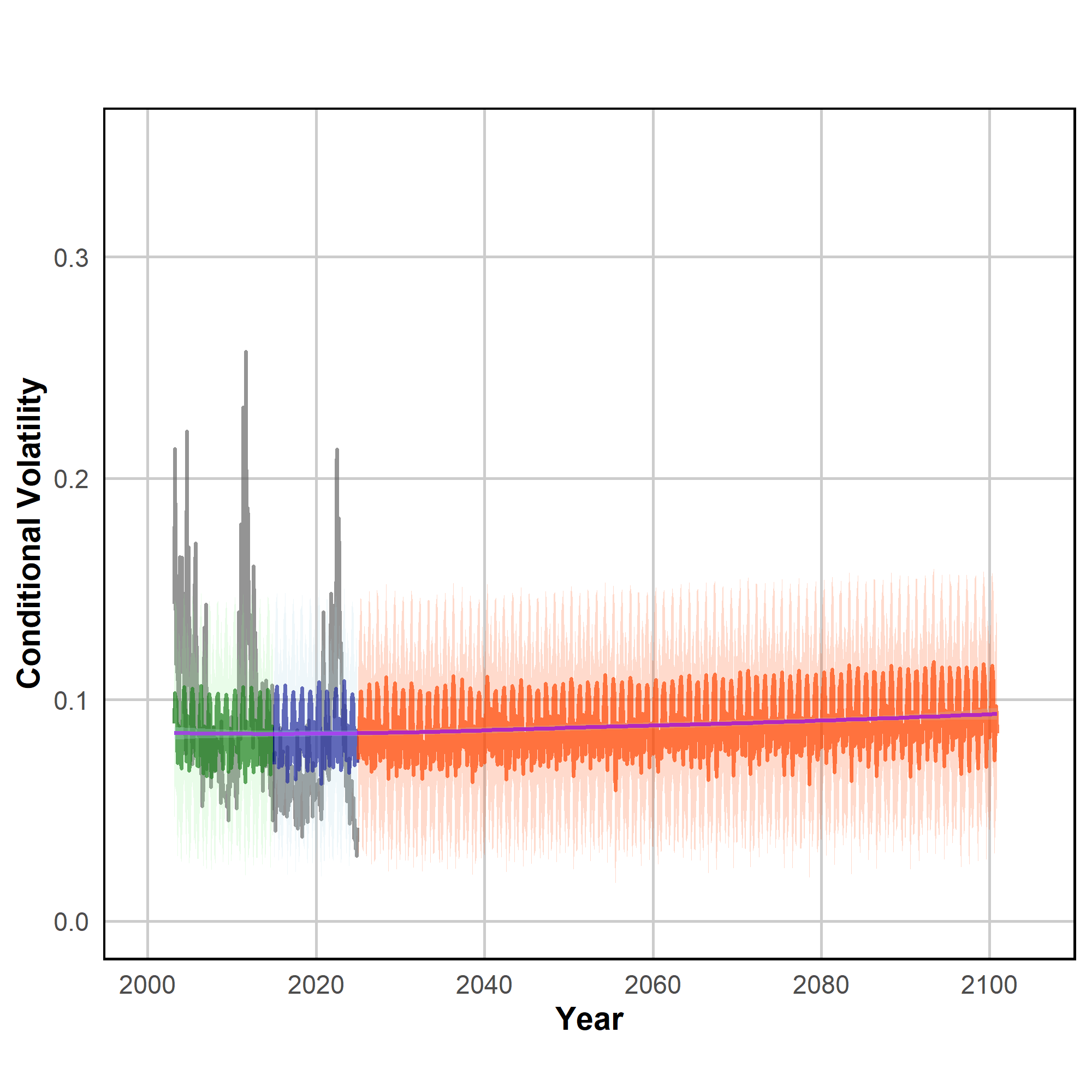}\\
    
     \caption{\textbf{Plots showing the conditional volatility estimates using SARIMAX with maximum temperature and precipitation as exogenous variables, for (left) soybean in Madhya Pradesh, (middle) rice in Assam and (right) cotton in Gujarat.} \textit{(A)} show the estimates with SSP2-4.5 scenario meteorological factors (maximum temperature and precipitation) as exogenous variables, while \textit{(B)} show the estimates with SSP5-8.5 scenario meteorological factors as exogenous variables. The grey lines represent the conditional volatility estimated using EGARCH while the dark green, dark blue and orange-red lines indicate predicted conditional volatility for historical, validation and forecast phases. Light green, light blue and light orange-red bands show the confidence intervals for these predictions. The purple lines show the smoothed trend summarising the overall pattern.}
    \label{fig:india_sarimax_subplots}
\end{figure*}

\begin{figure*}
   \centering
    \includegraphics[width=0.3\textwidth, height=0.3\textwidth]{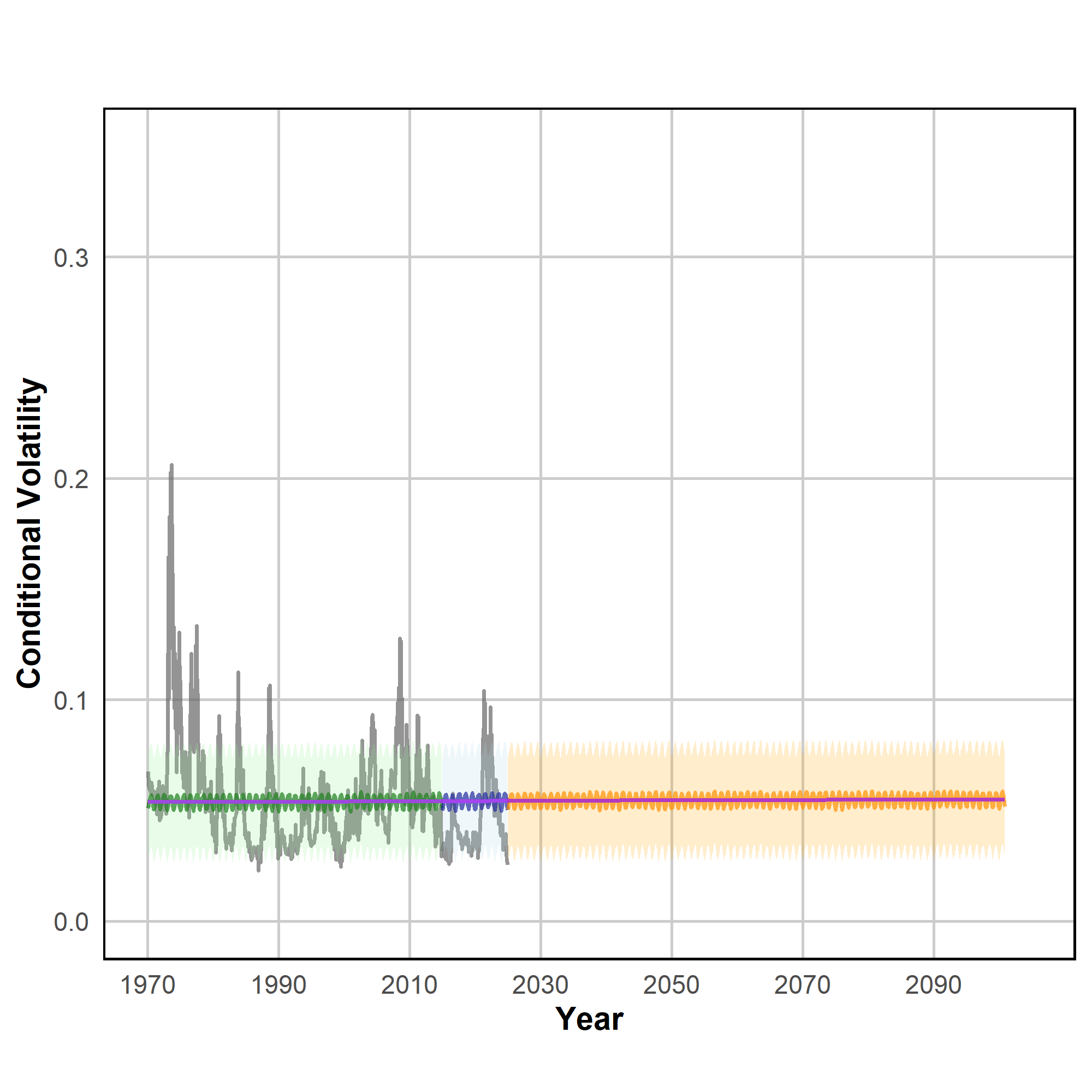}\llap{\parbox[b]{2.1in}{{\large\textsf{A}}\\\rule{0ex}{1.9in}}}
    \includegraphics[width=0.3\textwidth, height=0.3\textwidth]{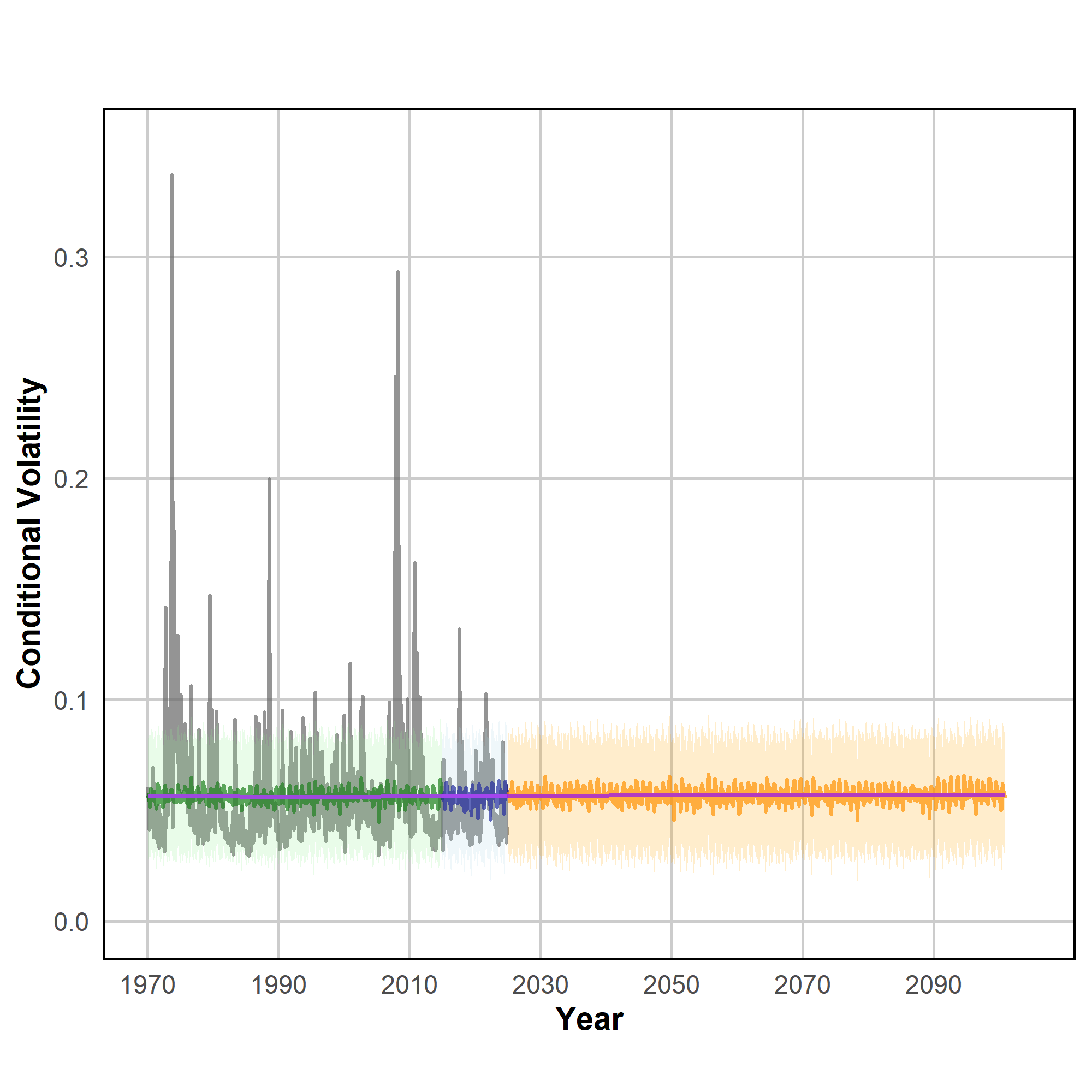}
    \includegraphics[width=0.3\textwidth, height=0.3\textwidth]{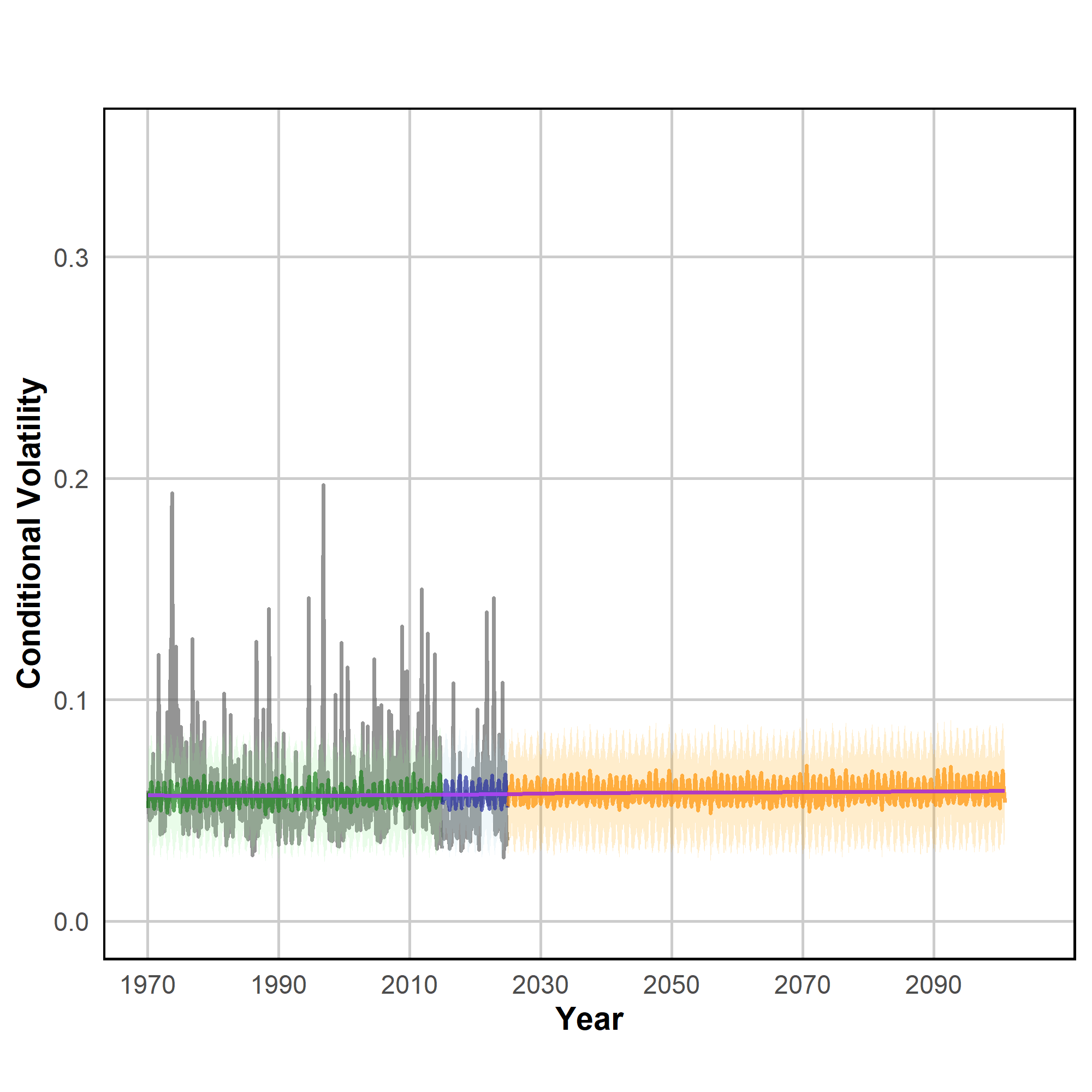}\\
    \includegraphics[width=0.3\textwidth, height=0.3\textwidth]{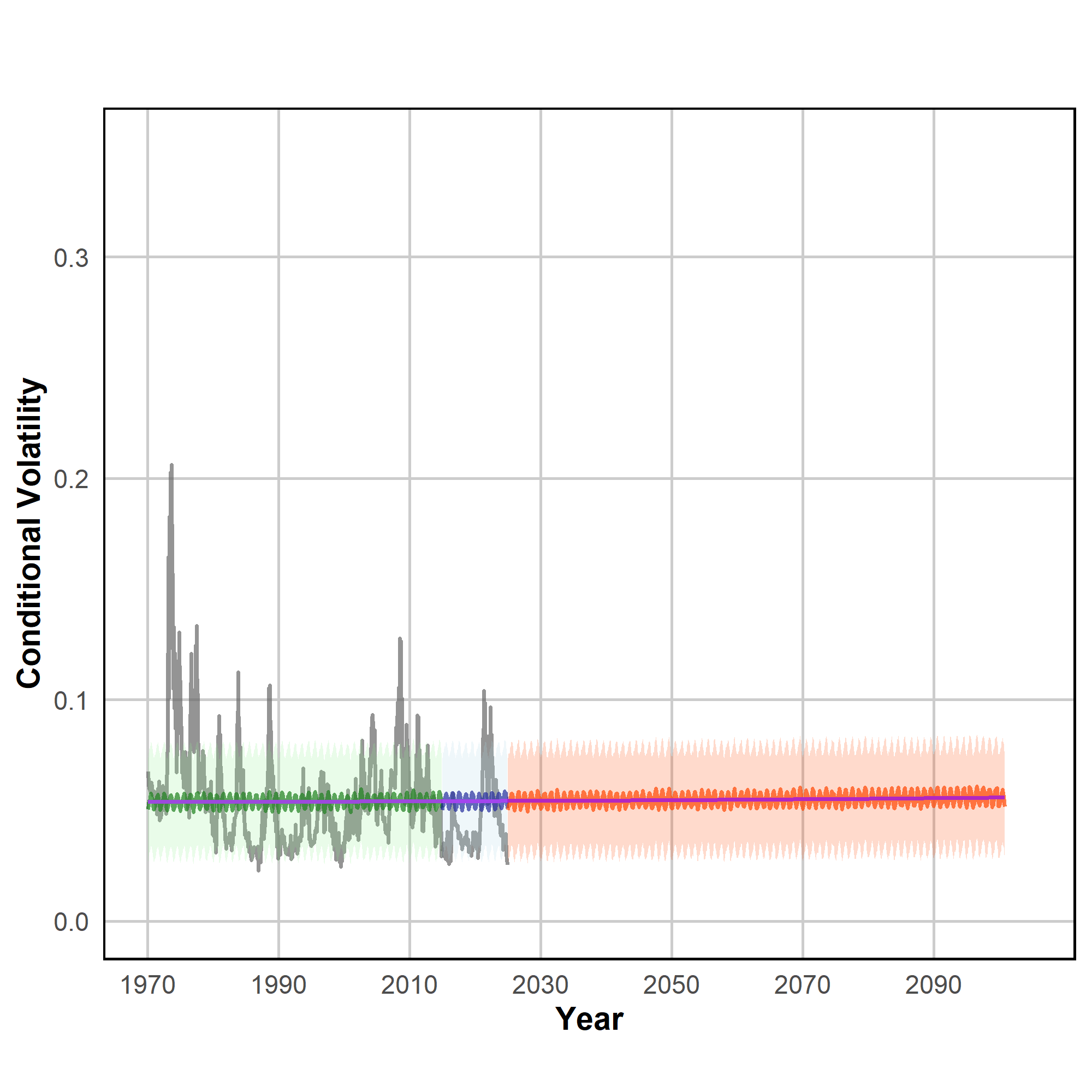}\llap{\parbox[b]{2.1in}{{\large\textsf{B}}\\\rule{0ex}{1.9in}}}
    \includegraphics[width=0.3\textwidth, height=0.3\textwidth]{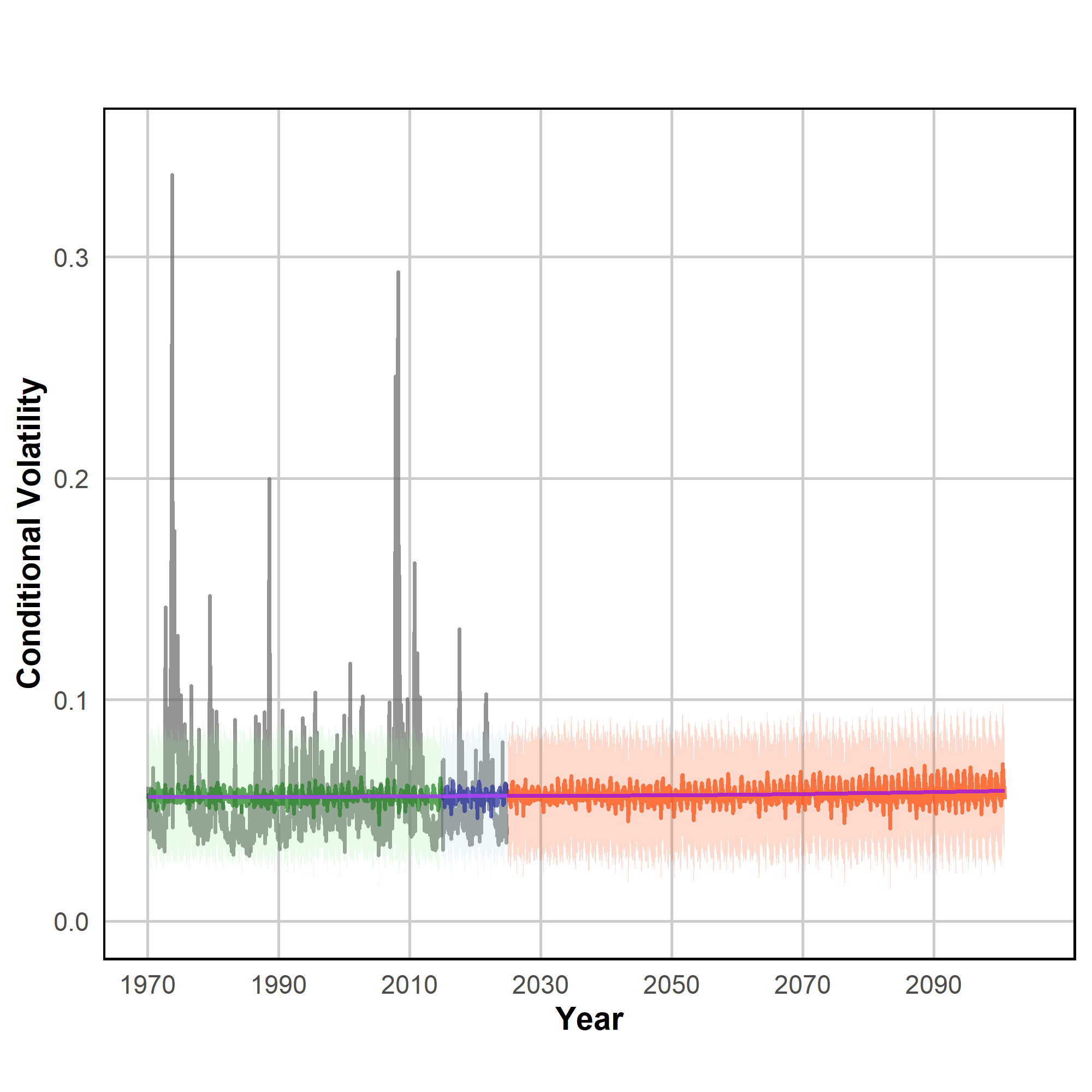}
    \includegraphics[width=0.3\textwidth, height=0.3\textwidth]{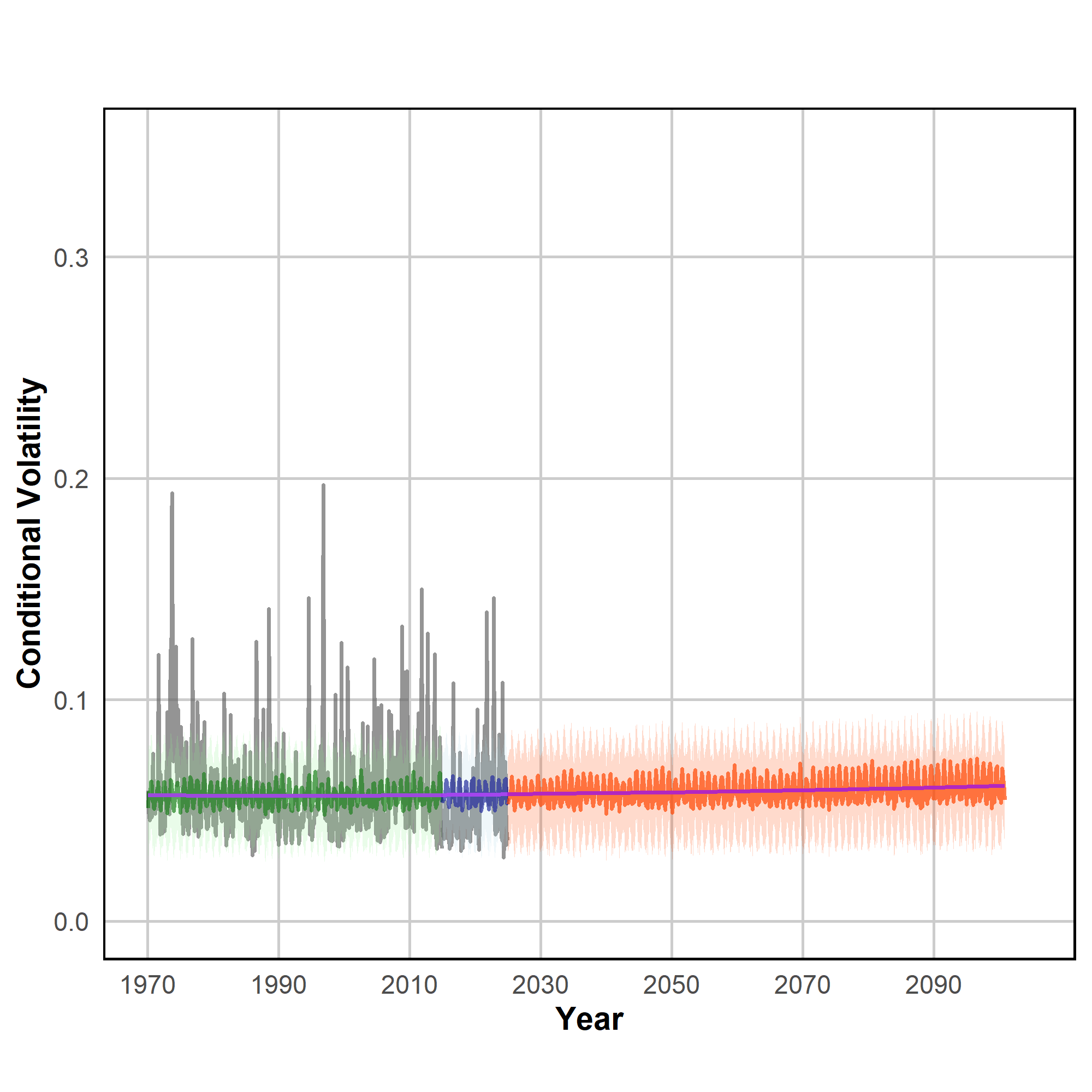}\\
    
     \caption{\textbf{Plots showing the conditional volatility estimates using SARIMAX with maximum temperature and precipitation as exogenous variables, for (left) soybean in Illinois, (middle) wheat in North Dakota and (right) corn in Iowa.} \textit{(A)} show the estimates with SSP2-4.5 scenario meteorological factors (maximum temperature and precipitation) as exogenous variables, while \textit{(B)} show the estimates with SSP5-8.5 scenario meteorological factors as exogenous variables. The grey lines represent the conditional volatility estimated using EGARCH while the dark green, dark blue and orange-red lines indicate predicted conditional volatility for historical, validation and forecast phases. Light green, light blue and light orange-red bands show the confidence intervals for these predictions. The purple lines show the smoothed trend summarising the overall pattern.}
    \label{fig:us_sarimax_subplots}
\end{figure*}

\noindent The fitted SARIMAX model is used to predict the conditional volatility (\(\hat{\sigma}_t\)) and construct 68\% prediction intervals:
\[
\text{Lower Bound: } \hat{\sigma}_t - 1 \cdot \text{SE}_t
\]
\[
\text{Upper Bound: } \hat{\sigma}_t + 1 \cdot \text{SE}_t,
\]

\noindent as shown in Figure~\ref{fig:india_sarimax_subplots} and \ref{fig:us_sarimax_subplots}.

One could directly incorporate exogenous climate variables into the variance equation of the EGARCH model. However, this method works best for short- and medium-term forecasting. On the other hand, the SARIMAX model is well-suited for long-term forecasting. Hence, we have adopted a two-step approach to fully utilise the SARIMAX model's potential for long-term volatility risk forecasting using meteorological predictions derived from Global Climate Models (GCM) as exogenous variables.

We first fit the EGARCH model using the log-returns (\(r_t\)) to estimate the conditional volatility. Next, the SARIMAX model uses external variables (\(\mathbf{x}_t\)) such as maximum temperature and precipitation to predict the estimated conditional volatility. This two-step approach combines financial time series modelling with meteorological data, providing a reliable framework for forecasting volatility across various time horizons and different climate scenarios.

%%%%%%%%%%%%%%%%%%%%%%%%%%%%%%%%%%%%%%%%%%%%%%%%%%%%%%%%%%%%%%%%%%%%%%%%%%%%%%
\subsection*{Black-Scholes put option framework for calculating premiums}

In this study, crop price insurance is conceptualized as a European put option, under which farmers are protected against adverse price movements. This framework allows the application of option pricing theory to quantify the premium of a contract that provides downside protection over a specified horizon.

We assume that the inflation-adjusted crop price $S_t$ follows a geometric Brownian motion under the risk-neutral measure with deterministic, time-varying volatility:
\begin{equation}
\frac{dS_t}{S_t} = r\,dt + \sigma(t)\,dW_t,
\label{eq:gbm_timevarying}
\end{equation}
where $r$ is the risk-free interest rate, $W_t$ is a standard Brownian motion, and $\sigma(t)$ is a deterministic volatility function specified ex ante using SARIMAX-based conditional volatility forecasts. This formulation generalizes the classical Black-Scholes model, allowing volatility to vary over time while retaining analytical tractability and incorporating forward-looking, climate-sensitive risk.

Integrating Eq.~\eqref{eq:gbm_timevarying} from $0$ to $T$ yields
\begin{equation}
\log S_T
=
\log S_0
+
\int_0^T \left(r - \tfrac{1}{2}\sigma(t)^2\right)\,dt
+
\int_0^T \sigma(t)\,dW_t.
\end{equation}
Since $\sigma(t)$ is deterministic, the stochastic integral satisfies
\begin{equation}
\int_0^T \sigma(t)\,dW_t
\sim
\mathcal{N}\!\left(0,\;\int_0^T \sigma(t)^2\,dt\right),
\end{equation}
implying that the terminal log-price is normally distributed:
\begin{equation}
\log S_T
\sim
\mathcal{N}\!\left(
\log S_0 + rT - \tfrac{1}{2} I_T,\;
I_T
\right),
\label{eq:lognormal_terminal}
\end{equation}
where the integrated variance is defined as
\begin{equation}
I_T = \int_0^T \sigma(t)^2\,dt.
\end{equation}
Thus, even in the presence of time-varying volatility, the terminal crop price remains lognormally distributed.

Defining the effective volatility over the contract horizon $[0,T]$ as
\begin{equation}
\sigma_{\mathrm{eff}}
=
\sqrt{\frac{1}{T}\int_0^T \sigma(t)^2\,dt},
\label{eq:effective_vol}
\end{equation}
Eq.~\eqref{eq:lognormal_terminal} coincides exactly with the standard Black-Scholes terminal distribution evaluated at volatility $\sigma_{\mathrm{eff}}$. Consequently, the price of the put option contract is given by the Black-Scholes formula,
\begin{equation}
P = e^{-rT} K \Phi(-d_2) - S_0 \Phi(-d_1),
\end{equation}
where
\begin{equation}
d_1 = \frac{\ln(S_0 / K) + (r + \tfrac{1}{2}\sigma_{\mathrm{eff}}^2)T}{\sigma_{\mathrm{eff}} \sqrt{T}},
\qquad
d_2 = d_1 - \sigma_{\mathrm{eff}} \sqrt{T}.
\end{equation}
Here, $P$ denotes the insurance premium, $S_0$ is the inflation-adjusted spot price at the beginning of the insurance period, $K$ is the strike price defined as a fixed fraction below the spot price, $r$ is the annualized risk-free interest rate, and $T$ is the contract maturity.

In the empirical implementation, we set the strike price as $K = (1 - 0.05)S_0$, representing a 5\% downside protection threshold. The risk-free rate is fixed at $r = 0.07$, and the maturity is set to $T = 12$ months. The volatility input $\sigma_{\mathrm{eff}}$ corresponds to the root-mean-square volatility derived from the integrated future variance obtained using SARIMAX-based conditional volatility forecasts. This formulation demonstrates that, for European-style insurance contracts, deterministic time-varying volatility affects pricing exclusively through the integrated variance, yielding an exact reduction to a constant-volatility Black-Scholes model with effective volatility $\sigma_{\mathrm{eff}}$.

Our proposed framework extends the classical Black-Scholes paradigm by allowing volatility to evolve deterministically over time, while retaining analytical tractability and avoiding reliance on option market data. The resulting insurance premium would reflect the cumulative future variance implied by climate-conditioned volatility forecasts, providing a transparent and economically interpretable measure of risk.
While this approach would not account for volatility risk premia or implied volatility smiles, would account for pricing European-style crop insurance contracts in data-scarce settings where forward-looking volatility forecasts are available but option market calibration is not feasible.
%%%%%%%%%%%%%%%%%%%%%%%%%%%%%%%%%%%%%%%%%%%%%%%%%%%%%%%%%%%%%%%%%%%%%%%%%%%%%

\section*{Results and Discussions}

\begin{figure*}
    \centering
     \includegraphics[width=0.3\linewidth]{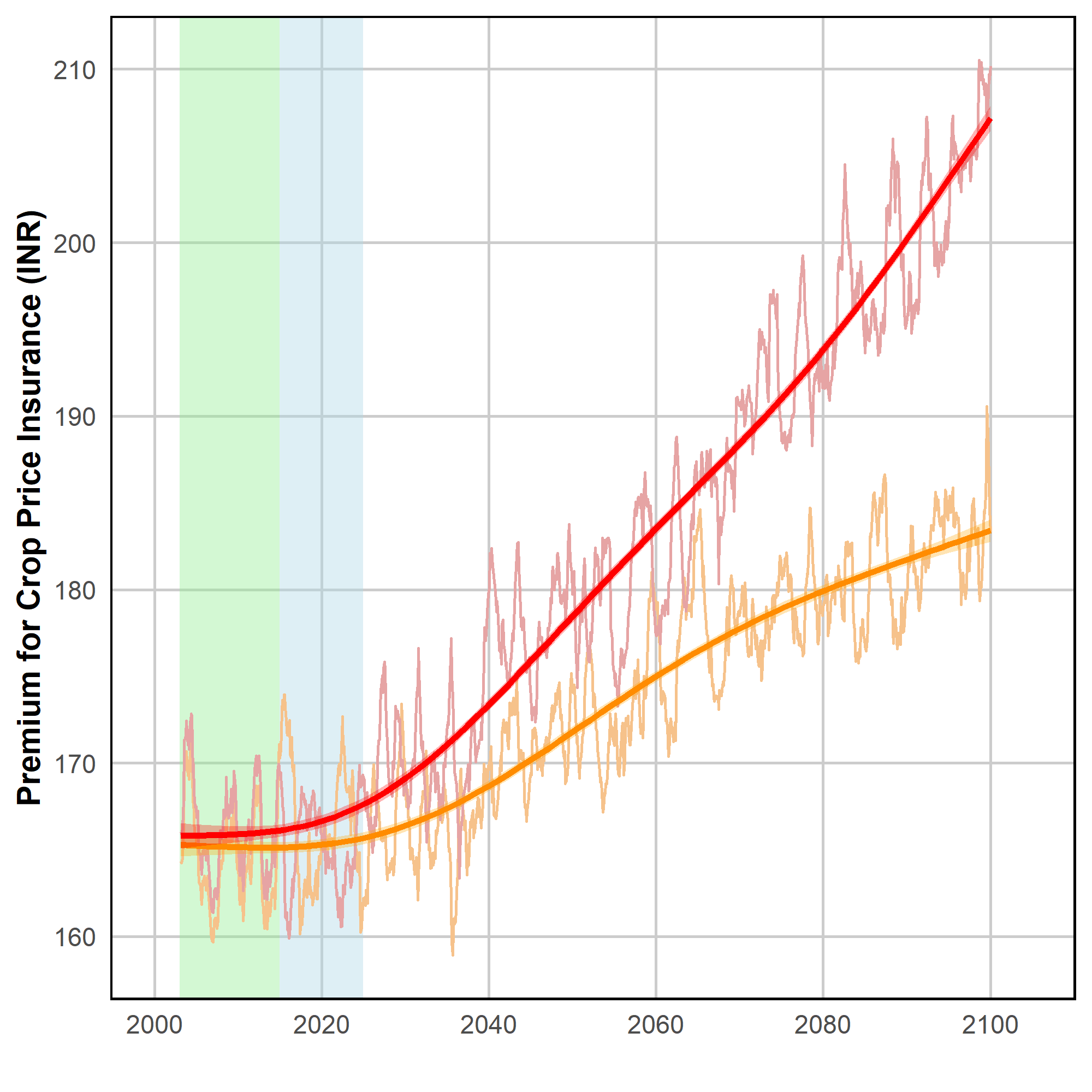}\llap{\parbox[b]{2.1in}{{\large\textsf{A}}\\\rule{0ex}{2.1in}}}
     \includegraphics[width=0.3\linewidth]{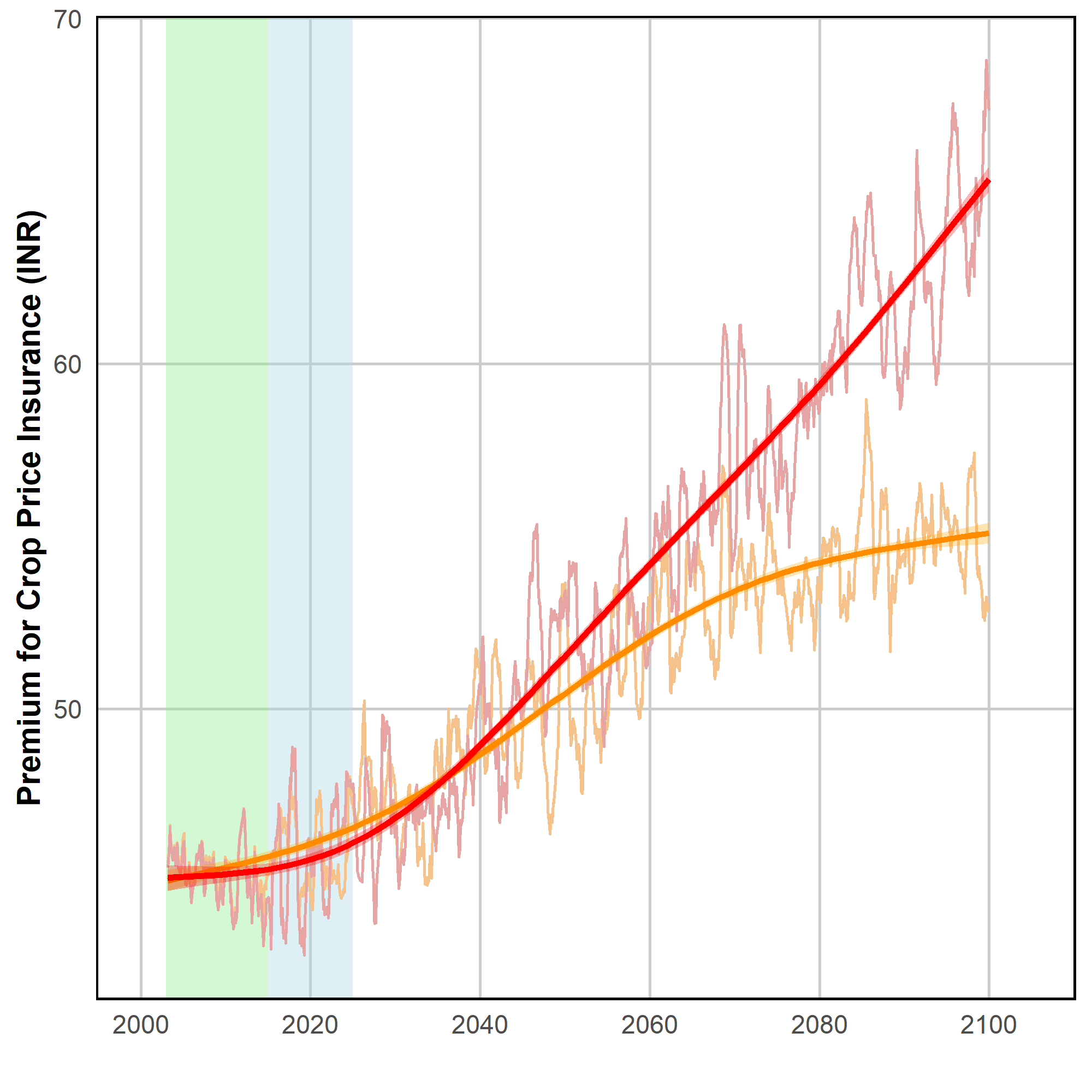}\llap{\parbox[b]{2.1in}{{\large\textsf{B}}\\\rule{0ex}{2.1in}}}
     \includegraphics[width=0.3\linewidth]{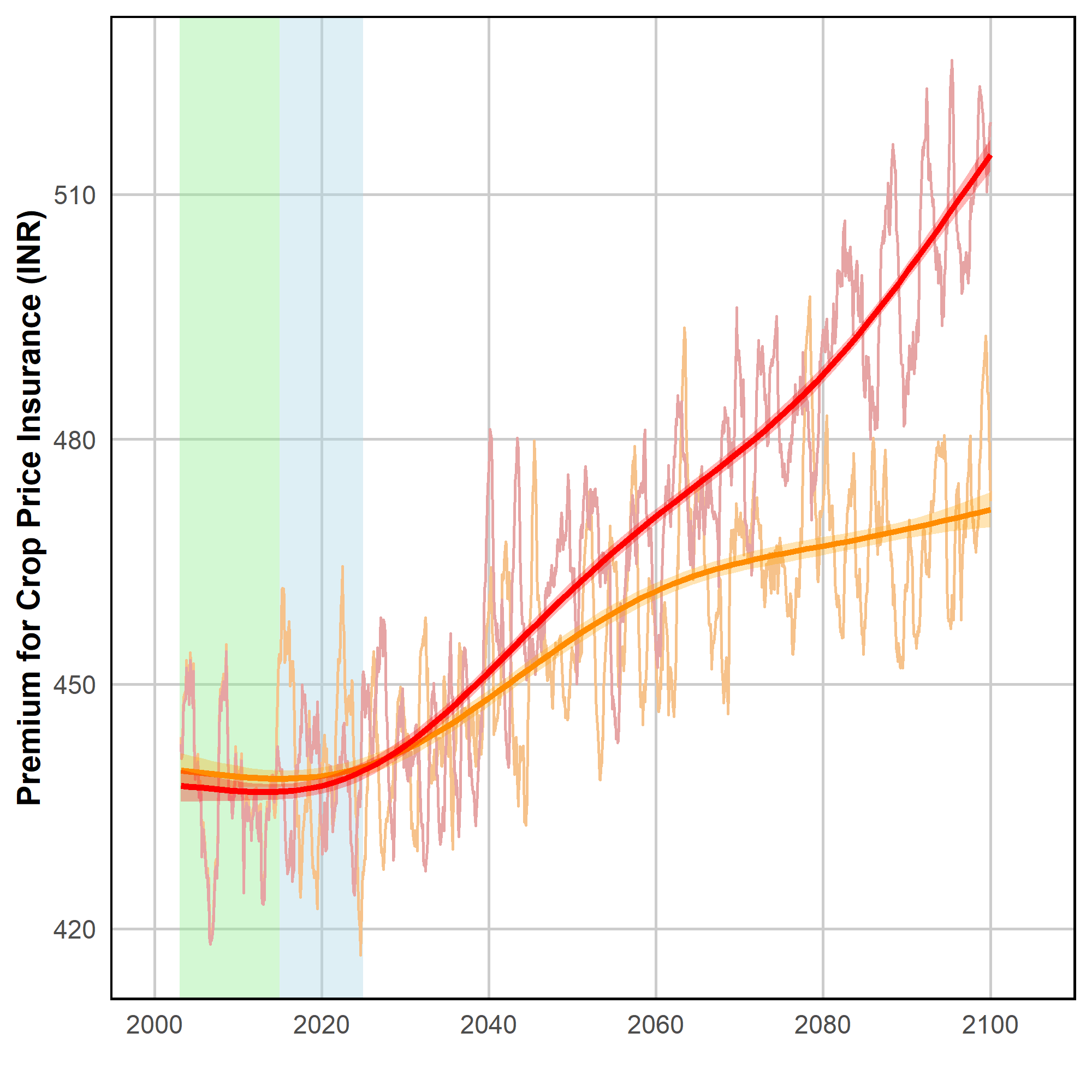}\llap{\parbox[b]{2.1in}{{\large\textsf{C}}\\\rule{0ex}{2.1in}}}
     \caption{\textbf{Comparative plots of crop insurance premiums under two scenarios SSP2-4.5 and SSP5-8.5, for \textit{(A)} soybean in Madhya Pradesh, \textit{(B)} rice in Assam and \textit{(C)} cotton in Gujarat.} The dark orange smoothed lines represent premiums under SSP2-4.5, while the red smoothed lines represent SSP5-8.5. The shaded green and blue regions correspond to the historical and validation phases, respectively.}
    \label{fig:india_premium_subplots}
\end{figure*}

\begin{figure*}
    \centering
     \includegraphics[width=0.3\linewidth]{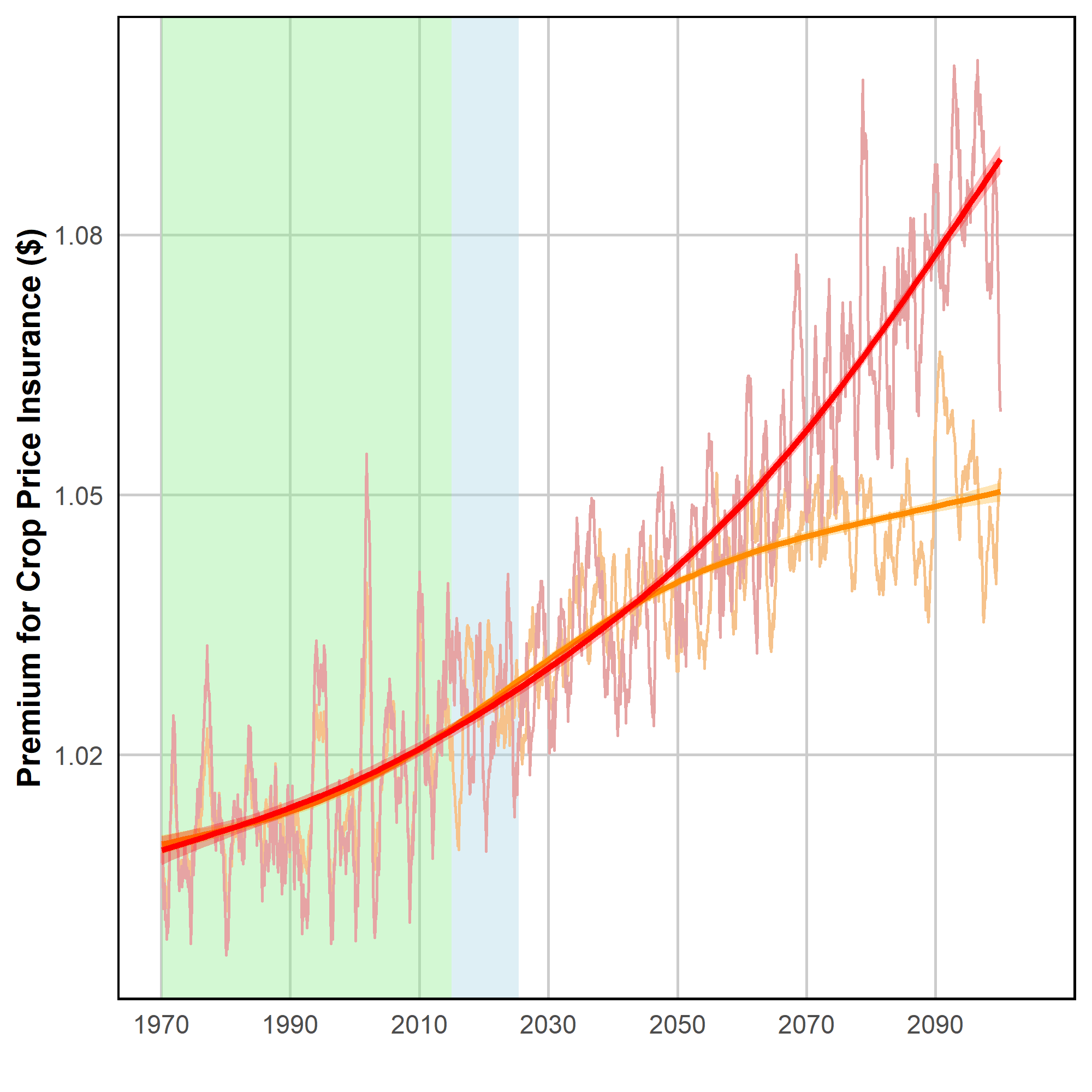}\llap{\parbox[b]{2.1in}{{\large\textsf{A}}\\\rule{0ex}{2.1in}}}
     \includegraphics[width=0.3\linewidth]{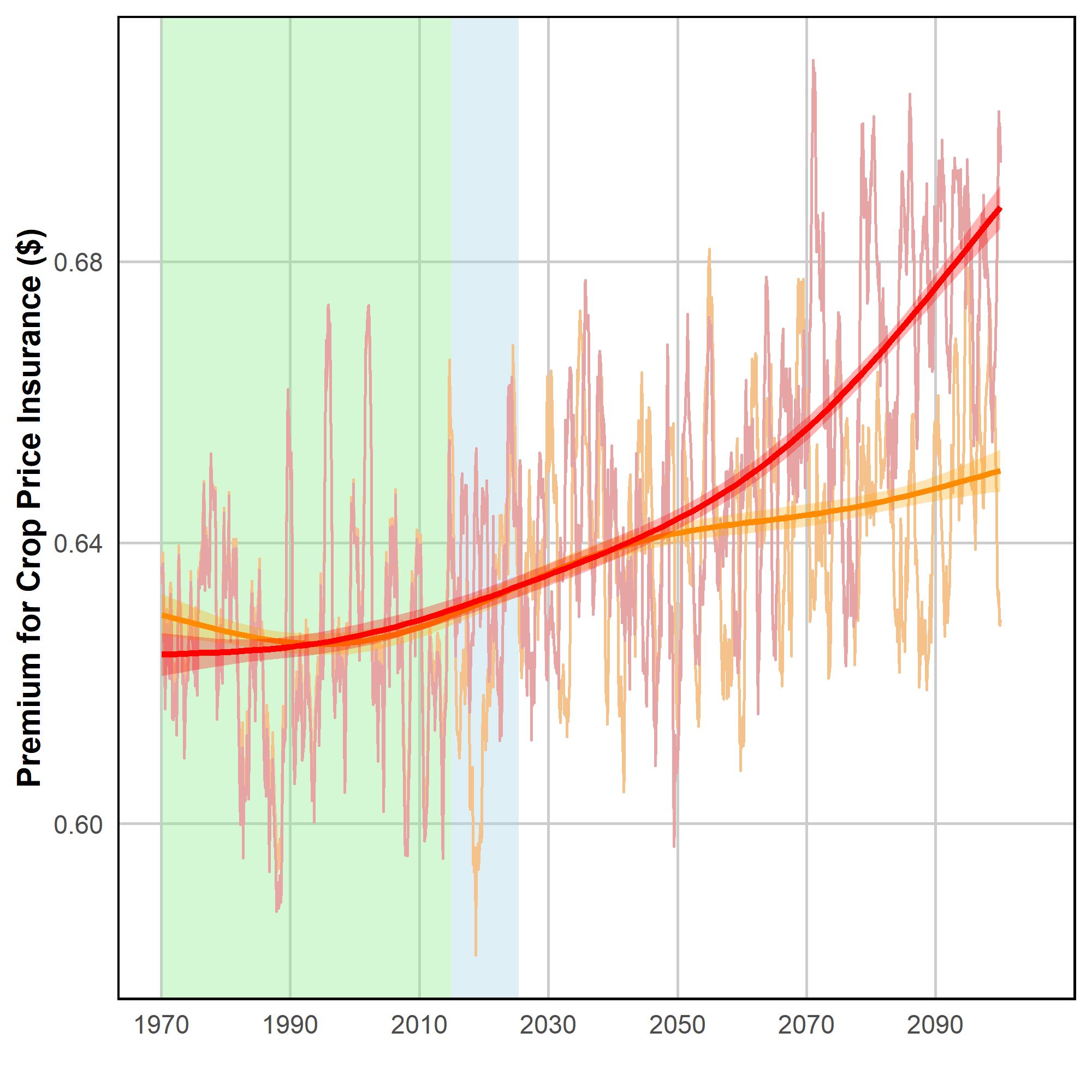}\llap{\parbox[b]{2.1in}{{\large\textsf{B}}\\\rule{0ex}{2.1in}}}
     \includegraphics[width=0.3\linewidth]{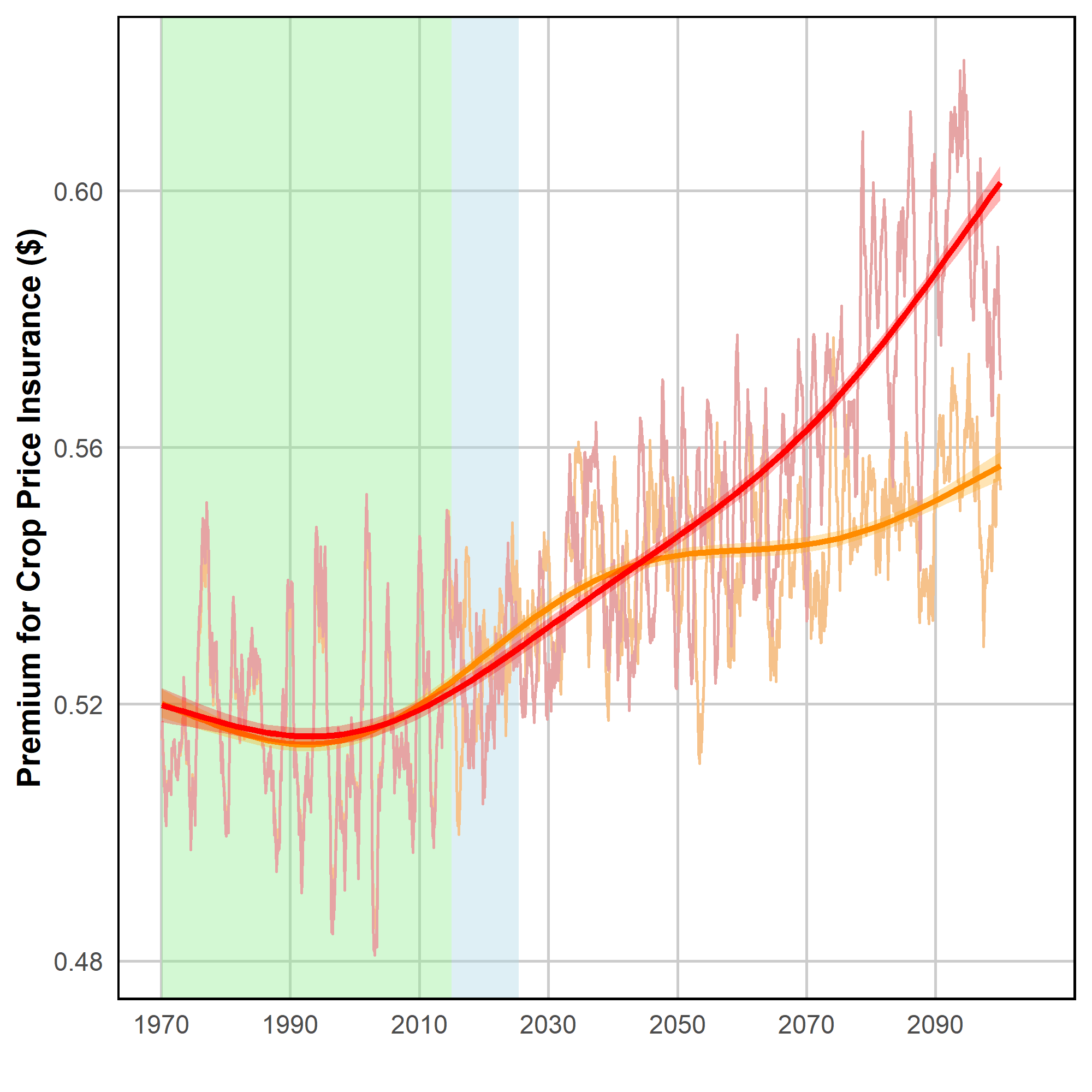}\llap{\parbox[b]{2.1in}{{\large\textsf{C}}\\\rule{0ex}{2.1in}}}
     \caption{\textbf{Comparative plots of crop insurance premiums under two scenarios SSP2-4.5 and SSP5-8.5, for \textit{(A)} soybean in Illinois, \textit{(B)} wheat in North Dakota and \textit{(C)} corn in Iowa.} The dark orange smoothed lines represent premiums under SSP2-4.5, while the red smoothed lines represent SSP5-8.5. The shaded green and blue regions correspond to the historical and validation phases, respectively.}
    \label{fig:us_premium_subplots}
\end{figure*}

%This section presents and interprets the key findings of our analysis, as illustrated in Figures~\ref{fig:india_price_subplots}–\ref{fig:us_price_subplots}. 
Figures~\ref{fig:india_price_subplots} and~\ref{fig:us_price_subplots} show the crop price time series with Bollinger bands, highlighting fluctuations in both Indian and U.S. markets. In India, cotton and soybean display stronger swings than rice, while in the U.S., corn and soybean are more volatile than wheat. Figures~\ref{fig:india_meteorological_subplots} and~\ref{fig:us_meteorological_subplots} present the historical and projected climate trends. Maximum temperatures rise more sharply under SSP5-8.5 than SSP2-4.5, while precipitation exhibits increasing variability in both India and the U.S. 
Figures~\ref{fig:india_returns_volatility_subplots} and~\ref{fig:us_returns_volatility_subplots} illustrate EGARCH-based conditional volatility estimates. Rice in Assam shows relatively low volatility, except during the 2011–12 price spike~\cite{AssamTribune2011}, while corn in Iowa displays strong fluctuations driven by climate sensitivity and demand shocks. 
Figures~\ref{fig:india_sarimax_subplots} and~\ref{fig:us_sarimax_subplots} depict SARIMAX volatility forecasts that incorporate temperature and precipitation as exogenous regressors. These forecasts consistently show higher volatility under SSP5-8.5. 
Finally, Figures~\ref{fig:india_premium_subplots} and~\ref{fig:us_premium_subplots} display smoothed insurance premiums derived from the Black–Scholes model, where scenario divergence occurs at different times across crops: after 2060 for soybean, 2080 for rice, and as early as 2030 for cotton in India; and earlier for corn in Iowa, with soybean in Illinois diverging mid-century and wheat in North Dakota showing later differences.

These results underscore the pivotal role of climate variability in shaping agricultural price behavior. However, this influence is not uniform across commodities or regions. Distinct crops exhibit markedly different volatility patterns: rice in Assam and wheat in North Dakota remain comparatively stable, while cotton in Gujarat and corn in Iowa display pronounced sensitivity to both climatic disturbances and market fluctuations. The contrast between the SSP2-4.5 and SSP5-8.5 scenarios is particularly evident in the SARIMAX-based projections, where the higher-emission pathway consistently produces elevated and more persistent volatility. This widening gap is directly reflected in the pricing of risk, as higher volatility under SSP5-8.5 translates into increased insurance premiums. Such increases capture the rising financial burden associated with safeguarding farmers against unfavorable price movements. Notably, the onset of divergence between the scenarios is crop-specific: cotton and corn respond earlier to changing climate conditions, whereas rice and wheat exhibit more delayed adjustments.

The framework further establishes that put-option based insurance provides a robust and operational mechanism for addressing climate-induced price uncertainty. By integrating climate-sensitive volatility estimates into option pricing, the study offers a transparent and quantitative approach to evaluating the cost of financial protection under different climate trajectories. This framework carries tangible benefits across stakeholders. Farmers gain access to tools that enable effective hedging against income variability, thereby enhancing livelihood stability. Insurers benefit from improved risk assessment, allowing for more accurate and responsive premium setting. Policymakers, in turn, are equipped with actionable insights to design region- and crop-specific interventions, particularly in areas where climate exposure is high and market risks are amplified.

Three key conclusions emerge from this work. First, climate variability is a primary driver of agricultural price volatility, though its magnitude and timing vary significantly across crops and geographic contexts. Second, the inclusion of climate variables within econometric frameworks (specifically through SARIMAX modeling) substantially enhances the accuracy and reliability of volatility forecasts. Third, converting these refined forecasts into option-based insurance premiums bridges the gap between predictive modeling and practical risk management, offering a scalable financial instrument for the agricultural sector.

Finally, the findings highlight the importance of a multidisciplinary approach that combines climate science, econometric modeling, and financial engineering. Such integration not only improves the understanding of climate–market interactions but also facilitates the development of adaptive mechanisms to manage risk. By aligning predictive insights with actionable financial tools, this approach contributes to strengthening the resilience of agricultural systems in two diverse countries/economies, India and the United States, under evolving climate conditions.

\section*{Acknowledgements}

Anish Rai is grateful to AlgoLabs for financial support and CMI for hospitality.

\bibliography{sample}

% \noindent LaTeX formats citations and references automatically using the bibliography records in your .bib file, which you can edit via the project menu. Use the cite command for an inline citation, e.g.  \cite{Hao:gidmaps:2014}.

% For data citations of datasets uploaded to e.g. \emph{figshare}, please use the \verb|howpublished| option in the bib entry to specify the platform and the link, as in the \verb|Hao:gidmaps:2014| example in the sample bibliography file.

% \section*{Author contributions statement}

% %Must include all authors, identified by initials, for example:
% A.A. conceived the experiment(s),  A.A. and B.A. conducted the experiment(s), C.A. and D.A. analysed the results.  All authors reviewed the manuscript. 

\end{document}